\documentclass[reprint,amsmath,amssymb,aps,twocolumn]{revtex4}

\usepackage{graphicx,color,setspace}
\usepackage[utf8]{inputenc}
\usepackage{mathtools}
\usepackage[normalem]{ulem}
\usepackage{tabularx}
\usepackage{enumerate}%
\usepackage{float}
\usepackage[normalem]{ulem}

\usepackage[caption=false]{subfig} %

\usepackage{ragged2e} %
\DeclareCaptionJustification{justified}{\justifying}

\usepackage{pifont}
\let\oldding\ding%
\renewcommand{\ding}[2][1]{\scalebox{#1}{\oldding{#2}}}%

\usepackage[labelfont=bf]{caption}

\begin{document}

\title{Engineering azeotropy to optimize the self-assembly of colloidal mixtures}

\author{Camilla Beneduce$^{1}$,Francesco Sciortino$^{1}$, Petr \v{S}ulc$^{2,3}$, John Russo$^{1}$}

\affiliation{$^{1}$ Dipartimento di Fisica, Sapienza Universit\`{a} di Roma, P.le Aldo Moro 5, 00185 Rome, Italy \\ $^{2}$ School of Molecular Sciences and Center for Molecular Design and Biomimetics, The Biodesign Institute, Arizona State University, 1001 South McAllister Avenue, Tempe, Arizona 85281, USA \\ $^{3}$ Life and Medical Sciences Institute (LIMES), University of Bonn, Bonn, Germany}

\begin{abstract}
The goal of inverse self-assembly is to design inter-particle interactions capable of assembling the units into a desired target structure. The effective assembly of complex structures often requires the use of multiple components, each new component increasing the thermodynamic degrees of freedom and hence the complexity of the self-assembly pathway. In this work we explore the possibility to use azeotropy, \textit{\textit{i.e.}} a special thermodynamic condition where the system behaves effectively as a one-component system, as a way to control the self-assembly of an arbitrarily  number of components. Exploiting the mass-balance equations we show how to select patchy particle systems that exhibit azeotropic points along the desired self-assembly pathway. As an example we map the phase diagram of a binary mixture that, by design, fully assembles into cubic (and only cubic) diamond crystal \textit{via} an azeotropic point.
The ability to explicitly include azeotropic points into artificial designs opens  novel pathways to   the self-assembly of complex structures.
\end{abstract}

\maketitle

\section*{}
When interactions between particles in a dilute fluid phase have strength comparable or larger than the thermal energy, the fluid becomes unstable and the particles condense searching for a lower free energy state. The spontaneous formation of inter-particle bonds, gives rise to aggregates whose final state can be either that of an ordered lattice, a connected percolating structure (\textit{e.g.} a liquid), or a collection of finite size clusters. When finite size or periodic structures are formed, this spontaneous search for the lowest free energy state is called {\it self-assembly\/}~\cite{whitelam2015statistical,kumar2017nanoparticle}.\\ 
While the computation of the free energy of a structure is a laborious but solved problem in statistical mechanics, several challenges hamper our understanding of self-assembly and our ability to mimic natural systems. In the {\it direct\/} self-assembly problem, one starts from a set of predetermined elementary units with known inter-particle interactions and is tasked with selecting structures that correspond to free energy minima. This is done either with intuition (for simple structures), with brute force approaches (direct molecular simulations), or with specialized algorithms~\cite{filion2009efficient,filion2009prediction}.
Even more challenging is the {\it inverse\/} self-assembly problem, where one is tasked with designing the inter-particle interactions that will self-assemble a desired target structure~\cite{jee2016nanoparticle,dijkstra2021predictive}. In this case the problems are two-fold: firstly designing an interaction-potential,  secondly   confirming that there are no alternative structures that preempt the formation of the target one~\cite{bupathy2022temperature}. So far, two types of approaches have been explored: optimisation algorithms and geometrical strategies. Optimisation algorithms allow one to design a pair potential whose free-energy minima is guaranteed to be the desired structure~\cite{rechtsman2005optimized,marcotte2011optimized,marcotte2013designeddiamond,zhang2013probing,miskin2016turning,lindquist2016communication,chen2018inverse,kumar2019inverse,dijkstra2021predictive,whitelam2021neuroevolutionary}. However, the inter-particle interactions that result from such procedures are often too complex and require a degree of precision that is out of reach for experimental realization.
In geometrical strategies, instead, one matches the geometric features of the target structure by tuning some interaction properties of the building units, \textit{e.g.} the shape and the directionality of the bonds, in order to match the geometric features of the target structure~\cite{ducrot2017colloidal,nelson2002toward,manoharan2003dense,zhang2005self,romano2014influence,halverson2013dna,romano2012patterning,tracey2019programming}. Although it is an experimentally feasible approach, it is system specific and it requires a high degree of geometrical intuition.

A different solution strategy to the inverse self-assembly problem is to extend the number of building blocks, going from single component systems to multi-component mixtures, shifting the problem of designing complex single particle potentials to that of optimizing simpler (and more geometrical) interactions between multiple components~\cite{bupathy2022temperature,russo2022sat}. Extending the {\it alphabet\/} of building blocks, \textit{\textit{i.e.}} the number of components, lowers the degree of symmetry in the final structure, allowing for a considerable reduction in competing structures, and an easier assembly pathway towards the target design. Compared to single-component mixtures, and leaving experimental challenges aside, two major problems are introduced by the increase in the number of components: a combinatorial problem and a thermodynamic problem.

The {\it combinatorial problem\/} arises from the fact that each new component increases exponentially the space of possible solutions, and with that the computational time required to find a solution. To tackle it, advanced optimization algorithms are necessary, such as genetic algorithms~\cite{srinivasan2013designing} or machine learning techniques~\cite{whitelam2020learning,whitelam2021neuroevolutionary}. Some of us have recently introduced a novel approach called {\it SAT-assembly\/}~\cite{romano2020designing,russo2022sat}, which encodes the bond topology of the target structure into a system of Boolean equations (a  satisfiability problem commonly named SAT) whose solution  gives the interaction matrix between different patches. The sophistication of modern SAT solvers \cite{een2005minisat} allows to effectively tackle the combinatorial problem for complex assemblies, including open crystalline structures, photonic crystals, and clathrate structures.

The {\it thermodynamic problem\/} arises instead because, according to Gibbs rule of phases~\cite{akahane2016possible}, each component represents an additional  thermodynamic degree of freedom of the system, extending the phase behaviour phenomenology in ways that can interfere with the self-assembly pathway.
No general strategy to tackle this problem has so far been proposed.
Full phase diagram calculations are in fact very time-consuming, and are often avoided in multi-component systems due to their complexity.
The goal of this article is to show how to overcome the thermodynamic difficulties associated with the use of multicomponent mixtures, by explicitly encoding azeotropic points in the self-assembly designs of patchy particles. The azeotropic point is a point where the free-energy of the mixture can be written as that of an effective one-component system (see Supplementary Materials I for a concise explanation of azeotropy), a condition that ensures that coexisting phases will have the same concentration as the parent homogeneous system. The ability to explicitly include azeotropic points along the self-assembly pathways of these systems  represents an attractive strategy to tame the complexity in phase behaviour usually associated with multi-component mixtures. Some of the advantages of combining azeotropic behaviour with self-assembly are listed here. i) The ability to (considerably) increase the reaction rates of the self-assembly process by quenching the system in a region of (liquid-gas) metastability:  in fact, it is well-established that for one-component systems nucleation rates increase in proximity of density fluctuations like the ones found near liquid-gas critical points~\cite{wolde1997enhancement} and spinodal loci~\cite{xu2012homogeneous}.  ii) Increase the kinetics of the self-assembly reaction: if the concentration of the azeotropic point is the same as the crystal composition, one can avoid slow diffusion-limited process, where the crystal nucleus has to wait for the concentration of the local environment to match the one of the target structure~\cite{russo2018glass}. iii) The yield of the self-assembly process can proceed theoretically until all components are exhausted (to 100\%), as the liquid phase will form at the same composition of the target crystalline structure.

In this article we will first show that it is indeed possible to effectively control the self-assembly of suitably  designed patchy particles by exploiting the encoded azeotropic properties. As a proof of concept, we then investigate in details a 
 binary mixture that is designed to form (only) the cubic diamond crystal.  This mixture also shows 
 a very interesting phase behaviour, where phase-separation only occurs for mixed states, and not for the pure components.

\begin{figure}[!t]
    \centering
    \includegraphics[width=0.33\textwidth]{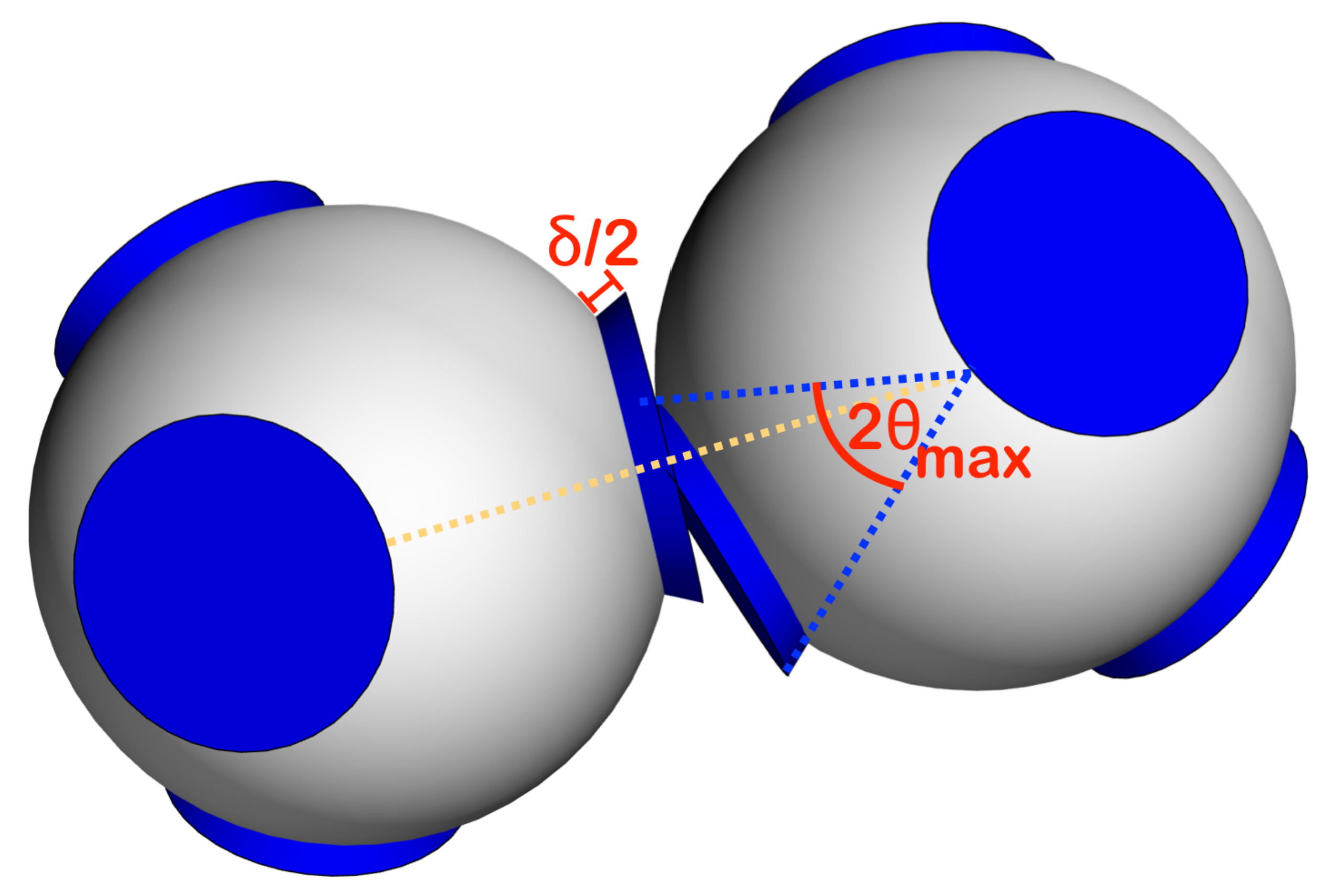}
    \caption{\textbf{Patchy particles schematisation.} Two patchy particles with four tetrahedrally arranged patches (in blue) interacting with the Kern-Frenkel potential defined in paragraph "Patchy Particles" in Materials and Methods section.}
    \label{fig_pp_kf}
\end{figure}

\section*{Results and discussion}
Our results pertain to systems whose components aggregate by forming bonds, \textit{\textit{i.e.}} to the vast class of associating systems~\cite{russo2021physics}. The main assumption is that the systems are in equilibrium, and that bond formation is controlled by a mass-balance equation.
We will propose general design rules that realize azeotropy in any system that satisfies these conditions. To demonstrate the effectiveness of our approach we will give a concrete example that considers mixtures of patchy particles (Fig.~\ref{fig_pp_kf}).
For these examples, thermodynamic properties will be computed both \textit{via} Wertheim’s first order perturbation theory\cite{wertheim1984fluids,chapman1988phase,de2011phase,bianchi2006empty,heras2011phase,rovigatti2013computing,seiferling2016percolation,teixeira2017phase,braz2021phase,russo2021physics}, and \textit{via} molecular simulations, both confirming the presence of the azeotropic point embedded in the phase diagram by design. In particular, to calculate the phase behaviour of the studied systems theoretically, we adopt the isochoric thermodynamic's framework while to study it numerically, we implement Monte Carlo simulations in the Gibbs ensemble. All these techniques are summarised in the Materials and Methods section.

In the rest of this article energy is measured in units of the square-well depth ($\epsilon$), distances in units of the patchy particle diameter ($\sigma$), pressure in units of $\epsilon/\sigma^3$ and $k_B=1$.

\subsection*{Law of mass action}
In deriving the azeotropy conditions we will make use of the law of mass action~\cite{chapman1988phase,heras2011phase,teixeira2017phase}, which quantifies the probability for a patch $\alpha$ to be non-bonded, and which we denote by $X_{\alpha}^i$, where the index $\alpha$ runs over all patches of species $i$

\begin{equation}
\label{eqn:X}
X_{\alpha}^{(i)}=\biggl[ 1+ \phi  \sum_{j=1,N_s} x^{(j)} \sum_{\gamma\in\Gamma(j)} X_{\gamma}^{(j)} \Delta_{\alpha\gamma} \biggr]^{-1}
\end{equation}

\noindent where $\Gamma(j)$ is the set of all patches in species $j$, and $\Delta_{\alpha\gamma}$ quantifies the strength of the interaction between patches $\alpha$ and $\gamma$. A detailed expression for $\Delta_{\alpha\gamma}$ is reported in the Methods section, but in the remainder we will consider the following simplification:
any pair of interacting patches forms bonds of the same type (bonding volume $V_b$ and energy $\epsilon$), so that $\Delta_{\alpha\gamma}=\Delta$ if $\alpha$ and $\gamma$ interact, or $\Delta_{\alpha\gamma}=0$ if they don't. We call 
 $\mathbf{\Upsilon}$ the {\it interaction matrix\/}, whose elements
$\Upsilon_{\alpha\gamma}=\Delta_{\alpha\gamma}/\Delta$ are ones if patches $\alpha$ and $\gamma$ interact, and zeros if they don't. By construction,   $\mathbf{\Upsilon}$ is a symmetric matrix (if patch $\alpha$ binds with patch $\gamma$, then also
patch $\gamma$ binds with patch $\alpha$).

One possible strategy to compute $\mathbf{\Upsilon}$, \textit{i.e.} to determine which pair of patches should interact, such that the particles will self-assemble into a desired structure,  is the {\it SAT-assembly\/} framework~\cite{russo2022sat}.
Here we focus on the general conditions one needs to impose on $\Upsilon_{\alpha\gamma}$ in order to obtain azeotropic mixtures, regardless of the desired target structure.

\subsection*{Azeotropy design rules}
We consider a $N_s$-component mixture with all species having the same diameter $\sigma$, the same number ($N_p$) and placements of patches, and differing only in the patches type (patches color). We first notice that a sufficient condition for azeotropy is obtained by imposing that all probabilities $X_{\alpha}^{(i)}$ in Eq.~\ref{eqn:X} are the same for all patches in the system, $X_\alpha^{(i)}=X$. In this way, all species will behave like an effective one-component system, where all bonds have the same probability to be formed. The same condition can be demonstrated to hold within Wertheim's perturbation theory: in paragraph "Wertheim perturbation theory" in Materials and Methods section we notice that the equality of all $X_{\alpha}^{(i)}$ implies that the Helmholtz bonding free energy (Eq.~\ref{eqn:betaf_bonding})  reduces to that of a one-component system.

In order to determine whether there is a thermodynamic point where all $X_{\alpha}^{(i)}$ have the same value, we turn to the mass balance condition, Eq.~\ref{eqn:X}, which is a set of $N_s\times N_p$ equations in the variables $X_{\alpha}^{(i)}$. Looking for the rules
under which all the mass balance equations become equivalent
provides a sufficient condition for the appearance of azeotropy in a multi-component mixture.

In the following we examine three families of rules that ensure azeotropy:

\begin{itemize}
\item{the \textit{bond exclusivity} condition. This rule  generates azeotropic points at equimolar conditions;}
\item{the \textit{bond multiplicity} condition. This rule  allows for azeotropic points at non-equimolar conditions;}
\item{the \emph{fully-connected bond} condition. This rule generates always-azeotropic mixtures, \textit{e.g.} where the concentration remains the same during demixing for every point in the coexistence region.}
\end{itemize}

\subsubsection*{Bond exclusivity condition}

\noindent One condition ensuring azeotropy is the \textit{bond exclusivity} constraint requiring that each patch has only one bonding partner (that can be itself in case of self-complementarity) among all patches of all species in the mixture. This implies that all patches are different and that  $\mathbf{\Upsilon}$  has a single one for each row, located at a different column for different rows. 
This  condition, with its  symmetric bonding rules, can be realized when 
  two species of particles are functionalized with complementary DNA strands, a system which has found great success in nanotechnology~\cite{nykypanchuk2008dna,park2008dna}.

We consider here the case where all bonds have the same bonding energy such that azeotropy appears at equimolar conditions: a $N_s$-component mixture will be azeotropic if it is prepared by mixing all the $N_s$ components at the equimolar concentration $1/N_s$.
To see this, we note that the bond exclusivity condition implies that the sum over the patches ($\sum_{\gamma\in\Gamma(j)}$) and the sum over the species ($\sum_{j=1,N_s}$) in Eq.~\ref{eqn:X} reduce to a single contribution since the patch $\alpha$ belonging to species $i$ can interact only with its partner patch $\gamma$ belonging to species $j$ ($j$ can be also equal to $i$ as well as $\alpha$ can be equal to $\gamma$). Therefore the $N_s\times N_p$ mass balance equations for $X_{\alpha}^{(i)}$ reduce all to equations of the form

\begin{equation}
\label{eqn:X_new}
X_{\alpha}^{(i)}=\biggl[ 1+\phi x^{(j)}X_{\gamma}^{(j)} \Delta_{\alpha\gamma}  \biggr]^{-1}
\end{equation}   
which couple only $X_{\alpha}^{(i)}$ with $X_{\gamma}^{(j)}$.
Moreover, by designing bonds with the same strength,   $\Delta_{\alpha \gamma}\equiv\Delta$ for all patches $\alpha$ and $\gamma$. 
By considering the pair of equations for $X_{\alpha}^{(i)}$ and  $X_{\gamma}^{(j)}$ 
one obtains, without knowing the exact patchy particles design,  that 
the $N_s\times N_p$ mass balance equations become all equivalent to
\begin{equation}
\label{eqn:XN}
	X_{\alpha}^{(i)}+ \phi \; x^{(i)}[X_{\alpha}^{(i)}]^{2} \Delta +\phi  (x^{(j)}-x^{(i)})X_{\alpha}^{(i)}\Delta -1=0
\end{equation}
\noindent With the equimolarity condition, $x^{(i)}=1/N_s$, the $N_s\times N_p$ equations above admit the azeotropic solution $X_\alpha^{(i)}=X$, where $X$ is the solution of

\begin{equation}
\label{eqn:XN_fin}
X+ \frac{\phi}{N_s} X^{2} \Delta-1=0
\end{equation}

Thus the bond exclusivity condition  generates an azeotrope at equimolar concentration, which can be exploited in self-assembly designs where the target structure is composed of an equal number of all species. An example of interaction matrix satisfying the bond exclusivity condition is given in the next section, where we will verify explicitly the presence of an equimolar azeotropic point not only with Wertheim's thermodynamic theory, but also explicitly with Monte Carlo simulation of a patchy particle realization of the interaction matrix.

The bond exclusivity condition is easily generalized to cases where multiple-bonding is allowed (one patch capable of bonding to more than one patch, a case which can be realized with DNA functionalization, as explained in Supplementary Materials IV
and/or when the patches are not distinct (when the interaction matrix has repeated columns or rows, \textit{i.e.} when its determinant is zero). In these cases, to have an equimolar azeotropy conditions one needs to ensure that every patch has the same total number ($m$) of bonding partners (distributed over one or more species). In this case the mass-balance equation admits the solution $X_\alpha^{(i)}=X$ (azeotropy), with $X$ satisfying the following equation

\begin{equation}
\label{eqn:XN_fin_multi}
X+\phi\frac{m}{N_s} X^{2} \Delta-1=0
\end{equation}

\subsubsection*{Bond multiplicity condition}
A simple generalization of the bond exclusivity condition allows to move the azeotropic point to off-equimolar conditions. Considering a binary mixture where the ratio between the two species (denoted as (1) and (2)) is $1:n$, in order to have an azeotrope at $x^{(2)}=nx^{(1)}$
(\textit{i.e.} $x^{(1)}=1/(n+1)$ and $x^{(2)}=n/(n+1)$)
it is sufficient to enforce
\begin{itemize}
\item bond exclusivity to all patches bonding to species (2), \textit{i.e.} each patch has a unique bonding partner with species (2)
\item $n$-bond multiplicity to all patches bonding to species (1), \textit{i.e.} each patch has $n$ bonding partners with species (1).
\end{itemize}

With these conditions, all mass balance equations, Eq.~\ref{eqn:X}, admit the azeotropic solution $X_\alpha^{(i)}=X$ with $X$ satisfying the following equation

\begin{equation}
X+ \frac{n}{1+n} \phi X^{2} \Delta-1=0
\end{equation}

The \textit{bond multiplicity} rule is a generalisation of the previous bond exclusivity case, that we recover if $n=1$.
This recipe is generalisable to multi-component mixtures with more than two species: the \textit{bond multiplicity} condition will require to establish a bond with $m$ patches belonging to certain species, where $m$ is the least common multiple between component ratios.

\noindent An explicit example of a binary system of patchy particles with bond multiplicity is reported in Supplementary Materials II.

In short bond multiplicity provides a way to shift the azeotropic point at a concentration different from the equimolar one. However we underline that, with the presented rules, once the number of species and of patches is set, it is not possible to design a mixture exhibiting azeotropy at arbitrary concentration. For instance for a binary mixture, with four patches tetrahedrally arranged, there is no design satisfying our bonding rules for the ratio $1:3$. More general conditions can be built by lifting the requirement that all bonds have the same energy, $\Delta_{\alpha\gamma}\neq\Delta$, but bearing in mind that a fine control over bonding energies represents a significant experimental challenge.

\subsubsection*{Fully-connected bond condition}

The \textit{fully-connected bond} condition introduces bonding rules that ensure full azeotropy at all concentrations without the need to tune bonding energies. In this case the concentration of the two coexisting phases is always constant during demixing.
For a general $N_s$-component mixture of patchy particles with $N_p$ patches, the \textit{fully-connected bond} condition is achieved
when each patch can bind with $N_s$ patches, each located on a different species. In this case the
 sum $\sum_{\gamma\in\Gamma(j)}$  in 
 the mass balance equation (Eq.~\ref{eqn:X}) drops out, as there is only one bonding partner on each species, becoming

\begin{equation}
X_{\alpha}^{(i)}=\biggl[ 1+ \phi  \sum_{j=1,N_s} x^{(j)} X_{\beta}^{(j)} \Delta_{\alpha\beta_j} \biggr]^{-1}
\end{equation}
\noindent where the patch $\beta$  on particle $j$ is the unique bonding partner  of patch $\alpha$ on species $i$. Now, assuming that all bonds are of the same type, $\Delta_{\alpha\gamma}=\Delta$, and remembering that $\sum_j x^{(j)}=1$, we see that the mass balance equation admits azeotropic solutions $X_\alpha^{(i)}=X$ where $X$ satisfies

\begin{equation}
X+X^2 \phi \Delta -1=0
\end{equation}
 
A possible interaction matrix for 
a binary mixture satisfying the fully-connected bond condition is reported in Supplementary Materials III and a DNA implementation in Supplementary Materials IV.

\begin{figure}[!t]\centering
\subfloat[]{\includegraphics[width=0.22\textwidth]{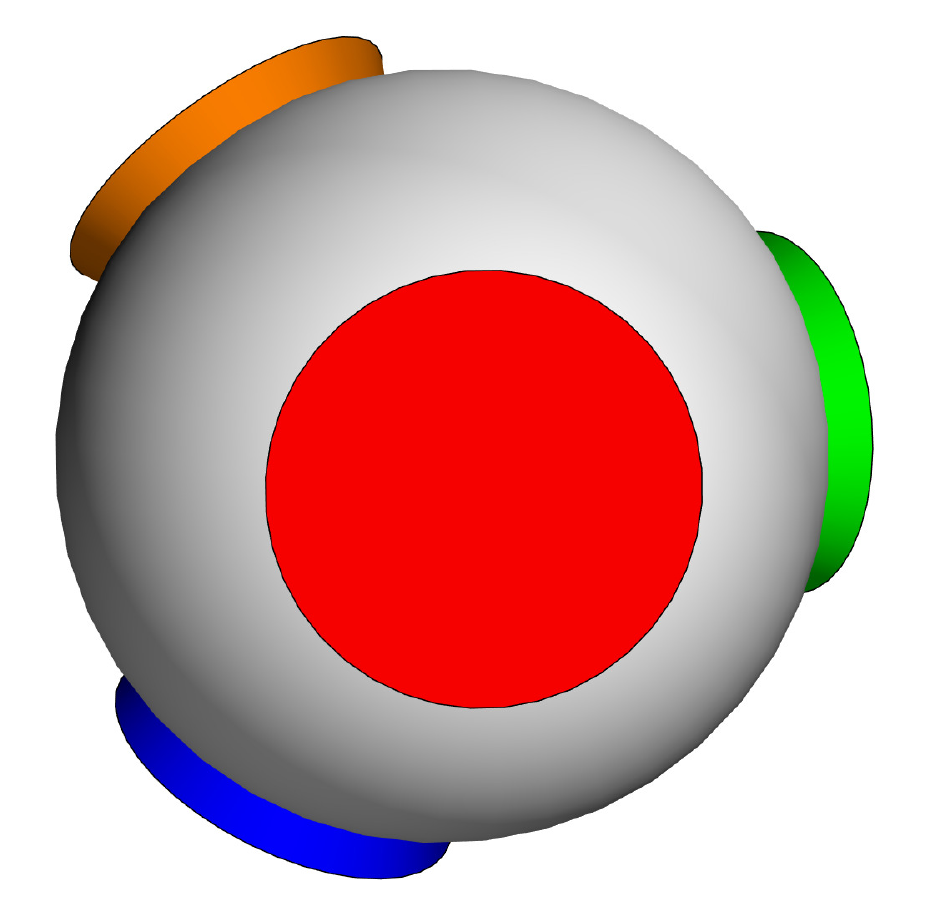}\label{fig:pp1}}
\hfil
\subfloat[]{\includegraphics[width=0.22\textwidth]{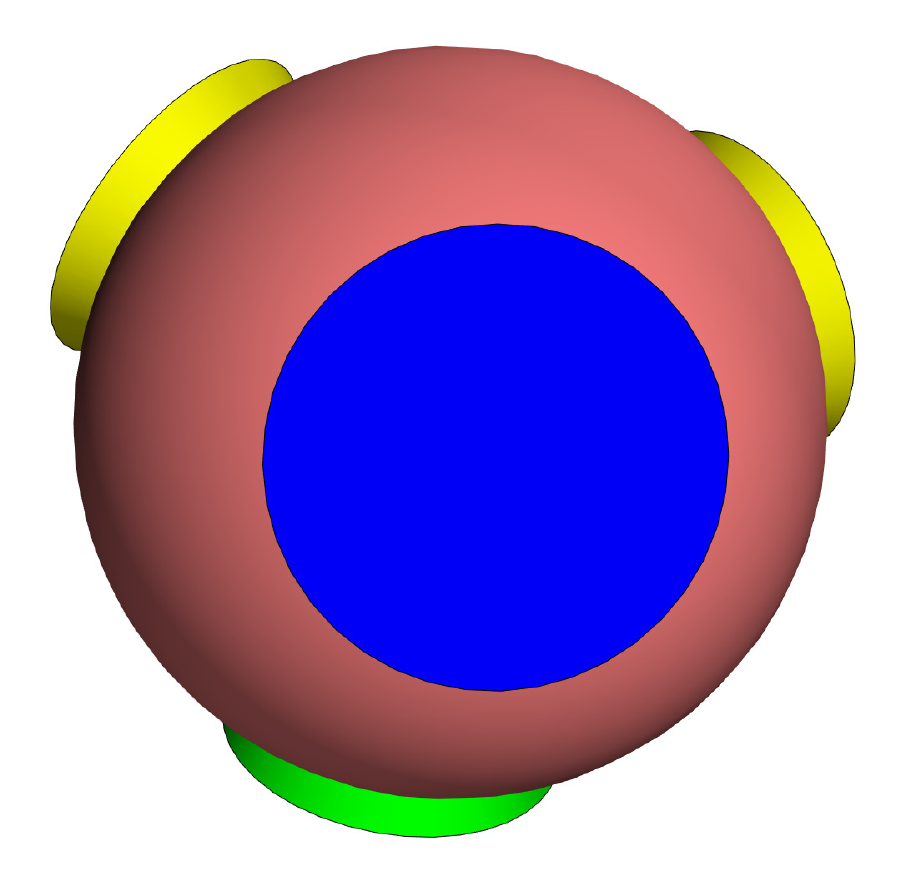}\label{pp2}}
 \caption{\textbf{3D representation of the two patchy particles species (a) and (b) of the SAT-designed N2c8s2 binary mixture.} Equal patches colors indicate which patches can bind to each other and the colors appearing only once are assigned to self-interacting patches.} 
\label{fig:pp_CD}
\end{figure}

\subsection*{Application to cubic diamond crystals}

One of the most interesting and challenging bottom-up realisation of a target structure is that of the cubic diamond~\cite{liu2016diamond,he2020colloidal}. Realizing a cubic diamond on colloidal scale  opens up the possibility of creating  a photonic crystal that allows for light manipulation in a controlled way~\cite{ho1990existence,soukoulis2010optical,ngo2006tetrastack}. The self-assembly of a cubic diamond is complex since its lattice is an open structure which competes with the hexagonal diamond structure, which prevents the cubic diamond from forming without defects such as stacking faults~\cite{romano2011crystallization}. Several studies have been performed to overcome these difficulties~\cite{romano2011crystallization,neophytou2021facile}, including solutions obtained within the SAT-assembly framework~\cite{romano2020designing,russo2022sat}. %
Because of the topology of the cubic diamond lattice, patchy particles of valence four with a tetrahedral  arrangement of the patches are used to self-assemble the crystal. 
The minimal SAT-designed solution (the one requiring the smallest number of distinct particles) is the so called N2c8s2 binary mixture~\cite{rovigatti2022simple} that uses two species (N2), eight patches types (colors) (c8) and two self-interacting colors (s2) and it is schematised in Fig.~\ref{fig:pp_CD} where colors identify the interacting (and not the different) patches according to the interaction matrix $\mathbf{\Upsilon}$. Note that the number of self-interacting colors is also the trace of the matrix $\mathbf{\Upsilon}$.

\noindent The N2c8s2 interaction matrix, encoding the design with 2 species and 8 distinct patches (or colors), is

\begin{equation}
\label{eqn:matrix_N2c8_phase}
\mathbf{\Upsilon}_\text{N2c8s2}=
	\begin{pmatrix}
	0 & 0 & 0 & 0 & 0 & 0 & 0 & 1\\
	0 & 1 & 0 & 0 & 0 & 0 & 0 & 0\\
	0 & 0 & 1 & 0 & 0 & 0 & 0 & 0\\
	0 & 0 & 0 & 0 & 0 & 1 & 0 & 0\\
	0 & 0 & 0 & 0 & 0 & 0 & 1 & 0\\
	0 & 0 & 0 & 1 & 0 & 0 & 0 & 0\\
	0 & 0 & 0 & 0 & 1 & 0 & 0 & 0\\
	1 & 0 & 0 & 0 & 0 & 0 & 0 & 0
\end{pmatrix}
\bigskip
\end{equation}

\noindent We notice that having a single one for each row, the bond exclusivity condition is satisfied, and we thus expect to find an azeotrope line at concentration $x^{(1)}=x^{(2)}=1/2$. The N2c8s2 mixture is thus an ideal candidate to test the appearance of azeotropy, and to investigate in detail its self-assembly properties.

\begin{figure}[!t]\centering
\subfloat{\includegraphics[width=0.44\textwidth]{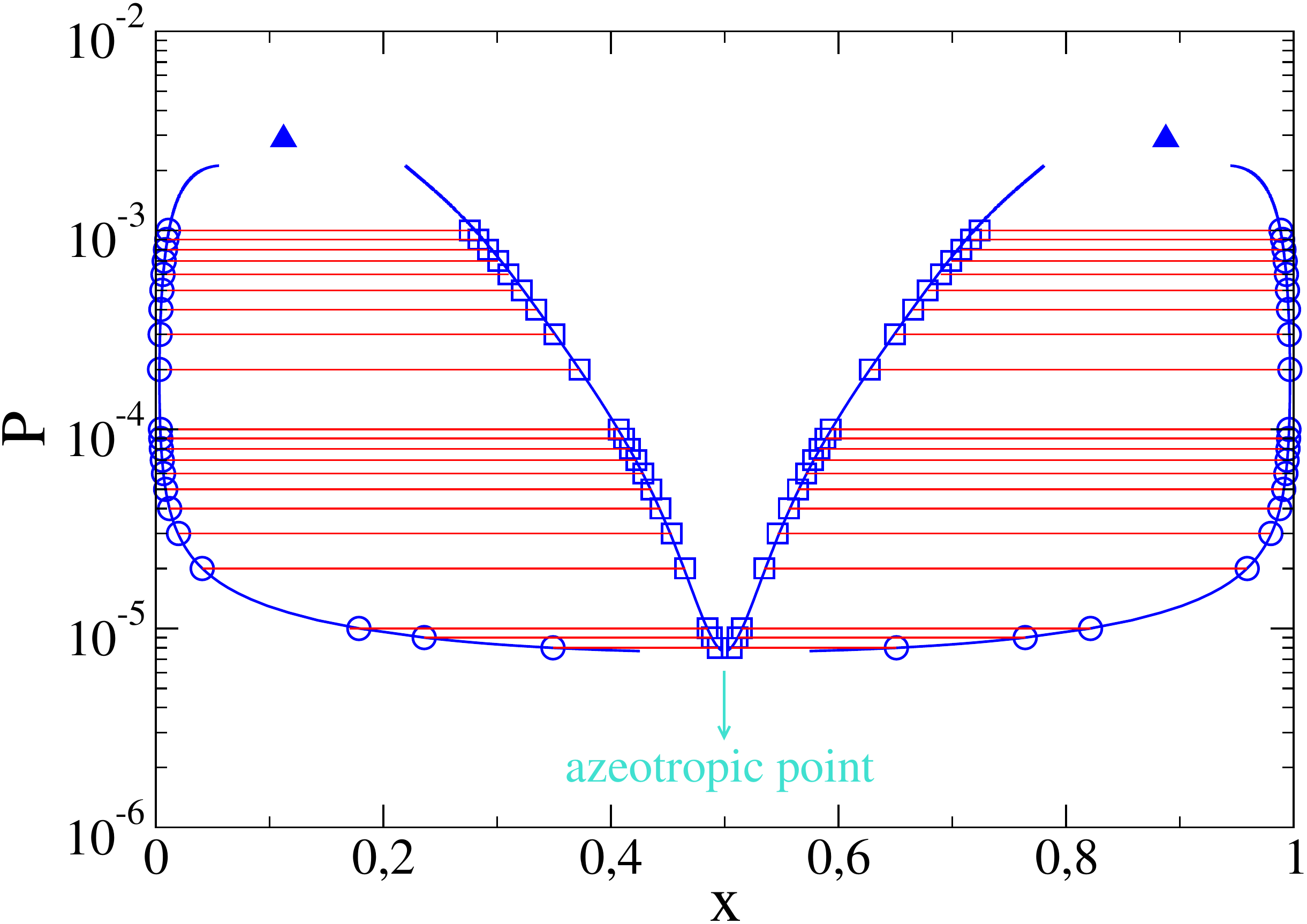}\llap{
  \parbox[b]{2.65in}{\textbf{\footnotesize{(a)}}\\\rule{0ex}{1.85in}
  }}\label{fig_coexistence_007}}
\hfill
\subfloat{\includegraphics[width=0.49\textwidth]{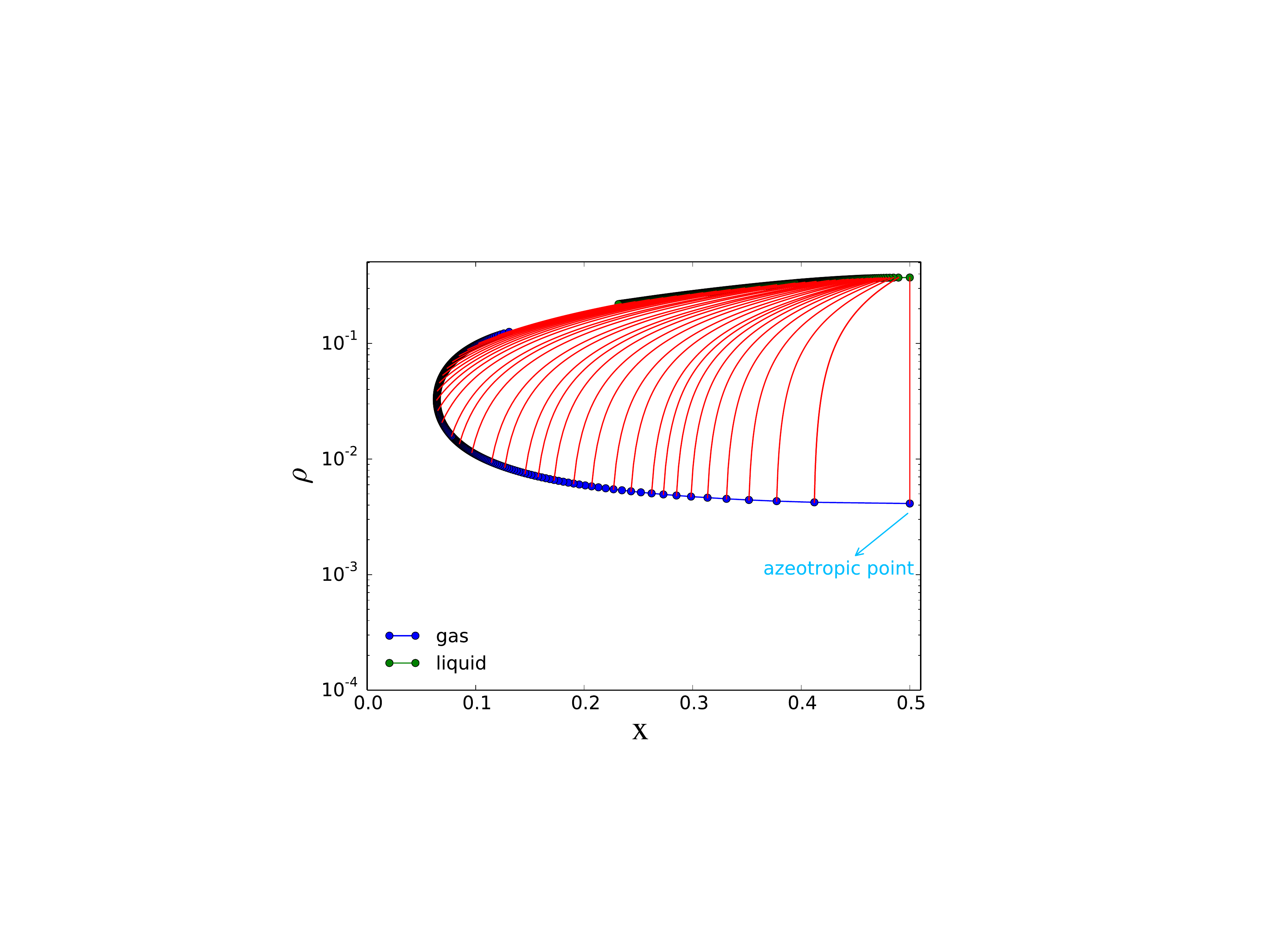}\llap{
  \parbox[b]{2.9in}{\textbf{\footnotesize{(b)}}\\\rule{0ex}{2.15in}
  }}\label{fig_coexistence_008_rho}}
 \caption{\textbf{Wertheim pressure-concentration (\textbf{a}) and density-concentration (\textbf{b}) phase diagrams for the N2c8s2 SAT-designed binary mixture.} The \textbf{(a)} phase diagram is computed at temperature $T=0.07$ while the \textbf{(b)} one at temperature $T=0.08$. In A circles and squares, connected by red tie-lines, represent the coexistence points obtained from the common tangent construction on the Gibbs free energy curve. Blue lines indicate the binodal curve computed by numerically integrating Eq.~\ref{eqn:rho_T}. Triangles are at the location of binary critical points. In \textbf{(b)} the only vertical tie-line is the one at the azeotropic concentration: only a binary mixture prepared in a homogeneous phase at the azeotropic concentration retains the original ratio between components when it phase separates. Tie-lines are not straight since the density axis is in logarithmic scale.} 
\label{fig:wertheim_pd}
\end{figure}

\noindent In order to verify the effective presence of an azeotrope when the two species are mixed at equal ratio, we first use Wertheim's theory~\cite{wertheim1984fluids} to determine the binodal curve in pressure-concentration and density-concentration phase diagrams. The thermodynamic conditions for a stable state of the mixture at constant pressure and temperature 
are found when the Gibbs free energy per particle $g$ has a minimum.  $g$, the Laplace transform of the Helmholtz free energy per particle $f$,  is defined as

\begin{equation}
\label{eqn:g}
g=\frac{P}{\rho}+f 
\end{equation}

\noindent where $P$ is the pressure and $\rho$ is the total number density. Since the same total density can be achieved by mixing species at more than one pair of concentrations $x_1\equiv x$ and $x_2=1-x_1$, firstly we must minimise $g$ for each fixed concentration $x$ with respect to the density $\rho$. In this way the Gibbs free energy becomes only a function of concentration.  Coexisting phases having the same temperature, pressure and chemical potential can be obtained by searching those points on $g(x)$ that are connected by a common tangent~\cite{de2011phase}.
Starting from a single pair of coexistence points found with the common tangent rule,
we use the isochoric thermodynamics equations (paragraph "Isochoric thermodynamics" in Materials and Methods section) to trace the coexistence lines as a function of concentration and pressure. For all the following numerical calculations we fix the  potential parameters to the values $\cos\theta_{max}=0.98$ and $\delta=0.2$. This choice follows from previous studies demonstrating that the nucleation is facilitated at small apertures of the angle $\theta_{max}$~\cite{romano2011crystallization,smallenburg2013limited,russo2021physics}.

\begin{figure}[!t]
    \centering
    \includegraphics[width=0.45\textwidth]{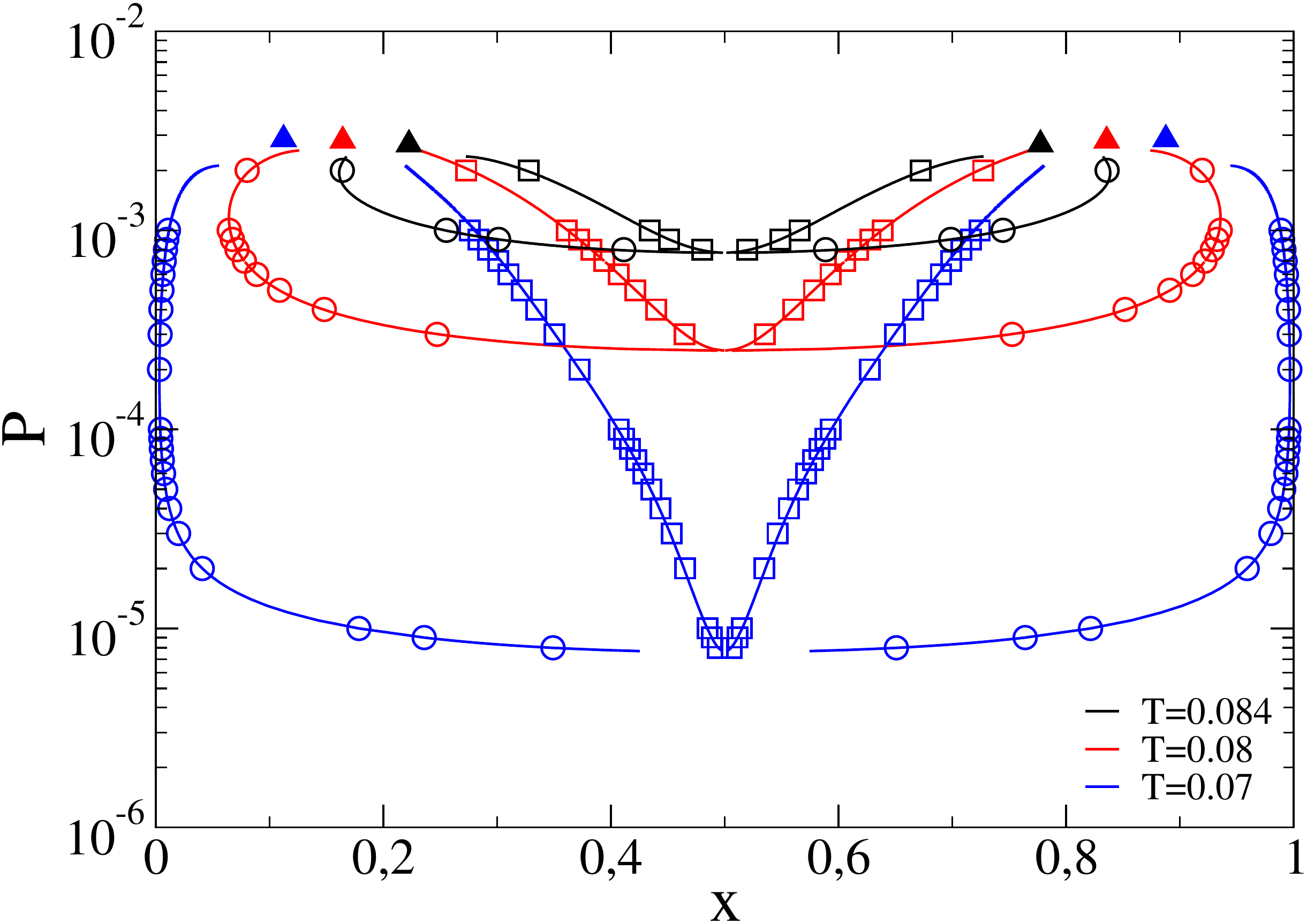}
    \caption{\textbf{Comparison of Wertheim pressure-concentration phase diagrams for the N2c8s2 SAT-designed binary mixture at temperatures $\mathbf{T=0.07}$, $\mathbf{T=0.08}$ and $\mathbf{T=0.084}$.} Circles represent points belonging to the dew point curve. Squares represent points belonging to the bubble point curve. Triangles indicate critical points.}
    \label{fig_coexistence}
\end{figure}

The pressure composition phase diagram obtained at $T=0.07$ is shown in Fig.~\ref{fig_coexistence_007}. This phase diagram confirms that the N2c8s2 has an azeotropic point at concentration equal to $x=0.5$: it is exactly at $x=0.5$ that the bubble point curve (where the  vapour phase first appears when pressure is lowered starting from a point greater than the total vapour pressure~\cite{smith1949introduction}) and the dew point curve (where the  liquid phase first originates when pressure in increased starting from a point in the vapour phase~\cite{smith1949introduction}) are tangent and the coexistence region reduces to a point. Moreover since the azeotrope is at the lower extremum in the pressure-concentration phase diagram, the N2c8s2 binary mixture is a negative azeotropic binary mixture~\cite{smith1949introduction,Moore1962Physical}.

In Fig.~\ref{fig_coexistence_008_rho} we plot the coexistence  region in the  density-concentration phase diagram. The azeotropic nature of the solution with $x=0.5$ is evident from the slope of the tie-lines: only at $x=0.5$ the tie-line is a vertical line indicating that only if the binary mixture is prepared by mixing together an equal concentration of the two species, then the coexisting phases will preserve the same concentration.

Unexpectedly, the shape of the coexistence regions in the $P-T$ plane (sometime called the phase diagram "topology"~\cite{smith1949introduction,Moore1962Physical}) indicates that the N2c8s2 mixture belongs to new type of binary phase diagram, in which the pure components ($x=0$ and $x=1$) do not have a liquid-gas transition but their mixture does. Fig.~\ref{fig_coexistence} shows that decreasing temperature the coexistence region becomes larger without ever crossing the limit concentrations $x=0$ and $x=1$.
The topology of the phase diagram is equivalent to that of an ordinary azeotropic binary mixture, but in which the binary critical point line goes to $(P,T)\rightarrow 0$ as the concentration goes to $x\rightarrow 0$ or $x\rightarrow 1$.

\begin{figure}[!t]
\subfloat[]{\includegraphics[width=0.22\textwidth]{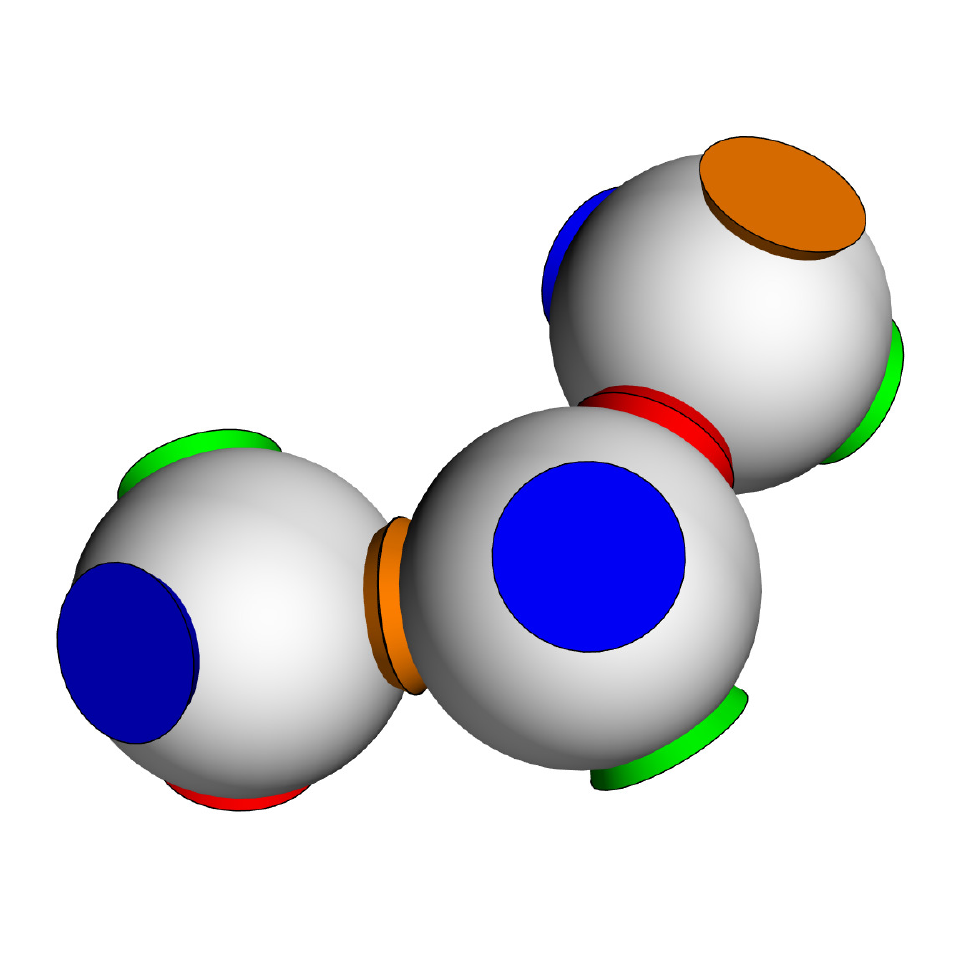}\label{chain_1}}
\hfil
\subfloat[]{\includegraphics[width=0.22\textwidth]{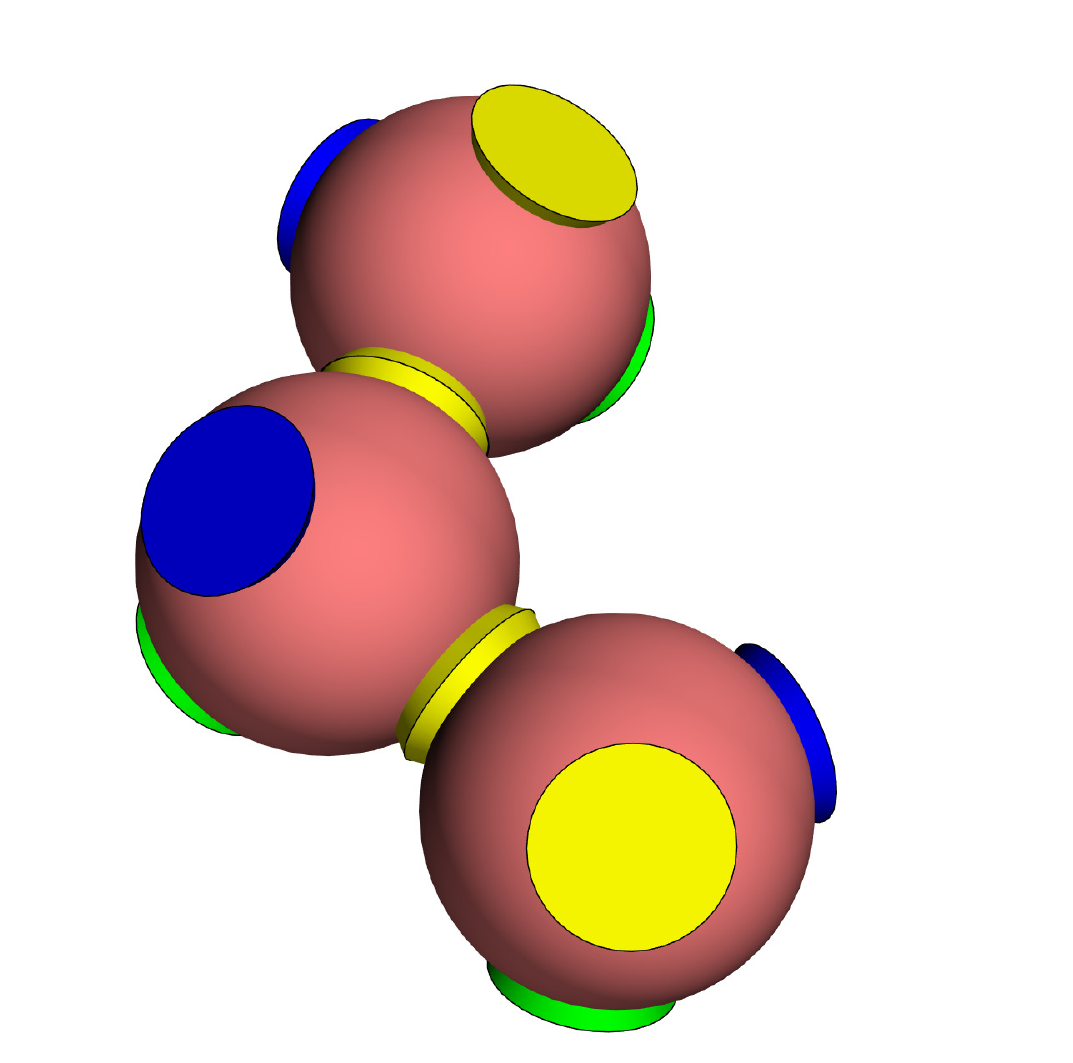}\label{chain_2}}   
\caption{\textbf{In single component systems only chain aggregates can form.} If the SAT-designed N2c8s2 binary mixture becomes a single component system, composed either just by patchy particles of the first species \textbf{(a)} or just by patchy particles of the second species \textbf{(b)}, patchy particles can aggregate only forming chains, \textit{i.e.} they behave like patchy particles with valence two.
} 
\label{fig:ppp}
\end{figure}

\noindent This unconventional behaviour is originated by the fact that  patchy particles of the same species can bind to each other with no more than two bonds, as encoded in the 
 interaction matrix~ (Eq.~\ref{eqn:matrix_N2c8_phase}). Hence, in pure component conditions, particles can aggregate only into chains as depicted in Fig.~\ref{fig:ppp}. Therefore even if particles have four patches, when $x=0$ or $x=1$ they behave like bi-functional particles and hence have no liquid-gas phase separation~\cite{bianchi2006empty}. We note that the idea that systems with two-patches have a hidden critical point at $P=0$ and $T=0$  has been recently revisited in Ref.~\cite{stopper2020remnants}, and generalized to colored patches in Ref.~\cite{tavares2020remnants}.

\begin{figure}[!t]
\centering
\subfloat{\includegraphics[width=0.44\textwidth]{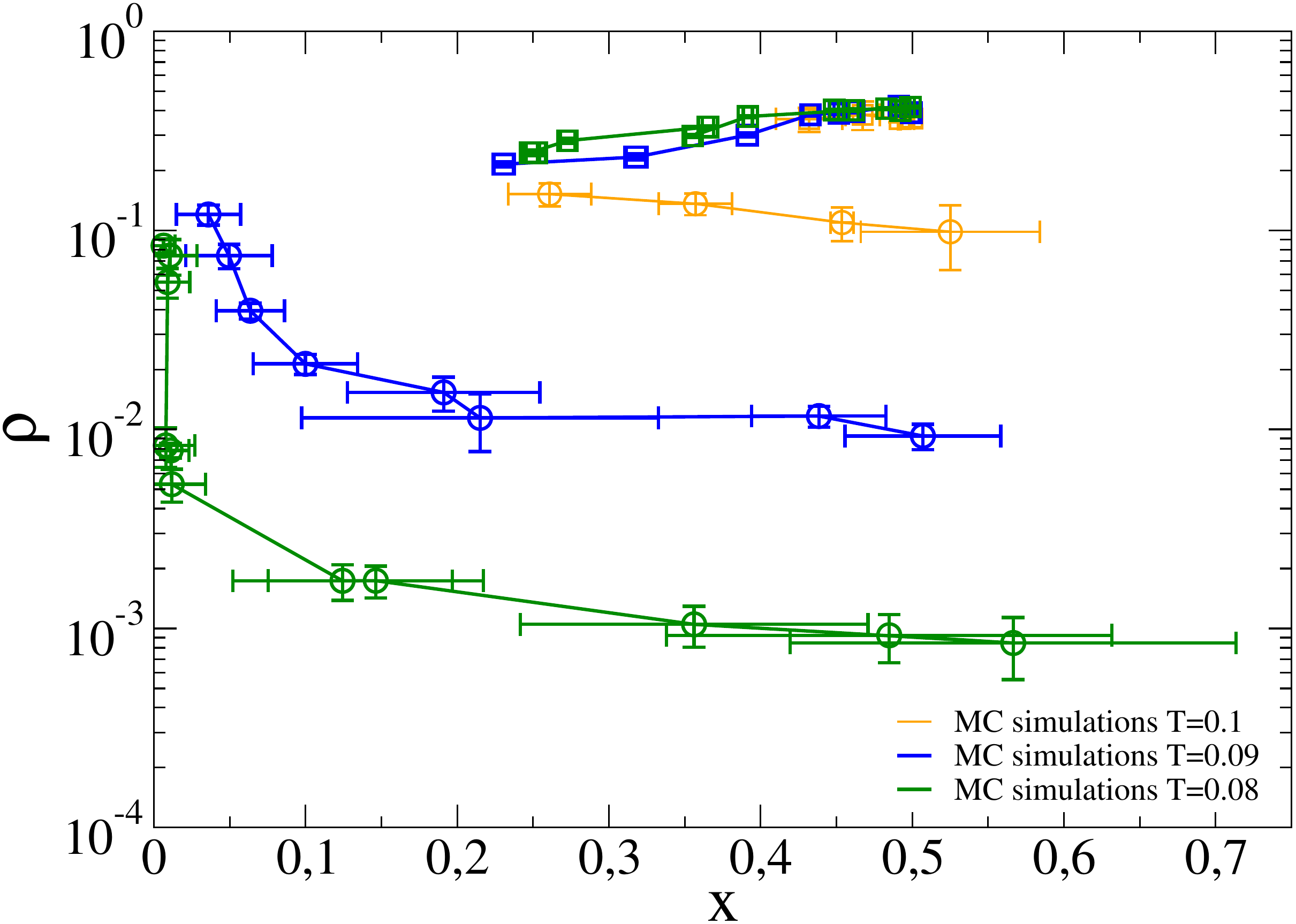}\llap{
  \parbox[b]{2.65in}{\textbf{\footnotesize{(a)}}\\\rule{0ex}{1.84in}
  }}\label{fig:confrontoMC}}
\hfil
\subfloat{\includegraphics[width=0.45\textwidth]{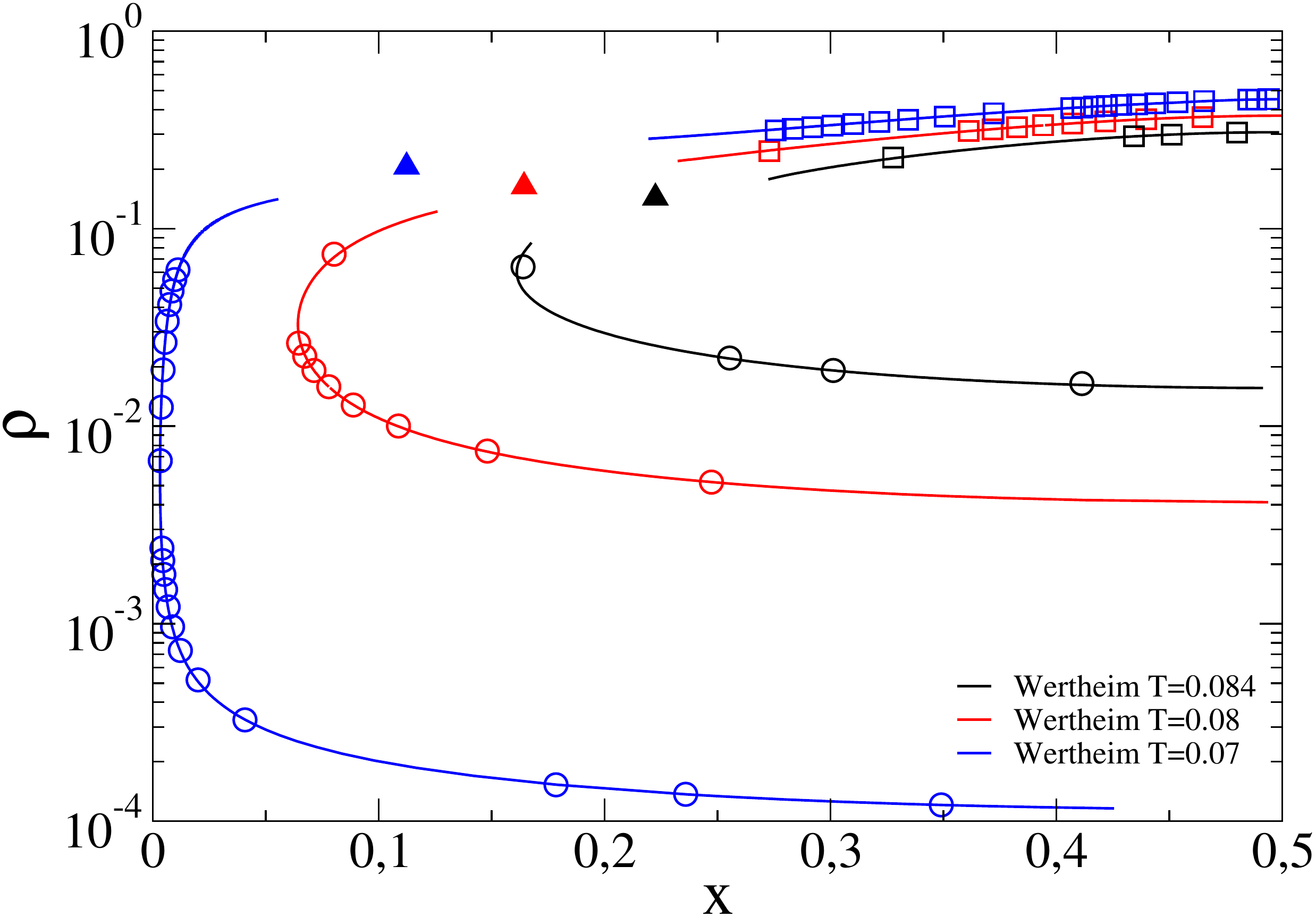}\llap{
  \parbox[b]{2.71in}{\textbf{\footnotesize{(b)}}\\\rule{0ex}{1.84in}
  }}\label{fig:confronto_w}}  
\caption{\textbf{SAT-designed N2c8s2 binary mixture density-concentration phase diagrams for different temperatures.} Comparison between the binodal curves obtained from Monte Carlo simulations \textbf{(a)} and the binodal curves computed within the Wertheim first order perturbation theory \textbf{(b)}. Circles represent points belonging to the dew point curve, while squares represent points belonging to the bubble point curve. Triangles indicate critical points.} 
\label{fig:confronto}
\end{figure}

Going beyond Wertheim's theory, we study the numerical phase behaviour of the N2c8s2 mixture \textit{\textit{via}} Monte Carlo simulations in the Gibbs ensemble. Simulations are performed at different temperatures ($T=0.1, T=0.09, T=0.08$) and, for each temperature, at different averaged (over the two boxes) densities and concentrations in order to compute the binodal curve in the 
density-concentration phase diagram, as shown in Fig.~\ref{fig:confrontoMC}. System size is fixed at $N=500$ particles for all simulations.

Equilibration of these systems at the (low) temperatures where phase separation is located is particularly challenging~\cite{rovigatti2018simulate}: this is reflected in the non-negligible error bars in  Fig.~\ref{fig:confrontoMC}. Nevertheless, the trend of the numerical computed binodal curves as well as the topology of the density-concentration phase diagrams are the same of the Wertheim ones as shown
in Fig.~\ref{fig:confronto_w}.
As commonly observed~\cite{russo2021physics}, Wertheim's theory tends to overestimate the size of the coexistence region. Still,  Monte Carlo simulations confirm the phase diagram topology with the presence of an azeotrope at  concentration $1/2$ in the N2c8s2 binary mixture.

\begin{figure}[!t]
    \centering
    \subfloat{\includegraphics[width=0.43\textwidth]{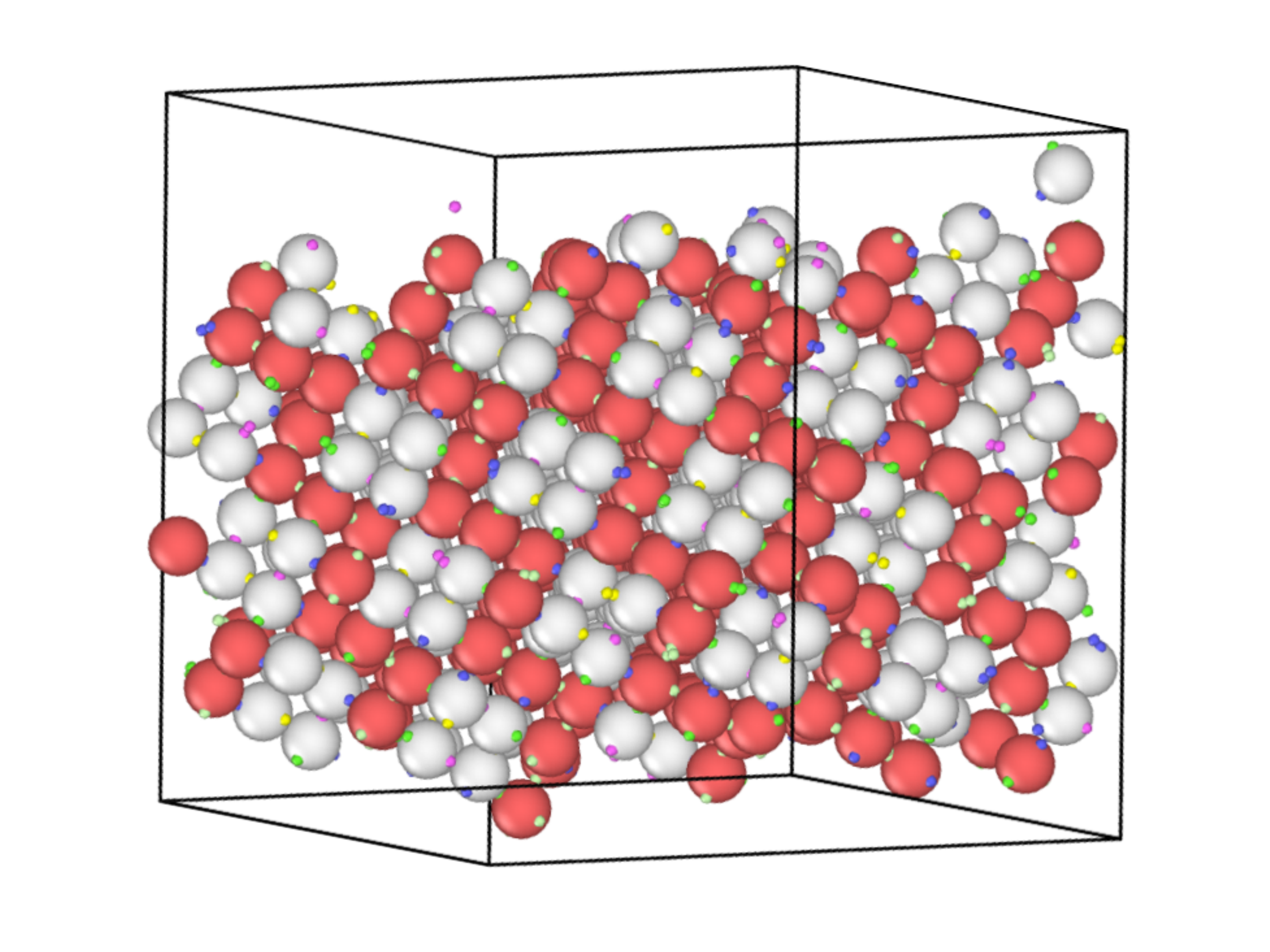}\llap{
  \parbox[b]{2.55in}{\textbf{\footnotesize{(a)}}\\\rule{0ex}{1.75in}
  }}\label{fig:dc_snap}}
\hfil
    \subfloat{\includegraphics[width=0.45\textwidth]{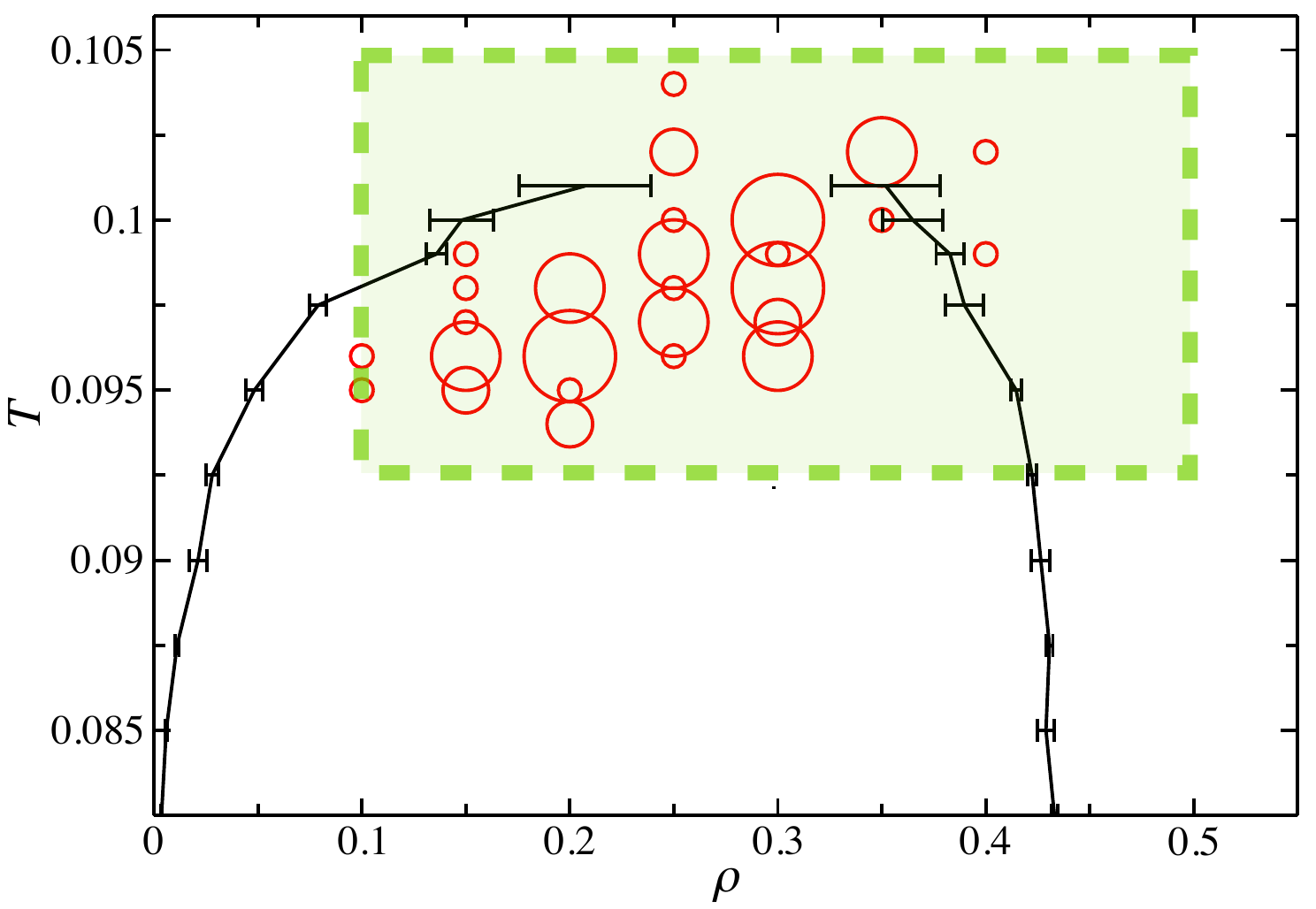}\llap{
  \parbox[b]{2.68in}{\textbf{\footnotesize{(b)}}\\\rule{0ex}{1.84in}
  }}\label{fig:nucl_plot}} 
    \caption{\textbf{Nucleation plots.} \textbf{(a)} Snapshot from a fully self-assembled solution, prepared from a random configuration at $T=0.1$ and $\rho=0.3$, with patchy parameters fixed at $\theta_{max}=0.98$ and $\delta=0.2$. Patchy particles are colored red or white according to their species. \textbf{(b)} $T-\rho$ phase diagram obtained from Gibbs ensemble simulations (black lines). The red circles are drawn in correspondence of the state points which nucleated. The radius of the red circles is proportional to the fraction of runs that successfully assembled within the simulation time of up to $5\,10^8$ MC sweeps.}
    \label{fig_assembly_runs}
\end{figure}

Next, we study the self-assembly process through the azeotropic point. We prepare disordered configurations at equimolar composition for different state points on a regular grid, with $\rho\in[0.1,0.5]$ and $\Delta\rho=0.05$,  $T\in[0.920,0.104]$ and $\Delta T=0.002$. For each $(\rho,T)$ state point we run $5$ independent trajectories in the NVT ensemble with $AVB$ biased moves~\cite{chen2000novel} (paragraph "Monte Carlo simulations: AVB moves and Gibbs ensemble" in Materials and Methods section). The state points considered are enclosed in the green shaded area in Fig.~\ref{fig:nucl_plot}, and each trajectory is run for $5\,10^8$ MC sweeps or until crystallization. The centers of the red circles in Fig.~\ref{fig:nucl_plot} represent the state points which crystallized within the simulation time. The diameter of each circle is proportional to the fraction of simulation runs (out of a total of $5$ runs) that have crystallized at the corresponding state point. To understand why crystallization occurs only at selected state points, we superimpose (black line) the results from Gibbs Ensemble simulations that have been initialized at equimolar conditions. Error bars are computed on 10 independent runs for each temperature, and the black lines connecting the points are guides to the eyes to help identifying the gas and liquid branches. We confirmed that once phase separation has occurred, both boxes (liquid and gas) are still found at equimolar concentration for all temperatures, \textit{i.e.} we are always at azeotropic conditions. From Fig.~\ref{fig:nucl_plot} it is clear that the self-assembly of the diamond cubic (red circles) occurs in correspondence of the phase separation boundaries. Self-assembly is aided by the formation of dense liquid regions during the phase-separation process. Interestingly, some state points in Fig.~\ref{fig:nucl_plot} have nucleated outside the binodal boundaries, but close to the critical temperature. The system thus represents an interesting example of nucleation aided by critical fluctuations, as first predicted in Ref.~\cite{wolde1997enhancement} for isotropic interactions.

To summarize, the self-assembly pathway at the azeotropic point is the following: an equimolar disordered solution first 
generates equimolar critical fluctuations or first demixes in an equimolar dense liquid, which then crystallizes in a equimolar crystalline structure. Self-assembly at azeotropic conditions has the advantage of bypassing the difficulties associated with concentration fluctuations, which could otherwise severely limit the nucleation rate.

\begin{figure}[!t]
    \centering
    \includegraphics[width=0.45\textwidth]{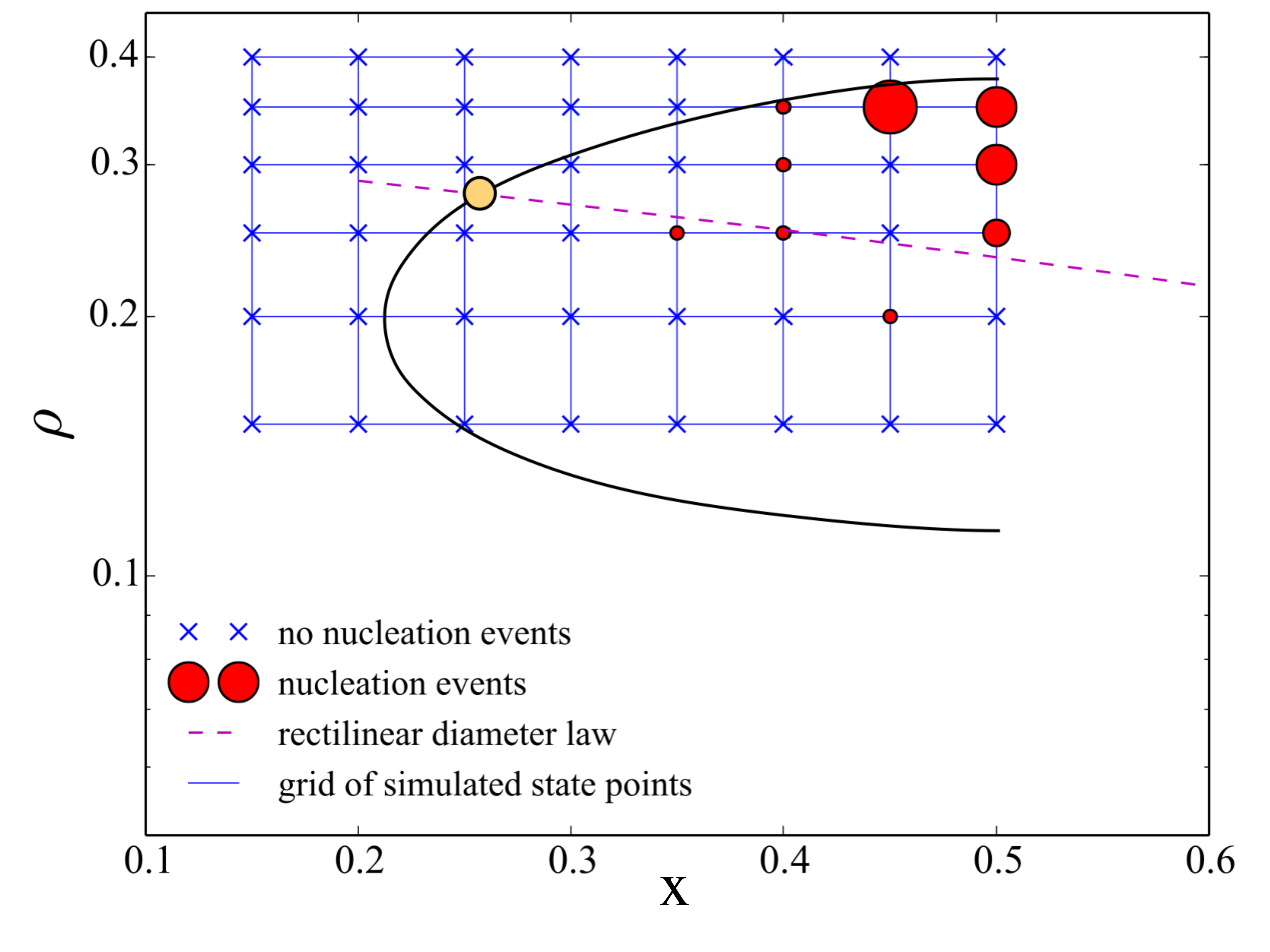}
    \caption{\textbf{Nucleation events at $\mathbf{T=0.1}$ superimposed to a schematic representation of the density-concentration phase diagram.} The binodal line is obtained from fitting the Gibbs ensemble results of Fig.~\ref{fig:confrontoMC}. The formation of crystals with a fraction of particles in the cubic diamond phase )equal or greater than $0.7$ occurs mostly near the liquid branch around $x=0.5$ (azeotropic condition). The blue grid defines all the state points considered; those showing no nucleation event are crossed out, while state points where at least one trajectory nucleated are represented with red circles. The radius of the circles is proportional to the fraction of trajectory that have nucleated within  $3.5\,10^8$ MC sweeps. The yellow circle indicates the critical point located at the intersection of the bindoal curve and the rectilinear diameter line, \textit{i.e.} the (dashed) straight line passing through the mid-point of the tie lines connecting each pair of coexisting points.}
    \label{fig_nucleation_T0.1}
\end{figure}

\noindent We further analyse the self-assembly process by studying the nucleation of solutions prepared at different densities and concentrations at temperature $T=0.1$. The considered state points are located on the regular grid $\rho \in [0.15,0.4]$ and $x \in [0.15,0.5]$ with $\Delta\rho=0.05$ and $\Delta x=0.05$, as shown with blue lines in Fig.~\ref{fig_nucleation_T0.1}. For each of these state points we run $10$ independent Monte Carlo simulations in the NVT ensemble with AVB dynamics and $500$ patchy particles for $3.5\,10^8$ MC sweeps. We look for state points exhibiting at least one nucleation event that gives rise to a cubic diamond with $350$ or more patchy particles. In Fig.~\ref{fig_nucleation_T0.1} we mark these state point with red circles with a radius proportional to the fraction of trajectories that have nucleated. The fraction of particles in the cubic diamond phase are identified with local bond-order analysis~\cite{tanaka2019revealing}. By superimposing the grid to the density concentration phase diagram, we can see that crystallisation occurs exclusively within the liquid-vapour coexistence region.
Fig.~\ref{fig_nucleation_T0.1} confirms extended crystals are formed only close to the azeotropic point. Indeed it is exactly at azeotropic condition that the ratio between the two components in the liquid phase is the same of that of the cubic diamond crystal.

\begin{figure}[!t]
    \centering
    \includegraphics[width=0.45\textwidth]{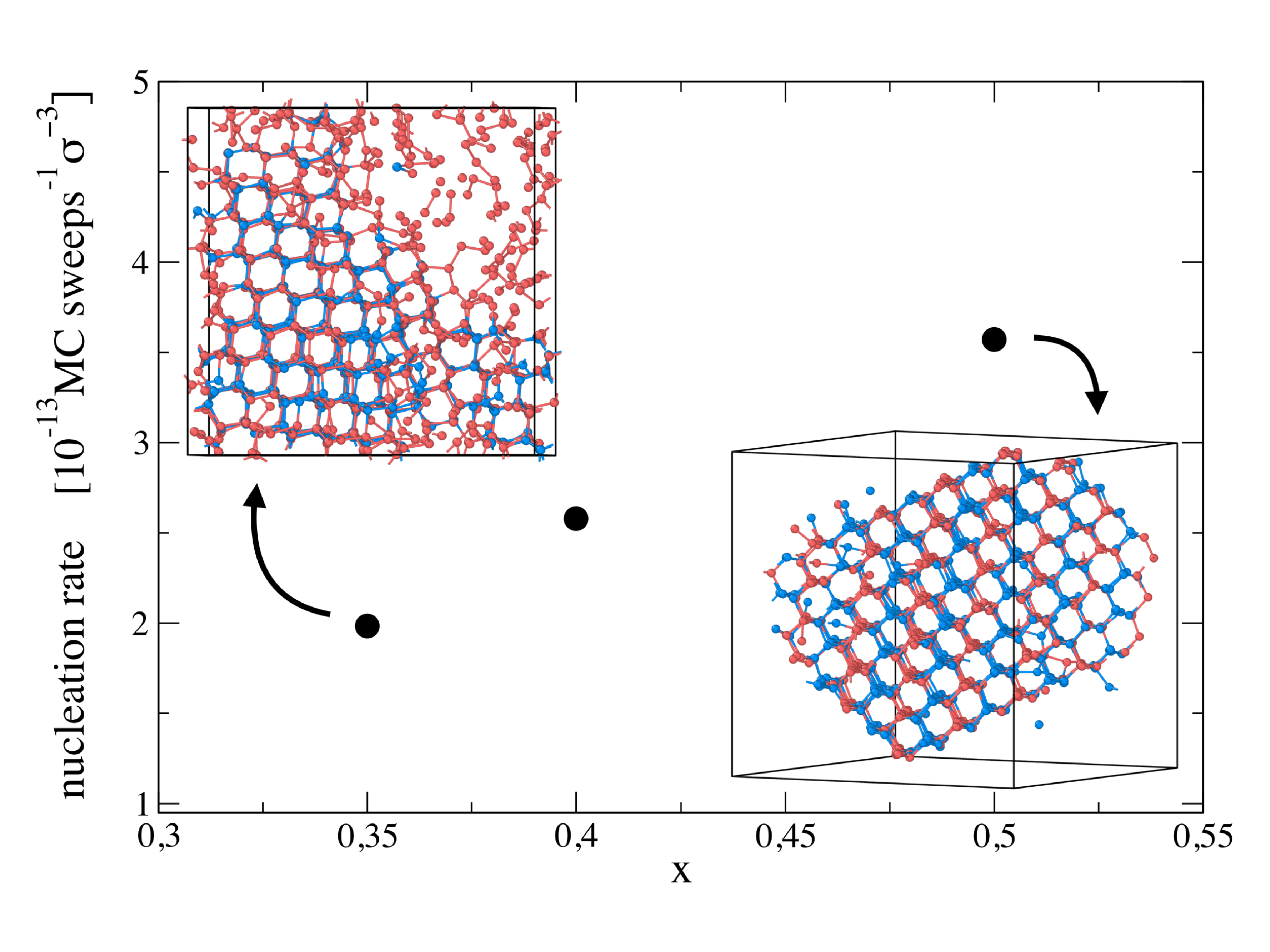}
    \caption{\textbf{Nucleation rate and final configuration snapshots.} Nucleation rate (black dots) as a function of concentration for systems of $1000$ particles at temperature $T=0.097$ and density $\rho=0.3$. The two snapshots display the last configuration of a trajectory at $x=0.35$ and at $x=0.5$. Red and blue colors indicate the species to which a particle belongs: blue for the minority component, and red for the majority component.}
    \label{fig_nuceation_rate&snap}
\end{figure}

In Fig.~\ref{fig_nuceation_rate&snap} we show the nucleation rate computed, for each $x$, from $56$ Monte Carlo trajectories ran at temperature $T=0.097$, density $\rho=0.3$ and with $N=1000$ particles for three concentrations: $x=0.35$, $x=0.4$, and $x=0.5$. The nucleation rate is estimated as the number of trajectory that successfully nucleate within $3.5\,10^8$ MC sweeps, per unit of time and volume. Also at this temperature we observe that the nucleation rate increases towards the azeotropic concentration. The snapshots display the last configuration of a mixture prepared at the azeotropic condition ($x=0.5$) and one away from it ($x=0.35$). A visual inspection of these snapshots highlights that crystal growth is limited when the concentration of the liquid phase is different from the stoichiometric ratio of the the target crystal components. Nucleation at the azeotropic point is advantageous as self-assembly can proceed up to $100\%$ without one component depleting before the other and an extended crystal can form. On the contrary, at off-azeotropic conditions, the self-assembled cubic diamond coexists with a gas phase composed of the majority component, that can only aggregate in chains. Going toward the azeotropic point, the density of the majority component diminishes until eventually all particles belong to the crystalline phase.  Finally, regarding the quality of the crystals we observe nuclei free from defects. The interaction matrix was indeed designed to avoid the hexagonal diamond phase, and this also forbids the formation of stacking faults that are the most common type of defects in diamond cubic crystals.

\section*{Conclusions}
Self-assembling complex structures requires designing complex interaction potentials, that not only need to have the target structure as a free energy minimum, but that also have to avoid competing local minima that can kinetically frustrate the assembly process. In recent years it has become increasingly clear that using multi-component mixtures can shift the problem from the need to accurately design the shape of the potential (\textit{e.g.} introducing torsional interactions to assemble cubic diamond and avoid hexagonal diamond~\cite{romano2012patterning}) to the optimization of a generic interaction matrix between different components. This last problem is amenable to an effective numerical solution \textit{via} the so-called SAT-assembly framework~\cite{romano2020designing}, where the interactions between the different components are found by solving satisfiability problems. But adding components increases the thermodynamic degrees of freedom, which considerably complicates the phase behaviour and the assembly pathway.

In this work we have shown that much of the thermodynamic difficulties can be removed by preparing the self-assembly pathway on an azeotropic point, where the system behaves effectively as a one-component mixture. We have then  shown under which conditions we can include azeotropy in self-assembly designs.

As a proof of concept, we have focused on the case of patchy particles, which represent a convenient model for systems whose interactions can be described by an isotropic repulsion and strong directional attractions. Exploting the laws of mass-action we have shown that in these systems azeotropy can be directly included in the interaction matrix. Different cases have been considered. The simplest condition, named \emph{bond exclusivity}, asserts that an equimolar azeotropic point can be obtained by imposing that each patch has a unique interaction partner. 
The equimolar condition can be relaxed and the azeotropic point can be located at a desired concentration vector $\mathbf{x}$, by considering the \emph{bond multiplicity} condition, which requires some patches to have more than one possible interaction partner. Finally, the \emph{fully-connected bond} condition, where each patch has one interaction partner on each of the species in the system, corresponds to a \emph{always azeotropic} mixture.

We have then provided a fully worked example of a binary mixture designed to self-assemble colloidal diamond while avoiding the hexagonal form, and that obeys the \emph{bond exclusivity} condition. We have explicitly derived its phase diagram, both within Wertheim's perturbation theory and \textit{via} Gibbs ensemble simulations, and shown that it contains the predicted negative azeotrope at equimolar conditions. The class of this phase diagram has never been reported to our knowledge.
It is  unique in the sense that the binary critical point line tends to $(P,T)\rightarrow 0$ for $x\rightarrow (0,1)$, meaning that the system phase separates only upon mixing. Finally we have analyzed the self-assembly pathway for systems prepared at azeotropic conditions, and shown that the pathway is the same as in one-component systems: more precisely an equimolar mixture condensates into an equimolar liquid, which, given the coincidence in concentration between the crystal and the melt, then nucleates into a crystal that grows without concentration defects.

We believe that the ability to explicitly include azeotropic points into artificial designs represents an exciting step towards a fully consistent framework for the self-assembly of arbitrary structures.
Efforts are now geared toward experimental realization of these designs, for example through wireframe DNA origami~\cite{nykypanchuk2008dna,liu2016diamond,kumar2017nanoparticle,bohlin2022design}, that naturally encode binding specificity.

\section*{Methods}

\subsection*{Patchy particles}
We consider multi-component mixtures of patchy particles. Patchy particles are spherical colloids whose surface is decorated by attractive site, named patches and different species of patchy particles can differ either by the number, the arrangement, and/or the type of the patches. To model their interaction we choose the Kern-Frenkel~\cite{bol1982monte,kern2003fluid} potential which describes hard-core spherical particles of diameter $\sigma$, interacting with an additional square well potential $V_{SW}$ of depth $\epsilon$ and width  $\delta$, modulated by a term $F$ depending on the patchy particles orientation.  Two patchy particles  attract in a strongly directional way if they are at distance between $\sigma$ and $\sigma+\delta$.
More precisely, the  interaction potential $V$ between particle $i$ and $j$, with a center to center distance $r_{ij}$ is

\begin{equation}
	\label{eqn:KF}
	V(\mathbf r_{ij},\hat{\mathbf r}_{\alpha,i},\hat{\mathbf r}_{\beta,j})= V_{SW}(r_{ij})F(\mathbf r	_{ij},\hat{\mathbf r}_{\alpha,i},\hat{\mathbf r}_{\beta,j})
\end{equation}

\noindent where  $\hat{\mathbf r}_{\alpha,i}$ ($\hat{\mathbf r}_{\beta,j}$) indicates the position of patch $\alpha$ ($\beta$) of particle $i$ ($j$), 
\noindent and

\begin{equation}
	\label{eqn:f_KF}
	F(\mathbf r_{ij},\hat{\mathbf r}_{\alpha,i},\hat{\mathbf r}_{\beta,j})=
	\begin{cases}
		1 &\text{if}\quad 
		\begin{array}{l}
			\hat{\mathbf r}_{ij} \cdot \hat{\mathbf r}_{\alpha,i} > \cos{(\theta_{max})}\\ 
			\hat{\mathbf r}_{ji} \cdot \hat{\mathbf r}_{\beta,j} > \cos{(\theta_{max})}
		\end{array} \\
		0 &\text{otherwise}
	\end{cases}
\end{equation}

\noindent For identical patches, the Kern-Frenkel potential is characterised by the two independent parameters $\delta$ and $\theta_{max}$ that specify the range and the angular width of the patches respectively (see Fig. 1) and that can be tuned giving rise to different phase diagrams~\cite{smallenburg2013limited}.

\subsection*{DNA-based implementation}

\begin{figure}[!t]
    \centering
    \includegraphics[width=0.52\textwidth]{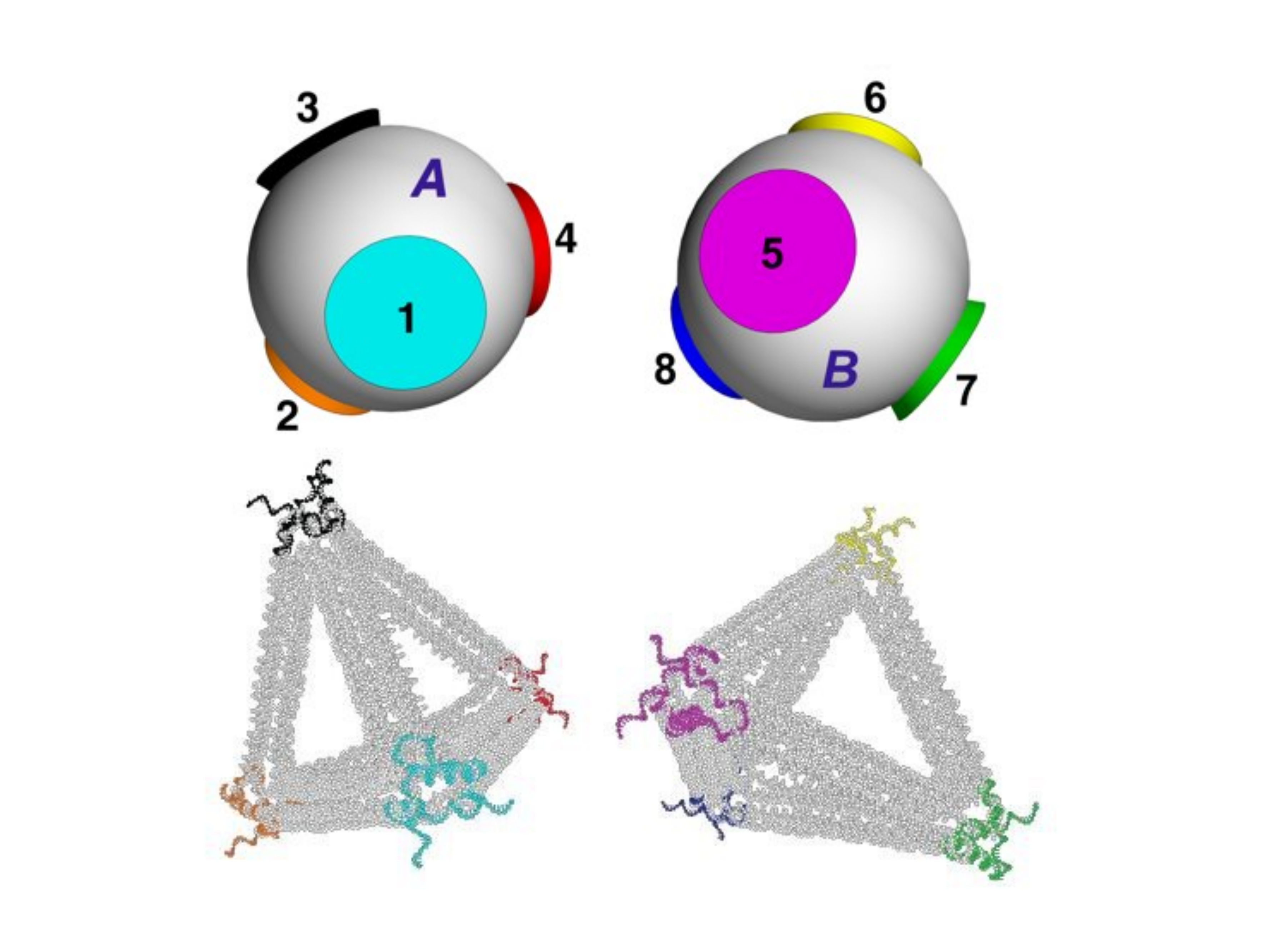}
    \caption{\textbf{Sketch of two patchy particles realised through DNA origami.} 
    The four patches tetrahedrally arranged are mapped in single-stranded overhangs at each vertex of a tetrahedron made by nanoscale folding of DNA. Different interacting patches correspond to complementary single DNA strands and the self-interacting ones to palindromic DNA strands.}
    \label{fig_pp_origami}
\end{figure}

Patchy particle models are particularly suited 
to tackle the inverse self-assembly task since it is possible to control the valence and to encode the desired topology in the number, the placement, and the type of patches. 
Apart from their computational convenience, patchy particles are also experimentally \textit{via}ble systems: short ranged anisotropic interactions between colloidal particles have in fact been achieved \textit{via} chemical patterning of their surfaces~\cite{zhang2004self,pawar2010fabrication,bianchi2011patchy,romano2011colloidal}, and \textit{via} modelling of their shape~\cite{van2013entropically}.

The most promising approach to realize specific interactions uses DNA nanotechnologies to create a selective binding between particles: matchable colors~\cite{vasilyev2015chromatic} correspond to complementary single DNA strands, equal colors to self-complementary sequences. Multiple color interactions can also be realized as discussed in Supplementary Materials IV.
Popular systems include DNA functionalised colloids~\cite{xiong2020three} or
DNA origami~\cite{Rothemund2006,liu2016diamond,zhang20183d,tian2020ordered,chakraborty2022self} where single strands of DNA are attached to well-defined positions on the particle surface~\cite{suzuki2009controlling,kim2011dna,wang2012colloids,feng2013dna,Rothemund2006,tian2020ordered}

Fig.~\ref{fig_pp_origami} shows a possible realisation of a binary mixture of patchy colloids with eight different patches (colors). The decorated hard-sphere colloidal model (which can  be closely
experimentally realized~\cite{gong2017patchy})  is displayed together with 
a DNA-origami implementation~\cite{liu2016diamond}. The tetrahedron  vertices are functionalised with DNA strands,  exploiting  DNA addressability to encode patch-patch interactions.  In Supplementary Materials IV we describe in full details
an algorithm which allows us to determine the sequences of DNA strands that satisfy predefined bonding rules, applicable to  both same- , distinct- and multiple-color interactions.  To apply the algorithm one needs to select  the total length $n_s$ of the oligomer grafted on each patch (for example an oligomer composed by six bases) and a rule quantifying the binding strength between any two oligomers (for example  the melting temperature, estimated according to  SantaLucia~\cite{santalucia1998unified}  or the number of consecutive paired bases).  
See Supplementary Materials IV for a full description of the algorithm.

\subsection*{Wertheim perturbation theory}

Here we report the results of the Wertheim first order perturbation theory~\cite{wertheim1984fluids} that was originally developed to derive a mean-field theory of associating fluids and that can be easily generalised to patchy particles~\cite{chapman1988phase,de2011phase}. Recently~\cite{bianchi2006empty,heras2011phase,rovigatti2013computing,seiferling2016percolation,teixeira2017phase,braz2021phase,russo2021physics}, the theory has been adopted to study in detail the static (\textit{e.g.} percolation) and thermodynamic (\textit{e.g.} phase behaviour) properties of patchy particle systems, both in pure components and in mixtures, showing excellent qualitative agreement with numerical simulations.
The main assumptions are that each attractive site cannot be engaged in more than one bond at the same time (one-bond-per-patch condition) and that a new bond occurs only between particles belonging to different clusters (loop formations are forbidden). Wertheim developed a perturbative method that, applied to patchy particles, estimates the effect of the attractive patches on the Helmholtz free energy of the reference system of hard spheres. The power of this theory is the chance to provide a good estimate of the Helmholtz free energy of a multicomponent system of patchy particles by only knowing the structure of the reference system and the interaction potential characterising patchy particles. Here we follow the conventions of Refs.~\cite{heras2011phase,teixeira2017phase}. The Helmholtz free energy per particle in units of $k_BT$ of a $n$-component mixture can be expressed as:

\begin{equation}
\label{eqn:betaf}
\beta f = \beta f_\text{reference} + \beta f_\text{bonding} 
\end{equation}

\noindent The reference free energy is the sum of the ideal gas contribution $\beta f_\text{ideal}$ and of the hard spheres excess term $\beta f_\text{HS}$. This hard spheres contribution takes into account the excluded volume of the patchy particles and it is given by the Carnahan-Starling formula~\cite{mansoori1971equilibrium} since the different species have all the same diameter. 

\begin{equation}
\label{eqn:betaf_reference}
	\begin{array}{l}
			\beta f_\text{reference}=\beta f_\text{ideal}+\beta f_\text{HS} \quad \text{with} \\[8pt]
			\beta f_\text{ideal}=\ln{\rho}-1+ \sum\limits_{i=1}^n x^{(i)}\ln{(x^{(i)}V_i)}\\[8pt]
			\beta f_\text{HS}=\frac{4\phi-3\phi^2}{(1-\phi)^2} 
		\end{array} \\
\end{equation}

\noindent where $\rho$ is the density, $x^{(i)}$ is the molar fraction of species i, $V_i$ is the thermal volume and $\phi$ is the packing fraction equal to $\rho V_s$ where $V_s=\sigma^3 \pi / 6$ is the volume of a single particle. \\ The bonding contribution contains the sum over the species $({\textstyle\sum}_{i=1}^n)$ and the sum over the patches of a certain species $i$ $({\textstyle\sum}_{\alpha\in\Gamma(i)})$; the number of patches of species i is denoted as $n(\Gamma(i))$.

\begin{equation}
\label{eqn:betaf_bonding}
\begin{split}
\beta f_{bonding}=&\sum\limits_{i=1}^n x^{(i)} \Biggl[\sum\limits_{\alpha\in\Gamma(i)}\bigg(\ln{X_{\alpha}^{(i)}}-\frac{X_{\alpha}^{(i)}}{2}\bigg)+ \\[2ex] +&\frac{1}{2}n(\Gamma(i))\Biggr]
\end{split}
\end{equation}

\noindent $X_{\alpha}^{(i)}$ is the probability that a patch $\alpha$ on a species i is not bonded and it is defined by the mass balance equation:

\begin{equation}
X_{\alpha}^{(i)}=\biggl[ 1+ \phi  \sum_{j=1,n} x^{(j)} \sum_{\gamma\in\Gamma(j)} X_{\gamma}^{(j)} \Delta_{\alpha\gamma}^{(ij)} \biggr]^{-1}
\end{equation}

\noindent where $\Delta_{\alpha\gamma}^{(ij)}$ does not depend on the species, since the diameter is always the same, and it is given by

\begin{equation}
\label{eqn:delta}
\Delta_{\alpha\gamma}^{(ij)}=\Delta_{\alpha\gamma}=\frac{1}{V_s}\int_{V_{\alpha\gamma}} g_{HS}(\mathbf r) (e^{\beta\epsilon_{\alpha\gamma}}-1) d\mathbf r
\end{equation}

\noindent where $g_{HS}$ is the radial distribution function of hard spheres, $V_{\alpha\gamma}$ is the bonding volume and $\epsilon_{\alpha\gamma}$ is the bonding energy both related to a bond between patches $\alpha$ and $\gamma$.
As for any short-ranged patchy potential (in the single-bond per patch condition), the static properties are controlled by the bonding volume~\cite{russo2021physics}, \textit{i.e.} the volume in which a particle can move while being bonded to another particle, which for the Kern-Frenkel potential assumes the following simple expression

\begin{equation}
V_b=\frac{4\pi}{3}((\sigma+\delta)^3-\sigma^3)\biggl[\frac{1-\cos{(\theta_{max})}}{2}\biggr]^2.
\end{equation}

 $\Delta_{\alpha\gamma}$ characterises the bond between the patch $\alpha$ on the patchy particle of species $i$ and the patch $\beta$ on the patchy particle of species $j$. Patches are in general different and therefore they can interact following different potentials (Kern-Frenkel in our case). In the following we consider that all bonds have the same bonding volume and we approximate the radial distribution function with an expansion around its value at contact, as detailed in Ref.~\cite{nezbeda1989primitive,sciortino2007self}. With these approximations, affecting the results only quantitatively, but not qualitatively, equation~\ref{eqn:delta} becomes:
\begin{equation}
\label{eqn:delta_corr}
\begin{split}
\Delta_{\alpha\gamma}=&\frac{1}{V_s} 4\pi\chi^2 \Biggl\{\biggl[\frac{(1+\delta)^3-1}{3}A_0\biggr] +\\[2ex] +&\biggl[\frac{(1+\delta)^4-1}{4}A_1\biggr] \Biggr\}(e^{\beta\epsilon_{\alpha\gamma}}-1)
\end{split}
\end{equation}

\noindent with

\begin{equation}
\label{eqn:delta_corr_2}
	\begin{array}{l}
		A_0=\frac{1-\frac{\phi}{2}+\frac{9\phi}{2}(1+\phi)}{(1-\phi)^3}\\[3ex]
		A_1=\frac{-\frac{9\phi}{2}(1+\phi)}{(1-\phi)^3}\\[3ex]
		\chi=\frac{1-\cos{\theta_{max}}}{2}
	\end{array}
\end{equation}

The theory allows the computation of the Helmholtz free energy for any state point.
Notice that solutions of the type $X_\alpha^{(i)}=X$ in Eq.~\ref{eqn:betaf_bonding} (remembering that $\sum_j x^{(j)}=1$) formally reduce the free energy of the mixture to that of a single component, \textit{i.e.} the solutions correspond to azeotropic points.

\subsection*{Isochoric thermodynamics}

One way to calculate the binodal curve for a single component system is offered by the integration of the Clausius-Clapeyron differential equation.  Also in the case of multi-components mixtures it is possible to define a set of differential equations that if integrated provides the binodal curve. Here we carry out the integration of these differential equations  in the isochoric thermodynamics framework~\cite{deiters2017differential,bell2018construction}. We provide here a short summary of this framework. In the canonical ensemble, the thermodynamic state of a $n$-component mixture is specified by temperature $T$, molar density $\rho$ and mole fractions $x_i$. However the mole fractions have some disadvantages: they are not independent variables and, conversely to density, they are dimensionless causing the density mole fractions space to have an ill defined metric. On the contrary, in the isochoric thermodynamics the independent variables are molar densities $\rho_i$ and the fundamental thermodynamic potential is the Helmholtz energy density $\Psi$. They are defined as:
\begin{equation}
\label{eqn:iso_variables}
\begin{array}{l}
	\rho_i=x_i \rho \\[2ex]
	\Psi(\boldsymbol{\rho},T)=\frac{A}{V}=a \rho
\end{array} \\
\end{equation}

\noindent where $A$ is the Helmholtz energy and $a$ is the molar Helmholtz energy, $\rho$  is the molar density of the $n$-component mixture $\rho=\sum_{i=1}^{n}\rho_i$ 
 while  $\boldsymbol{\rho}$ is the vector of molar densities $\boldsymbol{\rho}=(\rho_1,\rho_2,\ldots \rho_n)$. \\ The local curvature of the Helmholtz energy density is encoded in the hessian matrix:

\begin{equation}
\label{eqn:H}
H=\begin{bmatrix}
			 \biggl( \frac{\partial^2{\Psi}}{\partial{\rho_1}^2}\biggl)_T & \biggl( \frac{\partial^2{\Psi}}{\partial{\rho_1}\partial{\rho_2}}\biggl)_T & \cdots & \biggl( \frac{\partial^2{\Psi}}{\partial{\rho_1}\partial{\rho_n}}\biggl)_T \\[3ex]
			 \biggl( \frac{\partial^2{\Psi}}{\partial{\rho_2}\partial{\rho_1}}\biggl)_T &  \biggl( \frac{\partial^2{\Psi}}{\partial{\rho_2}^2}\biggl)_T & \cdots & \biggl( \frac{\partial^2{\Psi}}{\partial{\rho_2}\partial{\rho_n}}\biggl)_T \\[3ex]
			\vdots & \vdots & \ddots & \vdots  \\[3ex]
			\biggl( \frac{\partial^2{\Psi}}{\partial{\rho_n}\partial{\rho_1}}\biggl)_T &  \biggl( \frac{\partial^2{\Psi}}{\partial{\rho_n}\partial{\rho_2}}\biggl)_T & \cdots & \biggl( \frac{\partial^2{\Psi}}{\partial{\rho_n}^2}\biggl)_T
		\end{bmatrix}
\end{equation}

\noindent If it is positive defined, then the state is a stable state. We know that two phases (labeled $'$ and $''$ in the following) 
coexist in equilibrium at constant temperature if, along the phase boundary, the pressure and the chemical potentials of each component are equal for both phases. This means that the variation of the pressure and of the chemical potentials along the phase boundary must be the same for both phases:
\begin{equation}
\label{eqn:cond_coex}
\begin{array}{l}
	d \mu_i^{'}=d \mu_i^{''}  \quad \text{with} \; i=1,2, \ldots n \\[2ex]
	dP^{'}=dP^{''}
\end{array} \\
\end{equation}

\noindent having defined the chemical potentials and the pressure as $\mu_i= \partial{\Psi}/\partial{\rho_i}$ and $P=-\Psi + \sum_{i=1}^{n} \rho_i \mu_i$.

\noindent Integrating this system of first order differential equations allows us to numerically evaluate the coexistence region. For the isothermal phase equilibrium of a binary mixture we must solve:

\begin{equation}
\label{eqn:rho_T}
\begin{array}{l}
		\begin{bmatrix}
			H_{\Psi,1}^{'}\cdot \boldsymbol{\rho^{''}} & H_{\Psi,2}^{'}\cdot \boldsymbol{\rho^{''}}\\[2ex]
			H_{\Psi,1}^{'}\cdot \boldsymbol{\rho^{'}} &  H_{\Psi,2}^{'}\cdot \boldsymbol{\rho^{'}}
		\end{bmatrix}
		 \biggl( \frac{d \boldsymbol{\rho}}{d P}\biggl)_{T,\sigma}^{'} 
		 =
		 \begin{bmatrix}
			1\\
			1
		\end{bmatrix} \\[6ex]
		H_{\Psi}^{''}  \biggl( \frac{d \boldsymbol{\rho}}{d P}\biggl)_{T,\sigma}^{''} =H_{\Psi}^{'}  \biggl( \frac{d \boldsymbol{\rho}}{d P}\biggl)_{T,\sigma}^{'} 
\end{array} \\
\end{equation}

\noindent where $H_{\Psi,i}$ indicates the $i$-th row of the Hessian matrix with $n=2$ in Eq.~\ref{eqn:H} and the subscript $\sigma$ indicates that derivatives are calculated along the phase boundary. \\ By starting from available accurate initial values, the integration of the derivatives of the molar densities in the coexisting phases over the desired range of pressure predicts how molar densities of vapour and liquid change with pressure. This enables the construction of the binary mixture pressure-concentration and density-concentration binodal curves. In summary by knowing  one pair of coexisting points it is possible to determine the entire coexistence region by calculating how these coexisting points move along the binodal curve. Integration gets stiff and does not proceed further close to critical points, as the step-size of the adaptive step-size integrator~\cite{bell2018construction} progressively decreases as the hessian determinant vanishes at the critical points. Hence critical points, indicated in Fig. 3A by triangles, are computed by imposing the hessian determinant to be zero and the stability conditions.

\subsection*{Monte Carlo simulations: AVB moves and Gibbs ensemble}

When simulating patchy particle systems interacting \textit{via} anisotropic and short-ranged interactions, roto-translation moves are not always sufficient to ensure a good sampling of the phase space. Indeed patchy particles self-assembly occurs when the thermal energy is much smaller than the bonding energy $\epsilon$, which makes the Metropolis acceptance probability of a MC move that breaks a bond extremely low. Thus almost all moves that try to break a bond are rejected not allowing the system to equilibrate. To overcome this drawback, we have introduced aggregation-volume-bias-moves (AVB)~\cite{chen2000novel,rovigatti2018simulate} that facilitate  bond breaking by enhancing the acceptance probability. In particular, there are two types of AVB moves: the AVB-B move and the AVB-U move.
The AVB-B move attempts to create a bond by moving one patchy particle in the bonding volume ($V_b$) of another patchy particle, thus giving rise to a bond between two patchy particles that were not bonded to each other. Conversely, the AVB-U move tries to break a bond by taking one bonded patchy particle outside the bonding volume ($V_o=4\pi V-V_{b}$) of the patchy particle to which it is bonded, thus eliminating an existing bond between a patchy particles pair. These moves are biased, and their acceptance probabilities are

\begin{equation}
\label{eqn:A_AVB}
\begin{array}{l}
	A_{AVB-B}=\text{min}\biggl\{ 1,\frac{(N-N_i-1)V_{b}}{(N_i+1)V_o} e^{-\beta \Delta E} \biggr\} \\[3ex]
	A_{AVB-U}=\text{min}\biggl\{ 1,\frac{N_iV_o}{(N-N_i)V_{b}} e^{-\beta \Delta E} \biggr\}
\end{array} \\	
\end{equation}

\noindent where $N_i$ is the number of particles that are bonded to particle $i$. Importantly, the acceptance probability of breaking a bond is enhanced respect to the one of simple rototranslation move, as the ratio $V_o/V_{b}$ is much larger than one since the bonding volume $V_{b}$ is much smaller than its complementary volume $V_o=4\pi V-V_{b}$, where $V$ is the volume of the simulation box. 

\smallskip

\noindent In order to study the coexistence between two phases at a certain temperature, we employ Gibbs ensemble simulations~\cite{panagiotopoulos1987direct,panagiotopoulos1988phase}, where coexistence occurs between two simulation boxes that virtually interact among each other without an explicit interface. In addition to rototraslational moves, the Gibbs ensemble incorporates volume moves (which alter the size of the two boxes keeping the total volume fixed), and particle transfer moves (where a particle is moved from one simulation box to the other).

\subsection*{Acknowledgement } 
We thank Michael Matthies for help with designing DNA origami represenation of patchy particles.\\

We acknowledge the CINECA award under the ISCRA initiative, for the availability of high performance computing resources and support.
JR acknowledges support from the European Research Council Grant DLV-759187. P\v{S} acknowledges support from the ONR Grant N000142012094. P\v{S} further acknowledges the use of the Extreme Science and Engineering Discovery Environment (XSEDE), which is supported by National Science Foundation grant number TG-BIO210009.
This work is partially supported by ICSC – Centro Nazionale di Ricerca in High Performance Computing, Big Data and Quantum Computing, funded by European Union – NextGenerationEU.

\section*{Supporting Information Available} 
Supporting information available in the supplementary materials pdf: \\
explanation of azeotropy, example of a binary mixture satisfying the bond multiplicity condition, example of a binary mixture satisfying the fully-connected bond condition, description an algorithm to generate DNA strands from the interaction matrix.

\section*{Author information}
\subsection*{Author Contributions}
All authors contributed equally to the research and writing of the manuscript.

\subsection*{Notes}
The authors declare no competing financial interest.

\renewcommand\thefigure{S\arabic{figure}}
\setcounter{figure}{0}  

\numberwithin{equation}{subsection}
\renewcommand\theequation{S\arabic{equation}}

\onecolumngrid
\section*{Supplementary Materials}

\section{Azeotropy}

\begin{figure}[H]
    \centering
    \subfloat[]{\includegraphics[width=0.46\textwidth]{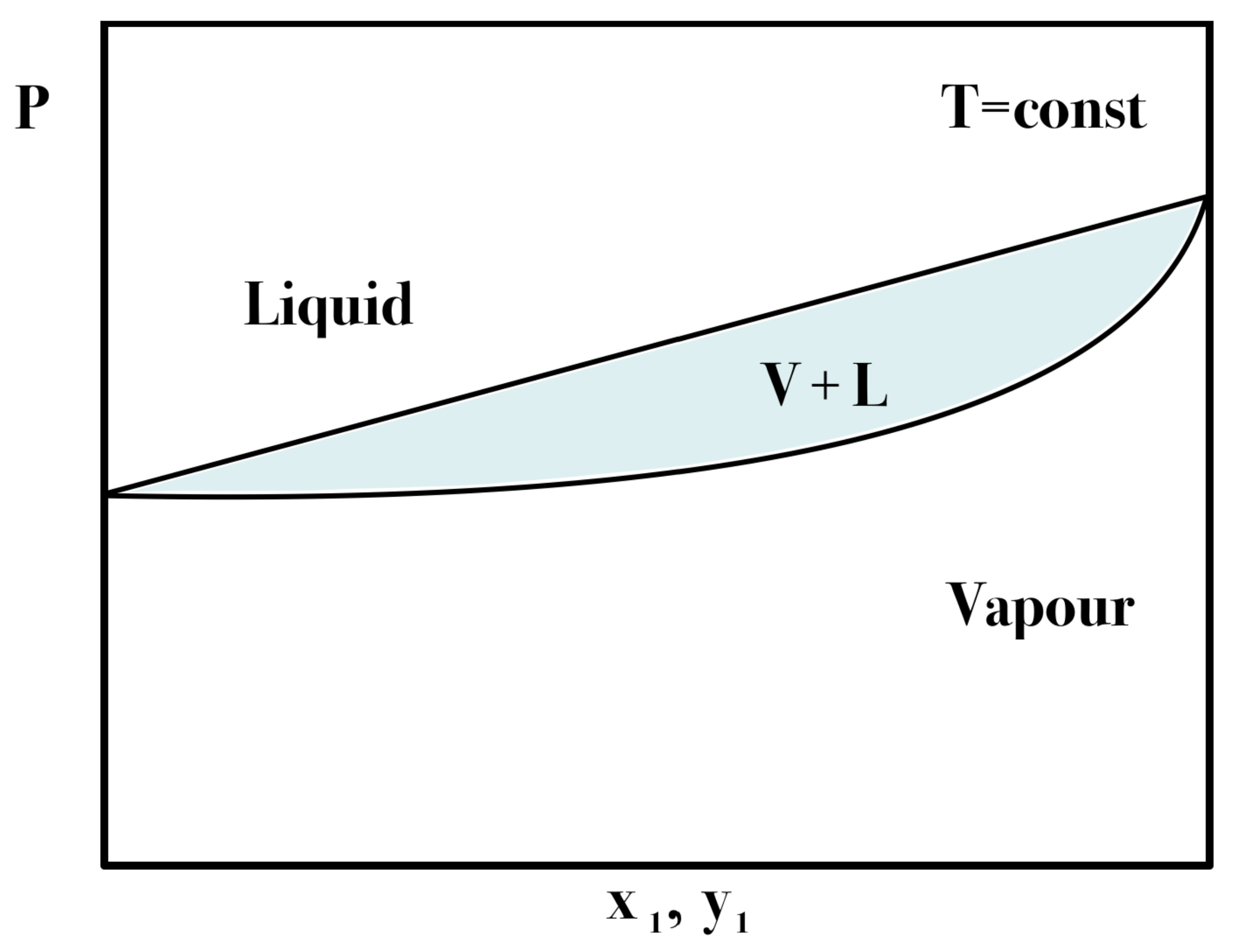}\label{fig:ideal_mixture}}
    \hfill
    \subfloat[]{\includegraphics[width=0.46\textwidth]{figS1b.pdf}\label{fig:azeotrope}}   
    \caption{\textbf{Pressure-concentration phase diagram for an ideal binary mixture (a) and for a binary mixture with a negative azeotrope (b).}
    The azeotropic point is located at the lowest pressure at which the liquid and the vapour phases can coexist at the specific temperature T. The inset highlights that at the azeotropic point the bubble point curve and the dew point curve are tangent; the bubble point curve is the locus of points where the first bubble of vapour appears when pressure is lowered starting from a point greater than the total vapour pressure, and the dew point curve is where the first liquid drop originates when pressure is increased starting from a point in the vapour phase.}
    
    \label{fig_azeo}
\end{figure}

In systems of two or more components, both the pressure-concentration and the temperature-concentration phase diagrams exhibit a coexistence region.
The presence of a coexistence region implies that the relative concentrations in the vapor and liquid phases are not the same.  Supplementary Fig.~\ref{fig_azeo} shows a qualitative pressure-concentration phase diagram for both an ideal (a) and a non ideal (b) binary mixture. In both cases, the coexistence pressures reduce to a single value when the first component concentration is equal to zero and to one, i.e. when the binary mixture becomes a one-component system. However, if the mixture strongly deviates from ideal behaviour, it can exist another point, named \emph{azeotropic point}~\cite{smith1949introduction,Moore1962Physical}, at concentration different from zero and one, where the coexistence region reduces to a single point. A multi-component mixture at the azeotropic point will separate into phases at the same azeotropic concentration, therefore behaving as a pure system.

\section{1:2 azeotropic mixture example}

\noindent We analyse the example of a binary mixture of patchy particles, with four patches tetrahedrally arranged, whose ratio is $1:2$. An interaction matrix satisfying the bond multiplicity rules is:

\begin{equation}
\label{matrix_deg}
\mathbf{\Upsilon}_\text{N2c6s3}=
	\begin{pmatrix}
	0 & 0 & 0 & 0 & 1 & 0 & 0 & 0\\
	0 & 0 & 0 & 0 & 1 & 0 & 0 & 0\\
	0 & 0 & 1 & 1 & 0 & 0 & 0 & 0\\
	0 & 0 & 1 & 1 & 0 & 0 & 0 & 0\\
	1 & 1 & 0 & 0 & 0 & 0 & 0 & 0\\
	0 & 0 & 0 & 0 & 0 & 0 & 1 & 0\\
	0 & 0 & 0 & 0 & 0 & 1 & 0&  0\\
	0 & 0 & 0 & 0 & 0 & 0 & 0 & 1
\end{pmatrix}
\bigskip
\end{equation}
\noindent Notice that the naming convention of $\mathbf{\Upsilon}_\text{N2c6s3}$, which refers to the fact that there are two different species, six independent patches (colors) (six distinct rows or columns) and three self-interacting colors, is not unique, i.e. there are different ways of arranging 6 colors over two species. For the design in the matrix above, we look at the number and the placement of each 1 at each row. Patches under \textit{bond exclusivity} constraint have a single one for each row, whereas patches obeying the \textit{bond multiplicity} condition have two one for each row, as expected being the components ratio $1:2$, i.e. $n=2$. \\ In this case, in the mass balance equations for $X_{\alpha}^{(i)}$ the sum over the species $(\sum_{j=1,2})$ still drops out while the sum over the patches $(\sum_{\gamma\in\Gamma(j)})$ reduces to one or to two terms depending if patch $\alpha$ satisfies the \textit{bond exclusivity} condition or the \textit{bond multiplicity} condition, respectively.
Together with the condition $\Delta_{\alpha \gamma}\equiv \Delta$ for all patches $\alpha$ and $\gamma$, this implies that the coefficient in front of $X_{\alpha}^{(i)}$ is the same for all species $i$.
Patches belonging to the same species and characterised by the same row in the interaction matrix are equal and therefore they share the same probability $X_{\alpha}^{(i)}$. Hence, unlike the bond exclusivity case, not all patches are different and the number of distinct mass balance equations is smaller than $N_s\times N_p$. In our example, $X_{1}^{(1)}=X_{2}^{(1)}$ and $X_{3}^{(1)}=X_{4}^{(1)}$ and with simple algebraic steps we obtain the four following types of equations: 
  
\begin{equation}
\label{eqn:X_deg}
\begin{split}
    &\text{if $\alpha=1,2$}\\
	&X_{\alpha}+ 2 \phi x^{(1)} {X_{\alpha}}^{2} \Delta + \phi X_{\alpha} \Delta(x^{(2)}-2x^{(1)})-1=0\\[2ex]
	&\text{if $\alpha=3,4$}\\
	&X_{\alpha}+ 2 \phi x^{(1)} {X_{\alpha}}^{2} \Delta-1=0\\[2ex]
	&\text{if $\alpha=5$}\\
	&X_{\alpha}+ \phi x^{(2)} {X_{\alpha}}^{2} \Delta + \phi X_{\alpha} \Delta(2x^{(1)}-x^{(2)})-1=0\\[2ex]
	&\text{if $\alpha=6,7,8$}\\
	&X_{\alpha}+ \phi x^{(2)} {X_{\alpha}}^{2} \Delta-1=0
\end{split}
\end{equation}

\noindent  We notice that if $x^{(2)}=2x^{(1)}$ all the equations in Eq.~\ref{eqn:X_deg} become equal to:

\begin{equation}
\label{eqn:X_deg_fin}
X+ \frac{2}{3} \phi X^{2} \Delta-1=0
\end{equation}

\noindent indeed $x^{(1)}+x^{(2)}=1$ and therefore $x^{(1)}=1/3$ and $x^{(2)}=2/3$. Hence the binary mixture displays an azeotrope behaving as a one component system at the specific non equimolar concentration of $x^{(1)}=1/3$ and $x^{(2)}=2/3$. 

\section{fully-connected bond example}

In the case of a binary mixture of patchy particles with four patches that differ only for their patch type, a possible interaction matrix satisfying the fully-connected bond recipe is: 

\begin{equation}
\mathbf{\Upsilon}_\text{N2c6s2}=
\label{matrix_cam}
	\begin{pmatrix}
	0 & 1 & 0 & 0 & 1 & 0 & 0 & 0\\
	1 & 0 & 0 & 0 & 0 & 1 & 0 & 0\\
	0 & 0 & 0 & 1 & 0 & 0 & 1 & 0\\
	0 & 0 & 1 & 0 & 0 & 0 & 0 & 1\\
	1 & 0 & 0 & 0 & 0 & 1 & 0 & 0\\
	0 & 1 & 0 & 0 & 1 & 0 & 0 & 0\\
	0 & 0 & 1 & 0 & 0 & 0 & 1&  0\\
	0 & 0 & 0 & 1 & 0 & 0 & 0 & 1
\end{pmatrix}
\bigskip
\end{equation}

\noindent Differently from the bond exclusivity interaction matrix that exhibits a single one for each row, this matrix has two ones for each row: the first is located among the first four columns (first species) and the other among the last four columns (second species). For a $N_s$-component mixture of patchy particles with $N_p$ patches we will have a $N_s\times N_p$ matrix with $N_s$ ones for each row: the first among the first group of $N_p$ columns, the second among the second group of $N_p$ columns, and so on.

\noindent In the following we demonstrate that with this binary mixture azeotropy is achieved without requiring equimolarity. Even better we show that this binary mixture exhibits azeotropy not only if the system is at a particular concentration, but whatever ratio the two species are mixed together. \\ In this case, in the mass balance equations for $X_{\alpha}^{(i)}$ the sum over the species $\sum_{j=1,2}$ does not drop out while the sum $\sum_{\gamma\in\Gamma(j)} X_{\gamma}^{(j)} \Delta_{\alpha\gamma}$  still reduces to one term as for the case where each patch can make a bond only with another patch. Indeed now, even if each patch makes a bond with two other patches, the patches involved in the bonds are located one on the first species and the other on the second species. Hence, for each patch $\alpha$, $\Delta_{\alpha\gamma}$ is different from zero only for two patches, $\gamma$ and $\delta$, not belonging to the same patchy particle species. Therefore $X_{\alpha}^{(i)}$ is recasted as

\begin{equation}
\label{eqn:X_cam_2}
X_{\alpha}^{(i)}=\frac{1}{1+\phi \biggl[x^{(i)}X_{\gamma}^{(i)} \Delta_{\alpha\gamma}+x^{(j)}X_{\delta}^{(j)} \Delta_{\alpha\delta}\biggr]}
\end{equation}   

\noindent Now we impose the equal bonding energy condition that allows to set $\Delta_{\alpha \gamma}$, for whatever $\alpha$ and $\gamma$, at the same value denoted as $\Delta$. In this way, for each patch $\alpha$, $X_{\alpha}^{(i)}$ becomes of the form

\begin{equation}
\label{eqn:X_cam_3}
X_{\alpha}^{(i)}=\frac{1}{1+\phi \biggl[x^{(i)}X_{\gamma}^{(i)}+x^{(j)}X_{\delta}^{(j)}\biggr]  \Delta}
\end{equation}   

\noindent In particular, considering the interaction matrix in Eq.~\ref{matrix_cam}, we have eight equations. For instance, the ones for the patches $1$ and $2$ are:
\begin{equation}
\label{eqn:8X_cam_1e2}
	\begin{array}{l}
			X_{1}^{(1)}=\frac{1}{1+\phi \biggl[x^{(1)}X_{2}^{(1)}+x^{(2)}X_{5}^{(2)}\biggr]  \Delta} \\[5ex]
			X_{2}^{(1)}=\frac{1}{1+\phi \biggl[x^{(1)}X_{1}^{(1)}+x^{(2)}X_{6}^{(2)}\biggr]  \Delta}
	\end{array} \\
\end{equation}

\noindent We notice that $X_1^{(1)}=X_6^{(2)}\equiv X$ and that $X_2^{(1)}=X_5^{(2)}\equiv X^{'}$. This implies that

\begin{equation}
\label{eqn:8X_cam_2_1}
	\begin{array}{l}
		X=\frac{1}{1+\phi [ x^{(1)}+x^{(2)}] X^{'} \Delta}=\frac{1}{1+\phi X^{'} \Delta} \\[3ex]
		X^{'}=\frac{1}{1+\phi [ x^{(1)}+x^{(2)}] X \Delta}=\frac{1}{1+\phi X \Delta} 
	\end{array} \\
\end{equation}  

\noindent By replacing the expression for $X^{'}$ in the equation for $X$ and vice-versa we obtain the two equal equations:

\begin{equation}
\label{eqn:8X_cam_2_1_bis}
	\begin{array}{l}
		X + X^2 \phi \Delta -1=0 \\[2ex]
		X^{'} + X^{'2} \phi \Delta -1=0
	\end{array} \\
\end{equation}  

\noindent Therefore, satisfying the same equations, $X_1^{(1)}=X_6^{(2)}=X_2^{(1)}=X_5^{(2)}$. We are left to demonstrate that also $X_3^{(1)}, X_4^{(1)}, X_7^{(2)}, X_8^{(2)}$ are defined by equations equal to the ones in Eq.~\ref{eqn:8X_cam_2_1_bis}. Firstly we notice that if $X_7^{(2)}=X_8^{(2)}$ then $X_3^{(1)}=X_4^{(1)}$ and this would imply that $X_7^{(2)}=X_8^{(2)}=X_3^{(1)}=X_4^{(1)}\equiv X^{''}$. Hence we can write

\begin{equation}
\label{eqn:8X_cam_2_2}
		X^{''}=\frac{1}{1+\phi [ x^{(1)}+x^{(2)}] X^{''} \Delta}=\frac{1}{1+\phi X^{''} \Delta}
\end{equation}  
 
\noindent which can be rewritten as 

\begin{equation}
\label{eqn:8X_cam_2_2_bis}
		X^{''} + X^{''2} \phi \Delta -1=0
\end{equation}  
i.e. the same equation as the ones reported in Eq.~\ref{eqn:8X_cam_2_1_bis}. Therefore if $X_7^{(2)}=X_8^{(2)}$ then all the $X_{\alpha}$, for whatever patch $\alpha$, are equal. The equalities of all the $X_{\alpha}$ are valid for whatever value $x^{(1)}$ (and so $x^{(2)}$) takes. This means that this binary mixture is always an azeotropic binary mixture.

\noindent Finally, the equality $X_7^{(2)}=X_8^{(2)}$ holds because, since the physics does not change if patch $7$ is replaced by patch $8$ and patch $3$ is replaced by patch $4$, then the equations must be invariant under these exchanges satisfying equalities $X_7^{(2)}=X_8^{(2)}$ and $X_3^{(1)}=X_4^{(1)}$.

\noindent In conclusion the bonding Helmholtz free energy of Eq.~\ref{eqn:betaf_bonding} is

\begin{equation}
\label{eqn:betaf_cam_2}
	\begin{split}
			\beta f_{bonding}=&x^{(1)}\biggl[ 4\biggl(\ln{X}-\frac{X}{2} \biggr) +\biggl(\frac{M}{2} \biggr) \biggr]+\\
			+& x^{(2)}\biggl[4\biggl(\ln{X}-\frac{X}{2} \biggr) +\biggl(\frac{M}{2} \biggr) \biggr] =\\
			=&\biggl( x^{(1)}+x^{(2)}\biggr) \biggl[ 4\biggl(\ln{X}-\frac{X}{2} \biggr) +\biggl(\frac{M}{2} \biggr) \biggr] =\\
			=& 4\biggl(\ln{X}-\frac{X}{2} \biggr) +\biggl(\frac{M}{2} \biggr)
		\end{split} 
\end{equation}
where $n(\Gamma(i))=M=4$ (with $i=1,2$) since we deal with patchy particles species having both four patches.  As expected, we notice that, for whatever concentration, $\beta f_{bonding}$ is equal to the free energy of a single component system. 

\section{An algorithm to generate DNA strands from the interaction matrix}
In the article we  introduced the interaction-matrix $\mathbf{\Upsilon}$, 
encoding the binding rules which must be satisfied by the patchy particles mixture.
We  also alluded to the possibility to use single-strand DNA  sequences
to encode the the binding rules, exploiting either  wireframe origami~\cite{liu2016diamond} or 
DNA-functionalized patchy  colloids~\cite{he2020colloidal}.  

In this appendix we present an algorithm to select sequences of single strands of DNA  to satisfy the desired interaction-matrix $\mathbf{\Upsilon}$.
We remember that DNA is a sequence of four types of nucleobases: adenine $A$, guanine $G$, thymine $T$ and cytosine $C$. The nucleobases can selectively bind to each other forming hydrogen bonds and the only possible base pairs are $A-T$ and $C-G$.  We also recall that the melting temperature of a DNA oligomer 
is a function of the length of the DNA complementary sequence.  For example, at a temperature at which  DNA complementary sequence of length  four are bonded,
DNA complementary sequence of length  two rarely bind. 

We focus here on the case relevant for this article, but the method can be
generalized to arbitrary binding rules.  Specifically, we focus on 
a binary mixture of particles with four patches each (see Supplementary Fig.~\ref{fig:cam_patches}),
interacting with the interaction matrix in Eq.~\ref{eq:matrix_cam_app} that satisfies the \textit{fully-connected} bond condition.

\begin{figure}[H]
	\centering
	\captionsetup{width=1\textwidth, justification=justified, singlelinecheck = false}
	\includegraphics[width=0.5\textwidth]{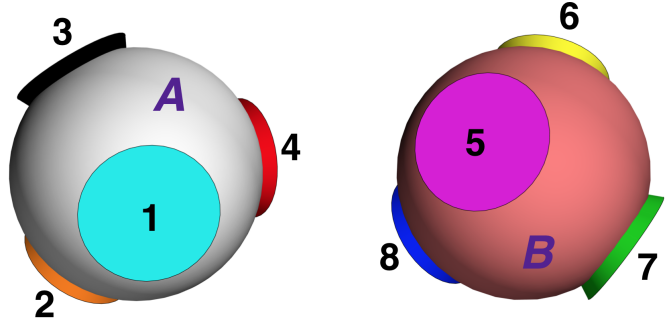}
	\caption{ \textbf{3D representation of patchy particles.} $A$ and $B$ are two patchy particles species with four patches indicated by numbers. They interact establishing bonds through patches according to the  interaction matrix in Eq.~\ref{eq:matrix_cam_app}.}
	\label{fig:cam_patches}
\end{figure}

\begin{equation}
\label{eq:matrix_cam_app}
	\begin{pmatrix}
	0 & 1 & 0 & 0 & 1 & 0 & 0 & 0\\
	1 & 0 & 0 & 0 & 0 & 1 & 0 & 0\\
	0 & 0 & 0 & 1 & 0 & 0 & 1 & 0\\
	0 & 0 & 1 & 0 & 0 & 0 & 0 & 1\\
	1 & 0 & 0 & 0 & 0 & 1 & 0 & 0\\
	0 & 1 & 0 & 0 & 1 & 0 & 0 & 0\\
	0 & 0 & 1 & 0 & 0 & 0 & 1&  0\\
	0 & 0 & 0 & 1 & 0 & 0 & 0 & 1
\end{pmatrix}
\end{equation}

The interaction matrix can also be represented as a list of nodes (the 
eight patches, labeled from 1 to 8 in Fig.~\ref{fig:cam_patches}
) connected by lines representing the 1s in the interaction matrix, resulting in the
connected graph  in Supplementary Fig.~\ref{fig:anello_e_catena}. 
We note that  in the "ring" forming graph (left graph in Supplementary Fig.~\ref{fig:anello_e_catena}), each patch binds to two different 
patches, while in the "chain" graph  (right graph), the first and the last node binds to one identical patches and to a different patch.

\begin{figure}[H]
	\centering
	\captionsetup{width=1\textwidth, justification=justified, singlelinecheck = false}
	\includegraphics[width=0.7\textwidth]{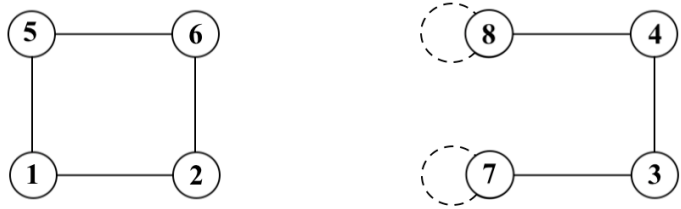}
	\caption{\textbf{Bond topology.} The  patch connections according to the color interaction matrix in Eq.~\ref{eq:matrix_cam_app} correspond to two disconnected bonded "clusters":
	 a ring of four patches and  a chain of four patches.}
	\label{fig:anello_e_catena}
\end{figure}

\noindent  For this specific case, one need to 
find DNA strands that represent, as close as possible,  the interaction matrix in Eq.~\ref{eq:matrix_cam_app} or equivalently the bond topology in Supplementary Fig.~\ref{fig:anello_e_catena}.   Specifically, a strand must
be able to form bonds with up to two different other strands. 
This "double bonding" condition can be realized by defining a bond
as a sequence  of $n_b$ consecutive base pairs (a realistic value could be
$n_b=4$) and a number of nucleotides in the DNA single strand $n_s$ larger than
$n_b$. An example of this type of double bonding, for  $n_s=6$,  is shown in Supplementary Fig.~\ref{fig:3strands}.

\begin{figure}[htbp] %
   \centering
   \includegraphics[width=3in]{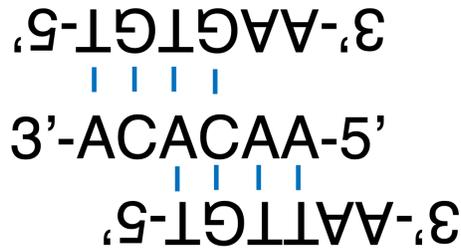} 
   \caption{\textbf{Schematic example of the double bonding of a single DNA strand.} Three single-strands DNA in which the central one is able to binds with four consecutive nucleotides with two other strands (reversed in their 3'-5' order).}
   \label{fig:3strands}
\end{figure}

To identify eight DNA single strand of length $n_s$ which satisfy   the interaction matrix in Eq.~\ref{eq:matrix_cam_app}, we propose the following algorithm

\begin{itemize}
\item
we generate all the $4^{n_s}$ different oligomers of length $n_s$ and
evaluate, for each pair of them the  maximum number of consecutive bases $n_{max}$ that bind to each other.  It is important to remind that DNA has directionality and the two complementary strands that form it have opposite directions: one goes from the five-prime end to the three-prime end $5^{'}\longrightarrow 3^{'}$ and the other one from the three-prime end to the five-prime end $3^{'}\longrightarrow 5^{'}$. 
We then construct a $4^{n_s} \times 4^{n_s}$ matrix  whose elements are the
strength of the binding between the two strands. For clarity, we identify here 
the strength with  the $n_{max}$ value. A more elaborate formulation could use the
binding Gibbs free energy or the strand pair melting temperature,  calculated for example
with the SantaLucia nearest-neighbor model~\cite{santalucia1998unified}.

\item Set a threshold $n_{\rm threshold}$  for the largest value of $n_{\rm max}$ which can be safely assumed as non-bonding.  In the case of $n_b=4$, this can be chosen as 2. Indeed at the melting temperature of sequences of length 4, the binding probability of sequences of length 2 is negligible. 
Put zero in the $4^{n_s} \times 4^{n_s}$ matrix of $n_{\rm max}$ for all 
elements  $n_{\rm max} \le n_{\rm threshold}$. In this way, the $4^{n_s} \times 4^{n_s}$ matrix has non-zero elements only for pair of sequences which bind to each other.

\item 

Eliminate from the set of all possible strands (randomly) one of the two strands for which $n_{\rm max}$ is larger than $n_b$ ($n_{\rm max}=5$ and 6 in our example).  This eliminates, among others, one of each pair of complementary strands and   one of the two self-complementary sequences (palindromic in the DNA nomenclature).  To eliminate the strand, it is sufficient to fill with zero 
the row and column associated to that strand. This makes sure that, among the remaining set of strands, the strongest binding is indeed the ones with $n_{max}=n_b$.

\item
By now, the $4^{n_s} \times 4^{n_s}$ matrix contains, weighted by their strength, all possible bonds between all possible strands, with values of $n_{\rm max}$ from 
$n_{\rm threshold}+1$ to $n_b$ (from 3 to 4 in the specific case). 
This matrix can also be seen as  a network of bonds (links) between the $4^{n_s}$ strands (nodes).  We then run a  search on this network to identify the 
desired bonding clusters.  In the specific case (Supplementary Fig.~\ref{fig:anello_e_catena})
we first identify all rings of four nodes connected by links with $n_{\rm max}=n_b$.
We then eliminate the rings which can be short-cut by a intra-ring bond and
all rings in which one of the node is able to self-bind.

\item
For  each of the remaining ring, we eliminate from the list of all possible
DNA strands, the strands (i.e. the nodes) which are connected to the ring strands.
This guarantee that the four strands defining the  ring do not 
interact with any other remaining strand. Among the remaining nodes, we search for
all chains of length four (in this specific case) on the network starting and ending with
a node which is able to self-bind. As for the ring, we eliminate all chains in which 
intra-chain bonds are present.

\item 
Iterating this procedure for each ring results in a list of 
DNA strands, all good candidates
 to experimentally realise the required interaction matrix.  In a more refined treatment, each of these possible solutions can be examined to select the
 smallest variance in the $\Delta G$  of binding between different patches
 (compensating the different strength of the A-T and G-C pairing), and further restrict sequence selection so that $\Delta G$ distance between pairs with $n_{b}$ and pairs with $n_{\rm threshold}$ is as large as possible to minimize crosstalk.
 $\Delta G$ of binding between any pair of sequences would be calculated using SantaLucia's model, and can be obtained from with available strand analysis and optimization tools such as NUPACK \cite{fornace2022nupack}.

\end{itemize}

\noindent In Eq.~\ref{tab:cam_sequences} we report one of the possible sets of $6$ bases to originate $8$ sequences that bind according to the request topology in Fig.~\ref{fig:anello_e_catena}.  

\begin{equation}
\label{tab:cam_sequences}
	\begin{array}{l}
		\text{\ding[1.5]{172}} \quad = \quad AAGGGG  \\[2ex]
		\text{\ding[1.5]{173}} \quad = \quad CCCCCC \\[2ex]
		\text{\ding[1.5]{174}} \quad = \quad  ACACAA \\[2ex]
		\text{\ding[1.5]{175}} \quad = \quad AAGTGT \\[2ex]
		\text{\ding[1.5]{176}} \quad = \quad CCCTCA \\[2ex]
		\text{\ding[1.5]{177}} \quad = \quad AGGGGA \\[2ex]
		\text{\ding[1.5]{178}} \quad = \quad AATTGT \\[2ex]
		\text{\ding[1.5]{179}} \quad = \quad CACTAG
	\end{array}\\ 
\end{equation} 

This sequence correspond to the following matrix of maximum number of
paired base pairs.  
\begin{equation}
\label{matrix_cam_sequences}
	\begin{pmatrix}
	       0 &           4 &           1 &           1 &           4 &           0 &           2 &           2 \\
           4 &           0 &           0 &           1 &           0 &           4 &           1 &           1 \\
           1 &           0 &           0 &           4 &           1 &           1 &           4 &           1 \\
           1 &           1 &           4 &           1 &           2 &           1 &           2 &           4 \\
           4 &           0 &           1 &           2 &           1 &           4 &           2 &           2 \\
           0 &           4 &           1 &           1 &           4 &           0 &           1 &           2 \\
           2 &           1 &           4 &           2 &           2 &           1 &           4 &           2 \\
           2 &           1 &           1 &           4 &           2 &           2 &           2 &           4 \end{pmatrix}
\end{equation}

Considering as bonded only the elements filled with a 4
(i.e. neglecting pairing of one or two nucleotides), this matrix 
 coincides with the matrix in
Eq.~\ref{eq:matrix_cam_app}.

For the interested reader, we call attention on the fact that Eq.~\ref{eq:matrix_cam_app} has two identical row/columns. This makes it possible to reduce, if needed, the number of distinct colors from eight to six without altering the connectivity table. Strand 	\text{\ding[1.5]{176}}  in Eq.~\ref{tab:cam_sequences} can be substituted with 
strand \text{\ding[1.5]{173}}  and  strand \text{\ding[1.5]{172}}  can be substituted with 
strand \text{\ding[1.5]{177}}  (or viceversa).


\begin{thebibliography}{89}
\expandafter\ifx\csname natexlab\endcsname\relax\def\natexlab#1{#1}\fi
\expandafter\ifx\csname bibnamefont\endcsname\relax
  \def\bibnamefont#1{#1}\fi
\expandafter\ifx\csname bibfnamefont\endcsname\relax
  \def\bibfnamefont#1{#1}\fi
\expandafter\ifx\csname citenamefont\endcsname\relax
  \def\citenamefont#1{#1}\fi
\expandafter\ifx\csname url\endcsname\relax
  \def\url#1{\texttt{#1}}\fi
\expandafter\ifx\csname urlprefix\endcsname\relax\def\urlprefix{URL }\fi
\providecommand{\bibinfo}[2]{#2}
\providecommand{\eprint}[2][]{\url{#2}}

\bibitem[{\citenamefont{Whitelam and Jack}(2015)}]{whitelam2015statistical}
\bibinfo{author}{\bibfnamefont{S.}~\bibnamefont{Whitelam}} \bibnamefont{and}
  \bibinfo{author}{\bibfnamefont{R.~L.} \bibnamefont{Jack}},
  \bibinfo{journal}{Annual review of physical chemistry}
  \textbf{\bibinfo{volume}{66}}, \bibinfo{pages}{143} (\bibinfo{year}{2015}).

\bibitem[{\citenamefont{Kumar et~al.}(2017)\citenamefont{Kumar, Kumaraswamy,
  Prasad, Bandyopadhyaya, Granick, Gang, Manoharan, Frenkel, and
  Kotov}}]{kumar2017nanoparticle}
\bibinfo{author}{\bibfnamefont{S.~K.} \bibnamefont{Kumar}},
  \bibinfo{author}{\bibfnamefont{G.}~\bibnamefont{Kumaraswamy}},
  \bibinfo{author}{\bibfnamefont{B.~L.} \bibnamefont{Prasad}},
  \bibinfo{author}{\bibfnamefont{R.}~\bibnamefont{Bandyopadhyaya}},
  \bibinfo{author}{\bibfnamefont{S.}~\bibnamefont{Granick}},
  \bibinfo{author}{\bibfnamefont{O.}~\bibnamefont{Gang}},
  \bibinfo{author}{\bibfnamefont{V.~N.} \bibnamefont{Manoharan}},
  \bibinfo{author}{\bibfnamefont{D.}~\bibnamefont{Frenkel}}, \bibnamefont{and}
  \bibinfo{author}{\bibfnamefont{N.~A.} \bibnamefont{Kotov}},
  \bibinfo{journal}{Current Science} pp. \bibinfo{pages}{1635--1641}
  (\bibinfo{year}{2017}).

\bibitem[{\citenamefont{Filion et~al.}(2009)\citenamefont{Filion, Marechal, van
  Oorschot, Pelt, Smallenburg, and Dijkstra}}]{filion2009efficient}
\bibinfo{author}{\bibfnamefont{L.}~\bibnamefont{Filion}},
  \bibinfo{author}{\bibfnamefont{M.}~\bibnamefont{Marechal}},
  \bibinfo{author}{\bibfnamefont{B.}~\bibnamefont{van Oorschot}},
  \bibinfo{author}{\bibfnamefont{D.}~\bibnamefont{Pelt}},
  \bibinfo{author}{\bibfnamefont{F.}~\bibnamefont{Smallenburg}},
  \bibnamefont{and} \bibinfo{author}{\bibfnamefont{M.}~\bibnamefont{Dijkstra}},
  \bibinfo{journal}{Physical review letters} \textbf{\bibinfo{volume}{103}},
  \bibinfo{pages}{188302} (\bibinfo{year}{2009}).

\bibitem[{\citenamefont{Filion and Dijkstra}(2009)}]{filion2009prediction}
\bibinfo{author}{\bibfnamefont{L.}~\bibnamefont{Filion}} \bibnamefont{and}
  \bibinfo{author}{\bibfnamefont{M.}~\bibnamefont{Dijkstra}},
  \bibinfo{journal}{Physical Review E} \textbf{\bibinfo{volume}{79}},
  \bibinfo{pages}{046714} (\bibinfo{year}{2009}).

\bibitem[{\citenamefont{Jee et~al.}(2016)\citenamefont{Jee, Lou, Jang,
  Nagamanasa, and Granick}}]{jee2016nanoparticle}
\bibinfo{author}{\bibfnamefont{A.-Y.} \bibnamefont{Jee}},
  \bibinfo{author}{\bibfnamefont{K.}~\bibnamefont{Lou}},
  \bibinfo{author}{\bibfnamefont{H.-S.} \bibnamefont{Jang}},
  \bibinfo{author}{\bibfnamefont{K.~H.} \bibnamefont{Nagamanasa}},
  \bibnamefont{and} \bibinfo{author}{\bibfnamefont{S.}~\bibnamefont{Granick}},
  \bibinfo{journal}{Faraday discussions} \textbf{\bibinfo{volume}{186}},
  \bibinfo{pages}{11} (\bibinfo{year}{2016}).

\bibitem[{\citenamefont{Dijkstra and Luijten}(2021)}]{dijkstra2021predictive}
\bibinfo{author}{\bibfnamefont{M.}~\bibnamefont{Dijkstra}} \bibnamefont{and}
  \bibinfo{author}{\bibfnamefont{E.}~\bibnamefont{Luijten}},
  \bibinfo{journal}{Nature Materials} \textbf{\bibinfo{volume}{20}},
  \bibinfo{pages}{762} (\bibinfo{year}{2021}).

\bibitem[{\citenamefont{Bupathy et~al.}(2022)\citenamefont{Bupathy, Frenkel,
  and Sastry}}]{bupathy2022temperature}
\bibinfo{author}{\bibfnamefont{A.}~\bibnamefont{Bupathy}},
  \bibinfo{author}{\bibfnamefont{D.}~\bibnamefont{Frenkel}}, \bibnamefont{and}
  \bibinfo{author}{\bibfnamefont{S.}~\bibnamefont{Sastry}},
  \bibinfo{journal}{Proceedings of the National Academy of Sciences}
  \textbf{\bibinfo{volume}{119}}, \bibinfo{pages}{e2119315119}
  (\bibinfo{year}{2022}).

\bibitem[{\citenamefont{Rechtsman et~al.}(2005)\citenamefont{Rechtsman,
  Stillinger, and Torquato}}]{rechtsman2005optimized}
\bibinfo{author}{\bibfnamefont{M.~C.} \bibnamefont{Rechtsman}},
  \bibinfo{author}{\bibfnamefont{F.~H.} \bibnamefont{Stillinger}},
  \bibnamefont{and} \bibinfo{author}{\bibfnamefont{S.}~\bibnamefont{Torquato}},
  \bibinfo{journal}{Physical review letters} \textbf{\bibinfo{volume}{95}},
  \bibinfo{pages}{228301} (\bibinfo{year}{2005}).

\bibitem[{\citenamefont{Marcotte et~al.}(2011)\citenamefont{Marcotte,
  Stillinger, and Torquato}}]{marcotte2011optimized}
\bibinfo{author}{\bibfnamefont{E.}~\bibnamefont{Marcotte}},
  \bibinfo{author}{\bibfnamefont{F.~H.} \bibnamefont{Stillinger}},
  \bibnamefont{and} \bibinfo{author}{\bibfnamefont{S.}~\bibnamefont{Torquato}},
  \bibinfo{journal}{Soft Matter} \textbf{\bibinfo{volume}{7}},
  \bibinfo{pages}{2332} (\bibinfo{year}{2011}).

\bibitem[{\citenamefont{Marcotte et~al.}(2013)\citenamefont{Marcotte,
  Stillinger, and Torquato}}]{marcotte2013designeddiamond}
\bibinfo{author}{\bibfnamefont{E.}~\bibnamefont{Marcotte}},
  \bibinfo{author}{\bibfnamefont{F.~H.} \bibnamefont{Stillinger}},
  \bibnamefont{and} \bibinfo{author}{\bibfnamefont{S.}~\bibnamefont{Torquato}},
  \bibinfo{journal}{The Journal of Chemical Physics}
  \textbf{\bibinfo{volume}{138}}, \bibinfo{pages}{061101}
  (\bibinfo{year}{2013}).

\bibitem[{\citenamefont{Zhang et~al.}(2013)\citenamefont{Zhang, Stillinger, and
  Torquato}}]{zhang2013probing}
\bibinfo{author}{\bibfnamefont{G.}~\bibnamefont{Zhang}},
  \bibinfo{author}{\bibfnamefont{F.}~\bibnamefont{Stillinger}},
  \bibnamefont{and} \bibinfo{author}{\bibfnamefont{S.}~\bibnamefont{Torquato}},
  \bibinfo{journal}{Physical Review E} \textbf{\bibinfo{volume}{88}},
  \bibinfo{pages}{042309} (\bibinfo{year}{2013}).

\bibitem[{\citenamefont{Miskin et~al.}(2016)\citenamefont{Miskin, Khaira,
  de~Pablo, and Jaeger}}]{miskin2016turning}
\bibinfo{author}{\bibfnamefont{M.~Z.} \bibnamefont{Miskin}},
  \bibinfo{author}{\bibfnamefont{G.}~\bibnamefont{Khaira}},
  \bibinfo{author}{\bibfnamefont{J.~J.} \bibnamefont{de~Pablo}},
  \bibnamefont{and} \bibinfo{author}{\bibfnamefont{H.~M.}
  \bibnamefont{Jaeger}}, \bibinfo{journal}{Proceedings of the National Academy
  of Sciences} \textbf{\bibinfo{volume}{113}}, \bibinfo{pages}{34}
  (\bibinfo{year}{2016}).

\bibitem[{\citenamefont{Lindquist et~al.}(2016)\citenamefont{Lindquist,
  Jadrich, and Truskett}}]{lindquist2016communication}
\bibinfo{author}{\bibfnamefont{B.~A.} \bibnamefont{Lindquist}},
  \bibinfo{author}{\bibfnamefont{R.~B.} \bibnamefont{Jadrich}},
  \bibnamefont{and} \bibinfo{author}{\bibfnamefont{T.~M.}
  \bibnamefont{Truskett}}, \bibinfo{journal}{The Journal of Chemical Physics}
  \textbf{\bibinfo{volume}{145}}, \bibinfo{pages}{111101}
  (\bibinfo{year}{2016}).

\bibitem[{\citenamefont{Chen et~al.}(2018)\citenamefont{Chen, Zhang, and
  Torquato}}]{chen2018inverse}
\bibinfo{author}{\bibfnamefont{D.}~\bibnamefont{Chen}},
  \bibinfo{author}{\bibfnamefont{G.}~\bibnamefont{Zhang}}, \bibnamefont{and}
  \bibinfo{author}{\bibfnamefont{S.}~\bibnamefont{Torquato}},
  \bibinfo{journal}{The Journal of Physical Chemistry B}
  \textbf{\bibinfo{volume}{122}}, \bibinfo{pages}{8462} (\bibinfo{year}{2018}).

\bibitem[{\citenamefont{Kumar et~al.}(2019)\citenamefont{Kumar, Coli, Dijkstra,
  and Sastry}}]{kumar2019inverse}
\bibinfo{author}{\bibfnamefont{R.}~\bibnamefont{Kumar}},
  \bibinfo{author}{\bibfnamefont{G.~M.} \bibnamefont{Coli}},
  \bibinfo{author}{\bibfnamefont{M.}~\bibnamefont{Dijkstra}}, \bibnamefont{and}
  \bibinfo{author}{\bibfnamefont{S.}~\bibnamefont{Sastry}},
  \bibinfo{journal}{The Journal of chemical physics}
  \textbf{\bibinfo{volume}{151}}, \bibinfo{pages}{084109}
  (\bibinfo{year}{2019}).

\bibitem[{\citenamefont{Whitelam and
  Tamblyn}(2021)}]{whitelam2021neuroevolutionary}
\bibinfo{author}{\bibfnamefont{S.}~\bibnamefont{Whitelam}} \bibnamefont{and}
  \bibinfo{author}{\bibfnamefont{I.}~\bibnamefont{Tamblyn}},
  \bibinfo{journal}{Physical review letters} \textbf{\bibinfo{volume}{127}},
  \bibinfo{pages}{018003} (\bibinfo{year}{2021}).

\bibitem[{\citenamefont{Ducrot et~al.}(2017)\citenamefont{Ducrot, He, Yi, and
  Pine}}]{ducrot2017colloidal}
\bibinfo{author}{\bibfnamefont{{\'E}.}~\bibnamefont{Ducrot}},
  \bibinfo{author}{\bibfnamefont{M.}~\bibnamefont{He}},
  \bibinfo{author}{\bibfnamefont{G.-R.} \bibnamefont{Yi}}, \bibnamefont{and}
  \bibinfo{author}{\bibfnamefont{D.~J.} \bibnamefont{Pine}},
  \bibinfo{journal}{Nature materials} \textbf{\bibinfo{volume}{16}},
  \bibinfo{pages}{652} (\bibinfo{year}{2017}).

\bibitem[{\citenamefont{Nelson}(2002)}]{nelson2002toward}
\bibinfo{author}{\bibfnamefont{D.~R.} \bibnamefont{Nelson}},
  \bibinfo{journal}{Nano Letters} \textbf{\bibinfo{volume}{2}},
  \bibinfo{pages}{1125} (\bibinfo{year}{2002}).

\bibitem[{\citenamefont{Manoharan et~al.}(2003)\citenamefont{Manoharan,
  Elsesser, and Pine}}]{manoharan2003dense}
\bibinfo{author}{\bibfnamefont{V.~N.} \bibnamefont{Manoharan}},
  \bibinfo{author}{\bibfnamefont{M.~T.} \bibnamefont{Elsesser}},
  \bibnamefont{and} \bibinfo{author}{\bibfnamefont{D.~J.} \bibnamefont{Pine}},
  \bibinfo{journal}{Science} \textbf{\bibinfo{volume}{301}},
  \bibinfo{pages}{483} (\bibinfo{year}{2003}).

\bibitem[{\citenamefont{Zhang et~al.}(2005)\citenamefont{Zhang, Keys, Chen, and
  Glotzer}}]{zhang2005self}
\bibinfo{author}{\bibfnamefont{Z.}~\bibnamefont{Zhang}},
  \bibinfo{author}{\bibfnamefont{A.~S.} \bibnamefont{Keys}},
  \bibinfo{author}{\bibfnamefont{T.}~\bibnamefont{Chen}}, \bibnamefont{and}
  \bibinfo{author}{\bibfnamefont{S.~C.} \bibnamefont{Glotzer}},
  \bibinfo{journal}{Langmuir} \textbf{\bibinfo{volume}{21}},
  \bibinfo{pages}{11547} (\bibinfo{year}{2005}).

\bibitem[{\citenamefont{Romano et~al.}(2014)\citenamefont{Romano, Russo, and
  Tanaka}}]{romano2014influence}
\bibinfo{author}{\bibfnamefont{F.}~\bibnamefont{Romano}},
  \bibinfo{author}{\bibfnamefont{J.}~\bibnamefont{Russo}}, \bibnamefont{and}
  \bibinfo{author}{\bibfnamefont{H.}~\bibnamefont{Tanaka}},
  \bibinfo{journal}{Physical review letters} \textbf{\bibinfo{volume}{113}},
  \bibinfo{pages}{138303} (\bibinfo{year}{2014}).

\bibitem[{\citenamefont{Halverson and Tkachenko}(2013)}]{halverson2013dna}
\bibinfo{author}{\bibfnamefont{J.~D.} \bibnamefont{Halverson}}
  \bibnamefont{and} \bibinfo{author}{\bibfnamefont{A.~V.}
  \bibnamefont{Tkachenko}}, \bibinfo{journal}{Physical Review E}
  \textbf{\bibinfo{volume}{87}}, \bibinfo{pages}{062310}
  (\bibinfo{year}{2013}).

\bibitem[{\citenamefont{Romano and Sciortino}(2012)}]{romano2012patterning}
\bibinfo{author}{\bibfnamefont{F.}~\bibnamefont{Romano}} \bibnamefont{and}
  \bibinfo{author}{\bibfnamefont{F.}~\bibnamefont{Sciortino}},
  \bibinfo{journal}{Nature communications} \textbf{\bibinfo{volume}{3}},
  \bibinfo{pages}{975} (\bibinfo{year}{2012}).

\bibitem[{\citenamefont{Tracey et~al.}(2019)\citenamefont{Tracey, Noya, and
  Doye}}]{tracey2019programming}
\bibinfo{author}{\bibfnamefont{D.~F.} \bibnamefont{Tracey}},
  \bibinfo{author}{\bibfnamefont{E.~G.} \bibnamefont{Noya}}, \bibnamefont{and}
  \bibinfo{author}{\bibfnamefont{J.~P.~K.} \bibnamefont{Doye}},
  \bibinfo{journal}{The Journal of Chemical Physics}
  \textbf{\bibinfo{volume}{151}}, \bibinfo{pages}{224506}
  (\bibinfo{year}{2019}).

\bibitem[{\citenamefont{Russo et~al.}(2022)\citenamefont{Russo, Romano, Kroc,
  Sciortino, Rovigatti, and {\v{S}}ulc}}]{russo2022sat}
\bibinfo{author}{\bibfnamefont{J.}~\bibnamefont{Russo}},
  \bibinfo{author}{\bibfnamefont{F.}~\bibnamefont{Romano}},
  \bibinfo{author}{\bibfnamefont{L.}~\bibnamefont{Kroc}},
  \bibinfo{author}{\bibfnamefont{F.}~\bibnamefont{Sciortino}},
  \bibinfo{author}{\bibfnamefont{L.}~\bibnamefont{Rovigatti}},
  \bibnamefont{and}
  \bibinfo{author}{\bibfnamefont{P.}~\bibnamefont{{\v{S}}ulc}},
  \bibinfo{journal}{Journal of Physics: Condensed Matter}
  (\bibinfo{year}{2022}).

\bibitem[{\citenamefont{Srinivasan et~al.}(2013)\citenamefont{Srinivasan, Vo,
  Zhang, Gang, Kumar, and Venkatasubramanian}}]{srinivasan2013designing}
\bibinfo{author}{\bibfnamefont{B.}~\bibnamefont{Srinivasan}},
  \bibinfo{author}{\bibfnamefont{T.}~\bibnamefont{Vo}},
  \bibinfo{author}{\bibfnamefont{Y.}~\bibnamefont{Zhang}},
  \bibinfo{author}{\bibfnamefont{O.}~\bibnamefont{Gang}},
  \bibinfo{author}{\bibfnamefont{S.}~\bibnamefont{Kumar}}, \bibnamefont{and}
  \bibinfo{author}{\bibfnamefont{V.}~\bibnamefont{Venkatasubramanian}},
  \bibinfo{journal}{Proceedings of the National Academy of Sciences}
  \textbf{\bibinfo{volume}{110}}, \bibinfo{pages}{18431}
  (\bibinfo{year}{2013}).

\bibitem[{\citenamefont{Whitelam and Tamblyn}(2020)}]{whitelam2020learning}
\bibinfo{author}{\bibfnamefont{S.}~\bibnamefont{Whitelam}} \bibnamefont{and}
  \bibinfo{author}{\bibfnamefont{I.}~\bibnamefont{Tamblyn}},
  \bibinfo{journal}{Physical Review E} \textbf{\bibinfo{volume}{101}},
  \bibinfo{pages}{052604} (\bibinfo{year}{2020}).

\bibitem[{\citenamefont{Romano et~al.}(2020)\citenamefont{Romano, Russo, Kroc,
  and {\v{S}}ulc}}]{romano2020designing}
\bibinfo{author}{\bibfnamefont{F.}~\bibnamefont{Romano}},
  \bibinfo{author}{\bibfnamefont{J.}~\bibnamefont{Russo}},
  \bibinfo{author}{\bibfnamefont{L.}~\bibnamefont{Kroc}}, \bibnamefont{and}
  \bibinfo{author}{\bibfnamefont{P.}~\bibnamefont{{\v{S}}ulc}},
  \bibinfo{journal}{Physical Review Letters} \textbf{\bibinfo{volume}{125}},
  \bibinfo{pages}{118003} (\bibinfo{year}{2020}).

\bibitem[{\citenamefont{Een}(2005)}]{een2005minisat}
\bibinfo{author}{\bibfnamefont{N.}~\bibnamefont{Een}}, in
  \emph{\bibinfo{booktitle}{Proc. SAT-05: 8th Int. Conf. on Theory and
  Applications of Satisfiability Testing}} (\bibinfo{year}{2005}), pp.
  \bibinfo{pages}{502--518}.

\bibitem[{\citenamefont{Akahane et~al.}(2016)\citenamefont{Akahane, Russo, and
  Tanaka}}]{akahane2016possible}
\bibinfo{author}{\bibfnamefont{K.}~\bibnamefont{Akahane}},
  \bibinfo{author}{\bibfnamefont{J.}~\bibnamefont{Russo}}, \bibnamefont{and}
  \bibinfo{author}{\bibfnamefont{H.}~\bibnamefont{Tanaka}},
  \bibinfo{journal}{Nature communications} \textbf{\bibinfo{volume}{7}},
  \bibinfo{pages}{1} (\bibinfo{year}{2016}).

\bibitem[{\citenamefont{Wolde and Frenkel}(1997)}]{wolde1997enhancement}
\bibinfo{author}{\bibfnamefont{P.~R.~t.} \bibnamefont{Wolde}} \bibnamefont{and}
  \bibinfo{author}{\bibfnamefont{D.}~\bibnamefont{Frenkel}},
  \bibinfo{journal}{Science} \textbf{\bibinfo{volume}{277}},
  \bibinfo{pages}{1975} (\bibinfo{year}{1997}).

\bibitem[{\citenamefont{Xu et~al.}(2012)\citenamefont{Xu, Buldyrev, Stanley,
  and Franzese}}]{xu2012homogeneous}
\bibinfo{author}{\bibfnamefont{L.}~\bibnamefont{Xu}},
  \bibinfo{author}{\bibfnamefont{S.~V.} \bibnamefont{Buldyrev}},
  \bibinfo{author}{\bibfnamefont{H.~E.} \bibnamefont{Stanley}},
  \bibnamefont{and} \bibinfo{author}{\bibfnamefont{G.}~\bibnamefont{Franzese}},
  \bibinfo{journal}{Physical Review Letters} \textbf{\bibinfo{volume}{109}},
  \bibinfo{pages}{095702} (\bibinfo{year}{2012}).

\bibitem[{\citenamefont{Russo et~al.}(2018)\citenamefont{Russo, Romano, and
  Tanaka}}]{russo2018glass}
\bibinfo{author}{\bibfnamefont{J.}~\bibnamefont{Russo}},
  \bibinfo{author}{\bibfnamefont{F.}~\bibnamefont{Romano}}, \bibnamefont{and}
  \bibinfo{author}{\bibfnamefont{H.}~\bibnamefont{Tanaka}},
  \bibinfo{journal}{Physical Review X} \textbf{\bibinfo{volume}{8}},
  \bibinfo{pages}{021040} (\bibinfo{year}{2018}).

\bibitem[{\citenamefont{Russo et~al.}(2021)\citenamefont{Russo, Leoni,
  Martelli, and Sciortino}}]{russo2021physics}
\bibinfo{author}{\bibfnamefont{J.}~\bibnamefont{Russo}},
  \bibinfo{author}{\bibfnamefont{F.}~\bibnamefont{Leoni}},
  \bibinfo{author}{\bibfnamefont{F.}~\bibnamefont{Martelli}}, \bibnamefont{and}
  \bibinfo{author}{\bibfnamefont{F.}~\bibnamefont{Sciortino}},
  \bibinfo{journal}{Reports on Progress in Physics}  (\bibinfo{year}{2021}).

\bibitem[{\citenamefont{Wertheim}(1984)}]{wertheim1984fluids}
\bibinfo{author}{\bibfnamefont{M.}~\bibnamefont{Wertheim}},
  \bibinfo{journal}{Journal of statistical physics}
  \textbf{\bibinfo{volume}{35}}, \bibinfo{pages}{19} (\bibinfo{year}{1984}).

\bibitem[{\citenamefont{Chapman et~al.}(1988)\citenamefont{Chapman, Jackson,
  and Gubbins}}]{chapman1988phase}
\bibinfo{author}{\bibfnamefont{W.~G.} \bibnamefont{Chapman}},
  \bibinfo{author}{\bibfnamefont{G.}~\bibnamefont{Jackson}}, \bibnamefont{and}
  \bibinfo{author}{\bibfnamefont{K.~E.} \bibnamefont{Gubbins}},
  \bibinfo{journal}{Molecular Physics} \textbf{\bibinfo{volume}{65}},
  \bibinfo{pages}{1} (\bibinfo{year}{1988}).

\bibitem[{\citenamefont{de~Las~Heras et~al.}(2011)\citenamefont{de~Las~Heras,
  Tavares, and da~Gama}}]{de2011phase}
\bibinfo{author}{\bibfnamefont{D.}~\bibnamefont{de~Las~Heras}},
  \bibinfo{author}{\bibfnamefont{J.~M.} \bibnamefont{Tavares}},
  \bibnamefont{and} \bibinfo{author}{\bibfnamefont{M.~M.~T.}
  \bibnamefont{da~Gama}}, \bibinfo{journal}{Soft Matter}
  \textbf{\bibinfo{volume}{7}}, \bibinfo{pages}{5615} (\bibinfo{year}{2011}).

\bibitem[{\citenamefont{Bianchi et~al.}(2006)\citenamefont{Bianchi, Largo,
  Tartaglia, Zaccarelli, and Sciortino}}]{bianchi2006empty}
\bibinfo{author}{\bibfnamefont{E.}~\bibnamefont{Bianchi}},
  \bibinfo{author}{\bibfnamefont{J.}~\bibnamefont{Largo}},
  \bibinfo{author}{\bibfnamefont{P.}~\bibnamefont{Tartaglia}},
  \bibinfo{author}{\bibfnamefont{E.}~\bibnamefont{Zaccarelli}},
  \bibnamefont{and}
  \bibinfo{author}{\bibfnamefont{F.}~\bibnamefont{Sciortino}},
  \bibinfo{journal}{Physical Review Letters} \textbf{\bibinfo{volume}{97}},
  \bibinfo{pages}{168301} (\bibinfo{year}{2006}).

\bibitem[{\citenamefont{Heras et~al.}(2011)\citenamefont{Heras, Tavares, and
  da~Gama}}]{heras2011phase}
\bibinfo{author}{\bibfnamefont{D.~d.~l.} \bibnamefont{Heras}},
  \bibinfo{author}{\bibfnamefont{J.~M.} \bibnamefont{Tavares}},
  \bibnamefont{and} \bibinfo{author}{\bibfnamefont{M.~M.~T.}
  \bibnamefont{da~Gama}}, \bibinfo{journal}{The Journal of chemical physics}
  \textbf{\bibinfo{volume}{134}}, \bibinfo{pages}{104904}
  (\bibinfo{year}{2011}).

\bibitem[{\citenamefont{Rovigatti et~al.}(2013)\citenamefont{Rovigatti, de~las
  Heras, Tavares, Telo~da Gama, and Sciortino}}]{rovigatti2013computing}
\bibinfo{author}{\bibfnamefont{L.}~\bibnamefont{Rovigatti}},
  \bibinfo{author}{\bibfnamefont{D.}~\bibnamefont{de~las Heras}},
  \bibinfo{author}{\bibfnamefont{J.~M.} \bibnamefont{Tavares}},
  \bibinfo{author}{\bibfnamefont{M.~M.} \bibnamefont{Telo~da Gama}},
  \bibnamefont{and}
  \bibinfo{author}{\bibfnamefont{F.}~\bibnamefont{Sciortino}},
  \bibinfo{journal}{The Journal of chemical physics}
  \textbf{\bibinfo{volume}{138}}, \bibinfo{pages}{164904}
  (\bibinfo{year}{2013}).

\bibitem[{\citenamefont{Seiferling et~al.}(2016)\citenamefont{Seiferling,
  de~Las~Heras, and Telo~da Gama}}]{seiferling2016percolation}
\bibinfo{author}{\bibfnamefont{F.}~\bibnamefont{Seiferling}},
  \bibinfo{author}{\bibfnamefont{D.}~\bibnamefont{de~Las~Heras}},
  \bibnamefont{and} \bibinfo{author}{\bibfnamefont{M.~M.} \bibnamefont{Telo~da
  Gama}}, \bibinfo{journal}{The Journal of Chemical Physics}
  \textbf{\bibinfo{volume}{145}}, \bibinfo{pages}{074903}
  (\bibinfo{year}{2016}).

\bibitem[{\citenamefont{Teixeira and Tavares}(2017)}]{teixeira2017phase}
\bibinfo{author}{\bibfnamefont{P.}~\bibnamefont{Teixeira}} \bibnamefont{and}
  \bibinfo{author}{\bibfnamefont{J.}~\bibnamefont{Tavares}},
  \bibinfo{journal}{Current Opinion in Colloid \& Interface Science}
  \textbf{\bibinfo{volume}{30}}, \bibinfo{pages}{16} (\bibinfo{year}{2017}).

\bibitem[{\citenamefont{Braz~Teixeira et~al.}(2021)\citenamefont{Braz~Teixeira,
  de~Las~Heras, Tavares, and Telo~da Gama}}]{braz2021phase}
\bibinfo{author}{\bibfnamefont{R.}~\bibnamefont{Braz~Teixeira}},
  \bibinfo{author}{\bibfnamefont{D.}~\bibnamefont{de~Las~Heras}},
  \bibinfo{author}{\bibfnamefont{J.~M.} \bibnamefont{Tavares}},
  \bibnamefont{and} \bibinfo{author}{\bibfnamefont{M.~M.} \bibnamefont{Telo~da
  Gama}}, \bibinfo{journal}{The Journal of Chemical Physics}
  \textbf{\bibinfo{volume}{155}}, \bibinfo{pages}{044903}
  (\bibinfo{year}{2021}).

\bibitem[{\citenamefont{Nykypanchuk et~al.}(2008)\citenamefont{Nykypanchuk,
  Maye, Van Der~Lelie, and Gang}}]{nykypanchuk2008dna}
\bibinfo{author}{\bibfnamefont{D.}~\bibnamefont{Nykypanchuk}},
  \bibinfo{author}{\bibfnamefont{M.~M.} \bibnamefont{Maye}},
  \bibinfo{author}{\bibfnamefont{D.}~\bibnamefont{Van Der~Lelie}},
  \bibnamefont{and} \bibinfo{author}{\bibfnamefont{O.}~\bibnamefont{Gang}},
  \bibinfo{journal}{Nature} \textbf{\bibinfo{volume}{451}},
  \bibinfo{pages}{549} (\bibinfo{year}{2008}).

\bibitem[{\citenamefont{Park et~al.}(2008)\citenamefont{Park, Lytton-Jean, Lee,
  Weigand, Schatz, and Mirkin}}]{park2008dna}
\bibinfo{author}{\bibfnamefont{S.~Y.} \bibnamefont{Park}},
  \bibinfo{author}{\bibfnamefont{A.~K.} \bibnamefont{Lytton-Jean}},
  \bibinfo{author}{\bibfnamefont{B.}~\bibnamefont{Lee}},
  \bibinfo{author}{\bibfnamefont{S.}~\bibnamefont{Weigand}},
  \bibinfo{author}{\bibfnamefont{G.~C.} \bibnamefont{Schatz}},
  \bibnamefont{and} \bibinfo{author}{\bibfnamefont{C.~A.}
  \bibnamefont{Mirkin}}, \bibinfo{journal}{Nature}
  \textbf{\bibinfo{volume}{451}}, \bibinfo{pages}{553} (\bibinfo{year}{2008}).

\bibitem[{\citenamefont{Liu et~al.}(2016)\citenamefont{Liu, Tagawa, Xin, Wang,
  Emamy, Li, Yager, Starr, Tkachenko, and Gang}}]{liu2016diamond}
\bibinfo{author}{\bibfnamefont{W.}~\bibnamefont{Liu}},
  \bibinfo{author}{\bibfnamefont{M.}~\bibnamefont{Tagawa}},
  \bibinfo{author}{\bibfnamefont{H.~L.} \bibnamefont{Xin}},
  \bibinfo{author}{\bibfnamefont{T.}~\bibnamefont{Wang}},
  \bibinfo{author}{\bibfnamefont{H.}~\bibnamefont{Emamy}},
  \bibinfo{author}{\bibfnamefont{H.}~\bibnamefont{Li}},
  \bibinfo{author}{\bibfnamefont{K.~G.} \bibnamefont{Yager}},
  \bibinfo{author}{\bibfnamefont{F.~W.} \bibnamefont{Starr}},
  \bibinfo{author}{\bibfnamefont{A.~V.} \bibnamefont{Tkachenko}},
  \bibnamefont{and} \bibinfo{author}{\bibfnamefont{O.}~\bibnamefont{Gang}},
  \bibinfo{journal}{Science} \textbf{\bibinfo{volume}{351}},
  \bibinfo{pages}{582} (\bibinfo{year}{2016}).

\bibitem[{\citenamefont{He et~al.}(2020)\citenamefont{He, Gales, Ducrot, Gong,
  Yi, Sacanna, and Pine}}]{he2020colloidal}
\bibinfo{author}{\bibfnamefont{M.}~\bibnamefont{He}},
  \bibinfo{author}{\bibfnamefont{J.~P.} \bibnamefont{Gales}},
  \bibinfo{author}{\bibfnamefont{{\'E}.}~\bibnamefont{Ducrot}},
  \bibinfo{author}{\bibfnamefont{Z.}~\bibnamefont{Gong}},
  \bibinfo{author}{\bibfnamefont{G.-R.} \bibnamefont{Yi}},
  \bibinfo{author}{\bibfnamefont{S.}~\bibnamefont{Sacanna}}, \bibnamefont{and}
  \bibinfo{author}{\bibfnamefont{D.~J.} \bibnamefont{Pine}},
  \bibinfo{journal}{Nature} \textbf{\bibinfo{volume}{585}},
  \bibinfo{pages}{524} (\bibinfo{year}{2020}).

\bibitem[{\citenamefont{Ho et~al.}(1990)\citenamefont{Ho, Chan, and
  Soukoulis}}]{ho1990existence}
\bibinfo{author}{\bibfnamefont{K.}~\bibnamefont{Ho}},
  \bibinfo{author}{\bibfnamefont{C.~T.} \bibnamefont{Chan}}, \bibnamefont{and}
  \bibinfo{author}{\bibfnamefont{C.~M.} \bibnamefont{Soukoulis}},
  \bibinfo{journal}{Physical Review Letters} \textbf{\bibinfo{volume}{65}},
  \bibinfo{pages}{3152} (\bibinfo{year}{1990}).

\bibitem[{\citenamefont{Soukoulis and Wegener}(2010)}]{soukoulis2010optical}
\bibinfo{author}{\bibfnamefont{C.~M.} \bibnamefont{Soukoulis}}
  \bibnamefont{and} \bibinfo{author}{\bibfnamefont{M.}~\bibnamefont{Wegener}},
  \bibinfo{journal}{Science} \textbf{\bibinfo{volume}{330}},
  \bibinfo{pages}{1633} (\bibinfo{year}{2010}).

\bibitem[{\citenamefont{Ngo et~al.}(2006)\citenamefont{Ngo, Liddell,
  Ghebrebrhan, and Joannopoulos}}]{ngo2006tetrastack}
\bibinfo{author}{\bibfnamefont{T.}~\bibnamefont{Ngo}},
  \bibinfo{author}{\bibfnamefont{C.}~\bibnamefont{Liddell}},
  \bibinfo{author}{\bibfnamefont{M.}~\bibnamefont{Ghebrebrhan}},
  \bibnamefont{and}
  \bibinfo{author}{\bibfnamefont{J.}~\bibnamefont{Joannopoulos}},
  \bibinfo{journal}{Applied physics letters} \textbf{\bibinfo{volume}{88}},
  \bibinfo{pages}{241920} (\bibinfo{year}{2006}).

\bibitem[{\citenamefont{Romano et~al.}(2011)\citenamefont{Romano, Sanz, and
  Sciortino}}]{romano2011crystallization}
\bibinfo{author}{\bibfnamefont{F.}~\bibnamefont{Romano}},
  \bibinfo{author}{\bibfnamefont{E.}~\bibnamefont{Sanz}}, \bibnamefont{and}
  \bibinfo{author}{\bibfnamefont{F.}~\bibnamefont{Sciortino}},
  \bibinfo{journal}{The Journal of chemical physics}
  \textbf{\bibinfo{volume}{134}}, \bibinfo{pages}{174502}
  (\bibinfo{year}{2011}).

\bibitem[{\citenamefont{Neophytou et~al.}(2021)\citenamefont{Neophytou,
  Chakrabarti, and Sciortino}}]{neophytou2021facile}
\bibinfo{author}{\bibfnamefont{A.}~\bibnamefont{Neophytou}},
  \bibinfo{author}{\bibfnamefont{D.}~\bibnamefont{Chakrabarti}},
  \bibnamefont{and}
  \bibinfo{author}{\bibfnamefont{F.}~\bibnamefont{Sciortino}},
  \bibinfo{journal}{Proceedings of the National Academy of Sciences}
  \textbf{\bibinfo{volume}{118}} (\bibinfo{year}{2021}).

\bibitem[{\citenamefont{Rovigatti et~al.}(2022)\citenamefont{Rovigatti, Russo,
  Romano, Matthies, Kroc, and {\v{S}}ulc}}]{rovigatti2022simple}
\bibinfo{author}{\bibfnamefont{L.}~\bibnamefont{Rovigatti}},
  \bibinfo{author}{\bibfnamefont{J.}~\bibnamefont{Russo}},
  \bibinfo{author}{\bibfnamefont{F.}~\bibnamefont{Romano}},
  \bibinfo{author}{\bibfnamefont{M.}~\bibnamefont{Matthies}},
  \bibinfo{author}{\bibfnamefont{L.}~\bibnamefont{Kroc}}, \bibnamefont{and}
  \bibinfo{author}{\bibfnamefont{P.}~\bibnamefont{{\v{S}}ulc}},
  \bibinfo{journal}{arXiv preprint arXiv:2205.10680}  (\bibinfo{year}{2022}).

\bibitem[{\citenamefont{Smallenburg and
  Sciortino}(2013)}]{smallenburg2013limited}
\bibinfo{author}{\bibfnamefont{F.}~\bibnamefont{Smallenburg}} \bibnamefont{and}
  \bibinfo{author}{\bibfnamefont{F.}~\bibnamefont{Sciortino}},
  \bibinfo{journal}{Nature Physics} \textbf{\bibinfo{volume}{9}},
  \bibinfo{pages}{554} (\bibinfo{year}{2013}).

\bibitem[{\citenamefont{Smith et~al.}(1949)\citenamefont{Smith, Van~Ness,
  Abbott, and Swihart}}]{smith1949introduction}
\bibinfo{author}{\bibfnamefont{J.~M.} \bibnamefont{Smith}},
  \bibinfo{author}{\bibfnamefont{H.~C.} \bibnamefont{Van~Ness}},
  \bibinfo{author}{\bibfnamefont{M.~M.} \bibnamefont{Abbott}},
  \bibnamefont{and} \bibinfo{author}{\bibfnamefont{M.~T.}
  \bibnamefont{Swihart}}, \emph{\bibinfo{title}{Introduction to chemical
  engineering thermodynamics}} (\bibinfo{publisher}{McGraw-Hill Singapore},
  \bibinfo{year}{1949}).

\bibitem[{\citenamefont{Moore}(1962)}]{Moore1962Physical}
\bibinfo{author}{\bibfnamefont{W.~J.} \bibnamefont{Moore}},
  \emph{\bibinfo{title}{Physical Chemistry}}
  (\bibinfo{publisher}{Prentice-Hall}, \bibinfo{year}{1962}).

\bibitem[{\citenamefont{Stopper et~al.}(2020)\citenamefont{Stopper,
  Hansen-Goos, Roth, and Evans}}]{stopper2020remnants}
\bibinfo{author}{\bibfnamefont{D.}~\bibnamefont{Stopper}},
  \bibinfo{author}{\bibfnamefont{H.}~\bibnamefont{Hansen-Goos}},
  \bibinfo{author}{\bibfnamefont{R.}~\bibnamefont{Roth}}, \bibnamefont{and}
  \bibinfo{author}{\bibfnamefont{R.}~\bibnamefont{Evans}},
  \bibinfo{journal}{The Journal of Chemical Physics}
  \textbf{\bibinfo{volume}{152}}, \bibinfo{pages}{111101}
  (\bibinfo{year}{2020}).

\bibitem[{\citenamefont{Tavares and Teixeira}(2020)}]{tavares2020remnants}
\bibinfo{author}{\bibfnamefont{J.}~\bibnamefont{Tavares}} \bibnamefont{and}
  \bibinfo{author}{\bibfnamefont{P.}~\bibnamefont{Teixeira}},
  \bibinfo{journal}{The Journal of Chemical Physics}
  \textbf{\bibinfo{volume}{153}}, \bibinfo{pages}{086101}
  (\bibinfo{year}{2020}).

\bibitem[{\citenamefont{Rovigatti et~al.}(2018)\citenamefont{Rovigatti, Russo,
  and Romano}}]{rovigatti2018simulate}
\bibinfo{author}{\bibfnamefont{L.}~\bibnamefont{Rovigatti}},
  \bibinfo{author}{\bibfnamefont{J.}~\bibnamefont{Russo}}, \bibnamefont{and}
  \bibinfo{author}{\bibfnamefont{F.}~\bibnamefont{Romano}},
  \bibinfo{journal}{The European Physical Journal E}
  \textbf{\bibinfo{volume}{41}}, \bibinfo{pages}{59} (\bibinfo{year}{2018}).

\bibitem[{\citenamefont{Chen and Siepmann}(2000)}]{chen2000novel}
\bibinfo{author}{\bibfnamefont{B.}~\bibnamefont{Chen}} \bibnamefont{and}
  \bibinfo{author}{\bibfnamefont{J.~I.} \bibnamefont{Siepmann}},
  \bibinfo{journal}{The Journal of Physical Chemistry B}
  \textbf{\bibinfo{volume}{104}}, \bibinfo{pages}{8725} (\bibinfo{year}{2000}).

\bibitem[{\citenamefont{Tanaka et~al.}(2019)\citenamefont{Tanaka, Tong, Shi,
  and Russo}}]{tanaka2019revealing}
\bibinfo{author}{\bibfnamefont{H.}~\bibnamefont{Tanaka}},
  \bibinfo{author}{\bibfnamefont{H.}~\bibnamefont{Tong}},
  \bibinfo{author}{\bibfnamefont{R.}~\bibnamefont{Shi}}, \bibnamefont{and}
  \bibinfo{author}{\bibfnamefont{J.}~\bibnamefont{Russo}},
  \bibinfo{journal}{Nature Reviews Physics} \textbf{\bibinfo{volume}{1}},
  \bibinfo{pages}{333} (\bibinfo{year}{2019}).

\bibitem[{\citenamefont{Bohlin et~al.}(2022)\citenamefont{Bohlin, Matthies,
  Poppleton, Procyk, Mallya, Yan, and {\v{S}}ulc}}]{bohlin2022design}
\bibinfo{author}{\bibfnamefont{J.}~\bibnamefont{Bohlin}},
  \bibinfo{author}{\bibfnamefont{M.}~\bibnamefont{Matthies}},
  \bibinfo{author}{\bibfnamefont{E.}~\bibnamefont{Poppleton}},
  \bibinfo{author}{\bibfnamefont{J.}~\bibnamefont{Procyk}},
  \bibinfo{author}{\bibfnamefont{A.}~\bibnamefont{Mallya}},
  \bibinfo{author}{\bibfnamefont{H.}~\bibnamefont{Yan}}, \bibnamefont{and}
  \bibinfo{author}{\bibfnamefont{P.}~\bibnamefont{{\v{S}}ulc}},
  \bibinfo{journal}{Nature Protocols} pp. \bibinfo{pages}{1--27}
  (\bibinfo{year}{2022}).

\bibitem[{\citenamefont{Bol}(1982)}]{bol1982monte}
\bibinfo{author}{\bibfnamefont{W.}~\bibnamefont{Bol}},
  \bibinfo{journal}{Molecular Physics} \textbf{\bibinfo{volume}{45}},
  \bibinfo{pages}{605} (\bibinfo{year}{1982}).

\bibitem[{\citenamefont{Kern and Frenkel}(2003)}]{kern2003fluid}
\bibinfo{author}{\bibfnamefont{N.}~\bibnamefont{Kern}} \bibnamefont{and}
  \bibinfo{author}{\bibfnamefont{D.}~\bibnamefont{Frenkel}},
  \bibinfo{journal}{The Journal of chemical physics}
  \textbf{\bibinfo{volume}{118}}, \bibinfo{pages}{9882} (\bibinfo{year}{2003}).

\bibitem[{\citenamefont{Zhang and Glotzer}(2004)}]{zhang2004self}
\bibinfo{author}{\bibfnamefont{Z.}~\bibnamefont{Zhang}} \bibnamefont{and}
  \bibinfo{author}{\bibfnamefont{S.~C.} \bibnamefont{Glotzer}},
  \bibinfo{journal}{Nano Letters} \textbf{\bibinfo{volume}{4}},
  \bibinfo{pages}{1407} (\bibinfo{year}{2004}).

\bibitem[{\citenamefont{Pawar and Kretzschmar}(2010)}]{pawar2010fabrication}
\bibinfo{author}{\bibfnamefont{A.~B.} \bibnamefont{Pawar}} \bibnamefont{and}
  \bibinfo{author}{\bibfnamefont{I.}~\bibnamefont{Kretzschmar}},
  \bibinfo{journal}{Macromolecular rapid communications}
  \textbf{\bibinfo{volume}{31}}, \bibinfo{pages}{150} (\bibinfo{year}{2010}).

\bibitem[{\citenamefont{Bianchi et~al.}(2011)\citenamefont{Bianchi, Blaak, and
  Likos}}]{bianchi2011patchy}
\bibinfo{author}{\bibfnamefont{E.}~\bibnamefont{Bianchi}},
  \bibinfo{author}{\bibfnamefont{R.}~\bibnamefont{Blaak}}, \bibnamefont{and}
  \bibinfo{author}{\bibfnamefont{C.~N.} \bibnamefont{Likos}},
  \bibinfo{journal}{Physical Chemistry Chemical Physics}
  \textbf{\bibinfo{volume}{13}}, \bibinfo{pages}{6397} (\bibinfo{year}{2011}).

\bibitem[{\citenamefont{Romano and Sciortino}(2011)}]{romano2011colloidal}
\bibinfo{author}{\bibfnamefont{F.}~\bibnamefont{Romano}} \bibnamefont{and}
  \bibinfo{author}{\bibfnamefont{F.}~\bibnamefont{Sciortino}},
  \bibinfo{journal}{Nature materials} \textbf{\bibinfo{volume}{10}},
  \bibinfo{pages}{171} (\bibinfo{year}{2011}).

\bibitem[{\citenamefont{van Anders et~al.}(2013)\citenamefont{van Anders,
  Ahmed, Smith, Engel, and Glotzer}}]{van2013entropically}
\bibinfo{author}{\bibfnamefont{G.}~\bibnamefont{van Anders}},
  \bibinfo{author}{\bibfnamefont{N.~K.} \bibnamefont{Ahmed}},
  \bibinfo{author}{\bibfnamefont{R.}~\bibnamefont{Smith}},
  \bibinfo{author}{\bibfnamefont{M.}~\bibnamefont{Engel}}, \bibnamefont{and}
  \bibinfo{author}{\bibfnamefont{S.~C.} \bibnamefont{Glotzer}},
  \bibinfo{journal}{Acs Nano} \textbf{\bibinfo{volume}{8}},
  \bibinfo{pages}{931} (\bibinfo{year}{2013}).

\bibitem[{\citenamefont{Vasilyev et~al.}(2015)\citenamefont{Vasilyev, Klumov,
  and Tkachenko}}]{vasilyev2015chromatic}
\bibinfo{author}{\bibfnamefont{O.~A.} \bibnamefont{Vasilyev}},
  \bibinfo{author}{\bibfnamefont{B.~A.} \bibnamefont{Klumov}},
  \bibnamefont{and} \bibinfo{author}{\bibfnamefont{A.~V.}
  \bibnamefont{Tkachenko}}, \bibinfo{journal}{Physical Review E}
  \textbf{\bibinfo{volume}{92}}, \bibinfo{pages}{012308}
  (\bibinfo{year}{2015}).

\bibitem[{\citenamefont{Xiong et~al.}(2020)\citenamefont{Xiong, Yang, Tian,
  Michelson, Xiang, Xin, and Gang}}]{xiong2020three}
\bibinfo{author}{\bibfnamefont{Y.}~\bibnamefont{Xiong}},
  \bibinfo{author}{\bibfnamefont{S.}~\bibnamefont{Yang}},
  \bibinfo{author}{\bibfnamefont{Y.}~\bibnamefont{Tian}},
  \bibinfo{author}{\bibfnamefont{A.}~\bibnamefont{Michelson}},
  \bibinfo{author}{\bibfnamefont{S.}~\bibnamefont{Xiang}},
  \bibinfo{author}{\bibfnamefont{H.}~\bibnamefont{Xin}}, \bibnamefont{and}
  \bibinfo{author}{\bibfnamefont{O.}~\bibnamefont{Gang}}, \bibinfo{journal}{ACS
  nano} \textbf{\bibinfo{volume}{14}}, \bibinfo{pages}{6823}
  (\bibinfo{year}{2020}).

\bibitem[{\citenamefont{Rothemund}(2006)}]{Rothemund2006}
\bibinfo{author}{\bibfnamefont{P.~W.~K.} \bibnamefont{Rothemund}},
  \bibinfo{journal}{Nature} \textbf{\bibinfo{volume}{440}},
  \bibinfo{pages}{297} (\bibinfo{year}{2006}), ISSN \bibinfo{issn}{1476-4687},
  \urlprefix\url{https://doi.org/10.1038/nature04586}.

\bibitem[{\citenamefont{Zhang et~al.}(2018)\citenamefont{Zhang, Hartl, Frank,
  Heuer-Jungemann, Fischer, Nickels, Nickel, and Liedl}}]{zhang20183d}
\bibinfo{author}{\bibfnamefont{T.}~\bibnamefont{Zhang}},
  \bibinfo{author}{\bibfnamefont{C.}~\bibnamefont{Hartl}},
  \bibinfo{author}{\bibfnamefont{K.}~\bibnamefont{Frank}},
  \bibinfo{author}{\bibfnamefont{A.}~\bibnamefont{Heuer-Jungemann}},
  \bibinfo{author}{\bibfnamefont{S.}~\bibnamefont{Fischer}},
  \bibinfo{author}{\bibfnamefont{P.~C.} \bibnamefont{Nickels}},
  \bibinfo{author}{\bibfnamefont{B.}~\bibnamefont{Nickel}}, \bibnamefont{and}
  \bibinfo{author}{\bibfnamefont{T.}~\bibnamefont{Liedl}},
  \bibinfo{journal}{Advanced Materials} \textbf{\bibinfo{volume}{30}},
  \bibinfo{pages}{1800273} (\bibinfo{year}{2018}).

\bibitem[{\citenamefont{Tian et~al.}(2020)\citenamefont{Tian, Lhermitte, Bai,
  Vo, Xin, Li, Li, Fukuto, Yager, Kahn et~al.}}]{tian2020ordered}
\bibinfo{author}{\bibfnamefont{Y.}~\bibnamefont{Tian}},
  \bibinfo{author}{\bibfnamefont{J.~R.} \bibnamefont{Lhermitte}},
  \bibinfo{author}{\bibfnamefont{L.}~\bibnamefont{Bai}},
  \bibinfo{author}{\bibfnamefont{T.}~\bibnamefont{Vo}},
  \bibinfo{author}{\bibfnamefont{H.~L.} \bibnamefont{Xin}},
  \bibinfo{author}{\bibfnamefont{H.}~\bibnamefont{Li}},
  \bibinfo{author}{\bibfnamefont{R.}~\bibnamefont{Li}},
  \bibinfo{author}{\bibfnamefont{M.}~\bibnamefont{Fukuto}},
  \bibinfo{author}{\bibfnamefont{K.~G.} \bibnamefont{Yager}},
  \bibinfo{author}{\bibfnamefont{J.~S.} \bibnamefont{Kahn}},
  \bibnamefont{et~al.}, \bibinfo{journal}{Nature materials}
  \textbf{\bibinfo{volume}{19}}, \bibinfo{pages}{789} (\bibinfo{year}{2020}).

\bibitem[{\citenamefont{Chakraborty et~al.}(2022)\citenamefont{Chakraborty,
  Pearce, Verweij, Matysik, Giomi, and Kraft}}]{chakraborty2022self}
\bibinfo{author}{\bibfnamefont{I.}~\bibnamefont{Chakraborty}},
  \bibinfo{author}{\bibfnamefont{D.~J.} \bibnamefont{Pearce}},
  \bibinfo{author}{\bibfnamefont{R.~W.} \bibnamefont{Verweij}},
  \bibinfo{author}{\bibfnamefont{S.~C.} \bibnamefont{Matysik}},
  \bibinfo{author}{\bibfnamefont{L.}~\bibnamefont{Giomi}}, \bibnamefont{and}
  \bibinfo{author}{\bibfnamefont{D.~J.} \bibnamefont{Kraft}},
  \bibinfo{journal}{ACS nano} \textbf{\bibinfo{volume}{16}},
  \bibinfo{pages}{2471} (\bibinfo{year}{2022}).

\bibitem[{\citenamefont{Suzuki et~al.}(2009)\citenamefont{Suzuki, Hosokawa, and
  Maeda}}]{suzuki2009controlling}
\bibinfo{author}{\bibfnamefont{K.}~\bibnamefont{Suzuki}},
  \bibinfo{author}{\bibfnamefont{K.}~\bibnamefont{Hosokawa}}, \bibnamefont{and}
  \bibinfo{author}{\bibfnamefont{M.}~\bibnamefont{Maeda}},
  \bibinfo{journal}{Journal of the American Chemical Society}
  \textbf{\bibinfo{volume}{131}}, \bibinfo{pages}{7518} (\bibinfo{year}{2009}).

\bibitem[{\citenamefont{Kim et~al.}(2011)\citenamefont{Kim, Kim, and
  Deaton}}]{kim2011dna}
\bibinfo{author}{\bibfnamefont{J.-W.} \bibnamefont{Kim}},
  \bibinfo{author}{\bibfnamefont{J.-H.} \bibnamefont{Kim}}, \bibnamefont{and}
  \bibinfo{author}{\bibfnamefont{R.}~\bibnamefont{Deaton}},
  \bibinfo{journal}{Angewandte Chemie International Edition}
  \textbf{\bibinfo{volume}{50}}, \bibinfo{pages}{9185} (\bibinfo{year}{2011}).

\bibitem[{\citenamefont{Wang et~al.}(2012)\citenamefont{Wang, Wang, Breed,
  Manoharan, Feng, Hollingsworth, Weck, and Pine}}]{wang2012colloids}
\bibinfo{author}{\bibfnamefont{Y.}~\bibnamefont{Wang}},
  \bibinfo{author}{\bibfnamefont{Y.}~\bibnamefont{Wang}},
  \bibinfo{author}{\bibfnamefont{D.~R.} \bibnamefont{Breed}},
  \bibinfo{author}{\bibfnamefont{V.~N.} \bibnamefont{Manoharan}},
  \bibinfo{author}{\bibfnamefont{L.}~\bibnamefont{Feng}},
  \bibinfo{author}{\bibfnamefont{A.~D.} \bibnamefont{Hollingsworth}},
  \bibinfo{author}{\bibfnamefont{M.}~\bibnamefont{Weck}}, \bibnamefont{and}
  \bibinfo{author}{\bibfnamefont{D.~J.} \bibnamefont{Pine}},
  \bibinfo{journal}{Nature} \textbf{\bibinfo{volume}{491}}, \bibinfo{pages}{51}
  (\bibinfo{year}{2012}).

\bibitem[{\citenamefont{Feng et~al.}(2013)\citenamefont{Feng, Dreyfus, Sha,
  Seeman, and Chaikin}}]{feng2013dna}
\bibinfo{author}{\bibfnamefont{L.}~\bibnamefont{Feng}},
  \bibinfo{author}{\bibfnamefont{R.}~\bibnamefont{Dreyfus}},
  \bibinfo{author}{\bibfnamefont{R.}~\bibnamefont{Sha}},
  \bibinfo{author}{\bibfnamefont{N.~C.} \bibnamefont{Seeman}},
  \bibnamefont{and} \bibinfo{author}{\bibfnamefont{P.~M.}
  \bibnamefont{Chaikin}}, \bibinfo{journal}{Advanced Materials}
  \textbf{\bibinfo{volume}{25}}, \bibinfo{pages}{2779} (\bibinfo{year}{2013}).

\bibitem[{\citenamefont{Gong et~al.}(2017)\citenamefont{Gong, Hueckel, Yi, and
  Sacanna}}]{gong2017patchy}
\bibinfo{author}{\bibfnamefont{Z.}~\bibnamefont{Gong}},
  \bibinfo{author}{\bibfnamefont{T.}~\bibnamefont{Hueckel}},
  \bibinfo{author}{\bibfnamefont{G.-R.} \bibnamefont{Yi}}, \bibnamefont{and}
  \bibinfo{author}{\bibfnamefont{S.}~\bibnamefont{Sacanna}},
  \bibinfo{journal}{Nature} \textbf{\bibinfo{volume}{550}},
  \bibinfo{pages}{234} (\bibinfo{year}{2017}).

\bibitem[{\citenamefont{SantaLucia}(1998)}]{santalucia1998unified}
\bibinfo{author}{\bibfnamefont{J.}~\bibnamefont{SantaLucia}},
  \bibinfo{journal}{Proceedings of the National Academy of Sciences}
  \textbf{\bibinfo{volume}{95}}, \bibinfo{pages}{1460} (\bibinfo{year}{1998}).

\bibitem[{\citenamefont{Mansoori et~al.}(1971)\citenamefont{Mansoori, Carnahan,
  Starling, and Leland~Jr}}]{mansoori1971equilibrium}
\bibinfo{author}{\bibfnamefont{G.}~\bibnamefont{Mansoori}},
  \bibinfo{author}{\bibfnamefont{N.~F.} \bibnamefont{Carnahan}},
  \bibinfo{author}{\bibfnamefont{K.}~\bibnamefont{Starling}}, \bibnamefont{and}
  \bibinfo{author}{\bibfnamefont{T.}~\bibnamefont{Leland~Jr}},
  \bibinfo{journal}{The Journal of Chemical Physics}
  \textbf{\bibinfo{volume}{54}}, \bibinfo{pages}{1523} (\bibinfo{year}{1971}).

\bibitem[{\citenamefont{Nezbeda et~al.}(1989)\citenamefont{Nezbeda, Kolafa, and
  Kalyuzhnyi}}]{nezbeda1989primitive}
\bibinfo{author}{\bibfnamefont{I.}~\bibnamefont{Nezbeda}},
  \bibinfo{author}{\bibfnamefont{J.}~\bibnamefont{Kolafa}}, \bibnamefont{and}
  \bibinfo{author}{\bibfnamefont{Y.~V.} \bibnamefont{Kalyuzhnyi}},
  \bibinfo{journal}{Molecular physics} \textbf{\bibinfo{volume}{68}},
  \bibinfo{pages}{143} (\bibinfo{year}{1989}).

\bibitem[{\citenamefont{Sciortino et~al.}(2007)\citenamefont{Sciortino,
  Bianchi, Douglas, and Tartaglia}}]{sciortino2007self}
\bibinfo{author}{\bibfnamefont{F.}~\bibnamefont{Sciortino}},
  \bibinfo{author}{\bibfnamefont{E.}~\bibnamefont{Bianchi}},
  \bibinfo{author}{\bibfnamefont{J.~F.} \bibnamefont{Douglas}},
  \bibnamefont{and}
  \bibinfo{author}{\bibfnamefont{P.}~\bibnamefont{Tartaglia}},
  \bibinfo{journal}{The Journal of chemical physics}
  \textbf{\bibinfo{volume}{126}}, \bibinfo{pages}{194903}
  (\bibinfo{year}{2007}).

\bibitem[{\citenamefont{Deiters}(2017)}]{deiters2017differential}
\bibinfo{author}{\bibfnamefont{U.~K.} \bibnamefont{Deiters}},
  \bibinfo{journal}{Fluid Phase Equilibria} \textbf{\bibinfo{volume}{447}},
  \bibinfo{pages}{72} (\bibinfo{year}{2017}).

\bibitem[{\citenamefont{Bell and Deiters}(2018)}]{bell2018construction}
\bibinfo{author}{\bibfnamefont{I.~H.} \bibnamefont{Bell}} \bibnamefont{and}
  \bibinfo{author}{\bibfnamefont{U.~K.} \bibnamefont{Deiters}},
  \bibinfo{journal}{AIChE Journal} \textbf{\bibinfo{volume}{64}},
  \bibinfo{pages}{2745} (\bibinfo{year}{2018}).

\bibitem[{\citenamefont{Panagiotopoulos}(1987)}]{panagiotopoulos1987direct}
\bibinfo{author}{\bibfnamefont{A.~Z.} \bibnamefont{Panagiotopoulos}},
  \bibinfo{journal}{Molecular Physics} \textbf{\bibinfo{volume}{61}},
  \bibinfo{pages}{813} (\bibinfo{year}{1987}).

\bibitem[{\citenamefont{Panagiotopoulos
  et~al.}(1988)\citenamefont{Panagiotopoulos, Quirke, Stapleton, and
  Tildesley}}]{panagiotopoulos1988phase}
\bibinfo{author}{\bibfnamefont{A.~Z.} \bibnamefont{Panagiotopoulos}},
  \bibinfo{author}{\bibfnamefont{N.}~\bibnamefont{Quirke}},
  \bibinfo{author}{\bibfnamefont{M.}~\bibnamefont{Stapleton}},
  \bibnamefont{and}
  \bibinfo{author}{\bibfnamefont{D.}~\bibnamefont{Tildesley}},
  \bibinfo{journal}{Molecular Physics} \textbf{\bibinfo{volume}{63}},
  \bibinfo{pages}{527} (\bibinfo{year}{1988}).

\bibitem[{\citenamefont{Fornace et~al.}(2022)\citenamefont{Fornace, Huang,
  Newman, Porubsky, Pierce, and Pierce}}]{fornace2022nupack}
\bibinfo{author}{\bibfnamefont{M.~E.} \bibnamefont{Fornace}},
  \bibinfo{author}{\bibfnamefont{J.}~\bibnamefont{Huang}},
  \bibinfo{author}{\bibfnamefont{C.~T.} \bibnamefont{Newman}},
  \bibinfo{author}{\bibfnamefont{N.~J.} \bibnamefont{Porubsky}},
  \bibinfo{author}{\bibfnamefont{M.~B.} \bibnamefont{Pierce}},
  \bibnamefont{and} \bibinfo{author}{\bibfnamefont{N.~A.} \bibnamefont{Pierce}}
  (\bibinfo{year}{2022}).

\end{thebibliography}
\end{document}